\pdfoutput=1
\documentclass[PAPER, atlasdraft=false, cernpreprint, USenglish, texmf, orcidlogo]{atlasdoc}
\usepackage{atlaspackage}
\usepackage{atlasbiblatex}

\usepackage[hion=true]{atlasphysics}
\graphicspath{{logos/}{figures/}}

\usepackage{ANA-HION-2025-07-PAPER-defs}

\AtlasTitle{Differential measurements of $\gamma\gamma\to\tau\tau$ and constraints on $\tau$-lepton electromagnetic moments in Pb+Pb collisions at $\sqrt{s_{_\text{NN}}} = 5.02$ TeV with ATLAS}

\AtlasAbstract{
This paper presents the first differential fiducial measurements of $\gamma\gamma\to\tau\tau$ using 1.93~nb$^{-1}$ of Pb+Pb data at $\sqrt{s_{_\text{NN}}} = 5.02$ TeV recorded by the ATLAS detector. Events in which one of the $\tau$-leptons decays into a muon and two neutrinos $\tau\to\nu_\tau\bar{\nu}_\mu\mu$ are selected and are categorized into three regions by the presence of an electron or either one or three charged-particle track(s) from the second $\tau$-lepton decay. The measurement is performed in events where both Pb ions remain intact and no neutrons are emitted. Differential cross-sections are measured for seven variables in three fiducial regions at particle level. The measurements are compared to theory predictions with different photon flux models and spin correlation effects. For the fiducial region with one muon and one electron in the final state, comparisons to next-to-leading-order electroweak predictions are also made. The transverse momentum ($p_\text{T}$) of the decay muon, the $p_\text{T}$ of the visible decay particles of the other $\tau$-lepton, the total $p_\text{T}$, invariant mass, and pseudorapidity of the visible particles from the di-$\tau$ system, and the rapidity and acoplanarity of the visible decay particles from either $\tau$-lepton are measured. A maximum-likelihood fit to the muon transverse-momentum distributions in the three regions before unfolding is performed to extract the $\tau$-lepton anomalous magnetic moment $a_{\tau}$ and electric dipole moment $d_{\tau}$, the latter for the first time in heavy ion collisions. The observed 95\% confidence level intervals are $-0.057 <a_{\tau}< 0.035$ and $|d_{\tau}|< 2.7 \times 10^{-16}~e\text{cm}$.}

\AtlasRefCode{HION-2025-07}

\PreprintIdNumber{CERN-EP-2026-135}

\AtlasJournal{JHEP}
\AtlasJournalRef{\JHEP 10 (2026) 053}
\AtlasDOI{10.1007/JHEP10(2026)053}
\hypersetup{pdftitle={ATLAS document},pdfauthor={The ATLAS Collaboration}}

\begin{document}

\maketitle

\section{Introduction}

Photon-induced processes occur in ultra-relativistic heavy-ion collisions at the LHC, in which the ion beams can be described as being accompanied by a large equivalent photon flux~\cite{BreitWheeler1934,HeisenbergEuler1936,SchwingerLimit}. For photons that are emitted coherently by an entire nucleus, the flux at matrix-element level is enhanced by a factor of $Z$ compared with the emission from a proton, resulting in a cross-section enhancement by $Z^2$ for single-photon processes, and $Z^4$ for di-photon processes, where $Z$ is the atomic number and $Z=82$ for lead. Depending on the closest distance of approach between the two nuclear centers-of-mass, defined as the impact parameter, photon-induced processes can become the dominant interaction mechanism. This is realized for impact parameters above twice the nuclear radius where hadronic interactions are suppressed. These events, referred to as ultra-peripheral collisions (UPC), can be used to study photon-photon and photonuclear interactions~\cite{Klein:2020fmr}. They typically have features such as regions of pseudorapidity devoid of produced particles (typically referred-to as "rapidity gaps"), the absence of neutrons in one or both beam directions, and low particle multiplicity or exclusive final states. These make them qualitatively different from typical high multiplicity inelastic nuclear collisions, where hadronic interactions occur~\cite{Baltz:2007kq,deFavereaudeJeneret:2009db}.

\begin{figure}[b]
\begin{center}
\includegraphics[width=0.33\textwidth]{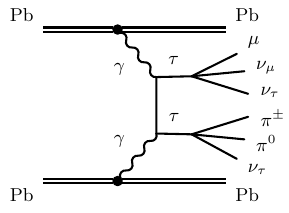}%
\includegraphics[width=0.33\textwidth]{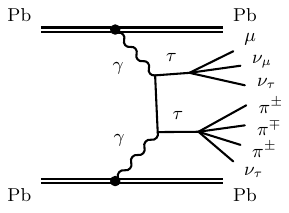}%
\includegraphics[width=0.33\textwidth]{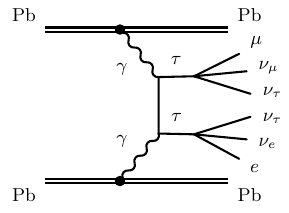}
\caption{Schematic diagrams of \ggtt production in UPC Pb+Pb collisions, with the $\tau$-leptons decaying into (left) one muon, one charged pion $\pi^{\pm}$, one neutral pion $\pi^{0}$ and neutrinos, (middle) one muon, three charged pions $\pi^{\pm}$ and neutrinos and (right) one muon, one electron and neutrinos.}
\label{fig:sig_diagram}
\end{center}
\end{figure}

Photon-induced processes can give rise to the production of oppositely-charged \tlep pairs, denoted by \ggtt\cite{Budnev:1974de,PhysRevD.7.3485}. In the Standard Model (SM) of particle physics, this process can be calculated using purely quantum electrodynamics (QED) at leading-order, and enables the measurement of electromagnetic properties of \tleps. Leading-order diagrams for \ggtt production and decay in lead-lead~(\PbPb) collisions, as considered in this analysis,  are shown in Figure~\ref{fig:sig_diagram}. The presence of $\gamma\tau\tau$ vertices in these diagrams provides sensitivity to the electromagnetic moments of the \tlep, corresponding to the $\tau$ anomalous magnetic moment \atau$= \frac{1}{2}(g_\tau-2)$, where $g_\tau$ is the \tlep g-factor, and electric dipole moment \dtau. In the SM, small non-zero values are predicted for \atau and \dtau due to higher order corrections to the calculations, resulting in a predicted value of $\atau = 0.001\,177\,21~(5)$~\cite{Eidelman:2007sb} and a limit on $|\dtau|\lesssim 10^{-37}e\text{cm}$~\cite{dtauTheo}. Measuring these moments with improved precision fundamentally tests the SM and can be sensitive to physics beyond the Standard Model (BSM). Some BSM models relate coupling strengths to the mass of the respective lepton, suggesting that their impact could be largest for the \tlep, compared to muons and electrons. A value of \dtau larger than predicted by the SM could indicate a new source of symmetry violation under charge-parity ($CP$) conjugation, in this case in the leptonic sector, which is of great interest for explaining the matter-antimatter-asymmetry in the universe~\cite{Sakharov:1967dj}. Models involving unparticles~\cite{AtauUnparticles} or leptoquarks~\cite{Iguro:2023rom} could produce enhancements to \atau or \dtau, or both.

At the LHC, the \ggtt process was first observed in \PbPb collisions by the ATLAS~\cite{STDM-2019-19} and CMS~\cite{CMS-HIN-21-009} Collaborations and in proton--proton ($pp$) collisions by the CMS Collaboration~\cite{CMS:2024qjo}. The integrated \ggtt fiducial cross-section was measured by the CMS Collaboration in \PbPb collisions~\cite{CMS-HIN-21-009} and $pp$ collisions~\cite{CMS:2024qjo}, and by the DELPHI~\cite{Abdallah:2003xd} and L3~\cite{L3:1997acq, L3:2004ful} Collaborations in electron-positron ($e^+e^-$) collisions at the Large Electron Positron~(LEP) collider. Both ATLAS and CMS set constraints on \atau using \PbPb collisions~\cite{STDM-2019-19, CMS-HIN-21-009} and the CMS Collaboration set constraints on \atau and \dtau using $pp$ collisions~\cite{CMS:2024qjo}. Most recently, ATLAS tested BSM contributions to the photon-\tlep interaction, using an effective field theory (EFT) approach in Drell--Yan (DY) production, $pp\to\gamma^*/Z\to\tau\tau$~\cite{EXOT-2022-42}. However, this does not directly probe the SM \atau component, and the SM \dtau component is far beyond current measurement sensitivities. At LEP, the DELPHI~\cite{Abdallah:2003xd}, OPAL~\cite{Ackerstaff:1998mt} and L3~\cite{Acciarri:1998iv,L3:2004ful} Collaborations set constraints on \atau and \dtau. For \dtau, constraints have also been set by the Belle~\cite{Belle:2021ybo} and ARGUS Collaborations~\cite{ARGUS:2000riz}.

Various strategies to measure \ggtt production using the high photon fluxes generated at the LHC have been proposed for \PbPb collisions~\cite{DELAGUILA1991256,Beresford:2019gww,Dyndal:2020yen,Bhide:2024nje,Shao:2023bga,Burmasov:2023cwv,Verducci:2023cgx,Burmasov:2022par,Goncalves:2020btj}, and $pp$ collisions~\cite{Atag:2010ja,Beresford:2024dsc}. In general, the structure and energy of the emitting particles, as well as the impact parameter between them influence the kinematics of the interaction. Measurements of \ggtt production in LHC Pb+Pb collisions cover a complementary phase space to LHC $pp$ collisions. The Pb+Pb measurements provide sensitivity in the low di-$\tau$ mass region, below $\sim$50~\GeV, due to the coherent photon-flux enhancement and lower trigger thresholds, resulting in a larger cross-section, and enabling greater precision on the cross-section measurement in that region. For the extraction of the electromagnetic moments, the phase space of \ggtt production is typically restricted to ensure photon virtualities $q^2\sim 0~\GeV^2$, since the electromagnetic moments are formally defined in the limit $q^2\to 0~\GeV^2$. Alternative approaches for extracting \atau and \dtau, not using \ggtt, have been proposed for $pp$ collisions~\cite{Haisch:2023upo,Galon:2016ngp,Cao2021}, $e^+e^-$ collisions~\cite{BERNABEU2008160,TauSpinCorrsInee,Eidelman:2016aih}, and for fixed target experiments with bent crystals~\cite{Fomin:2018ybj,BentCrystal}. Some of these have $q^2\gg0~\GeV^2$, e.g. in DY production, which means that the SM form factors are suppressed. Different processes and beam types typically provide statistically independent datasets with interesting future opportunities for combining their respective constraints on \tlep electromagnetic moments.

This paper presents differential fiducial cross-section measurements of photon-induced \tlep pair production \ggtt as well as constraints on the \tlep \atau and \dtau. The results are obtained using UPC lead-lead collisions recorded in 2015 and 2018 at $\sqrt{s_{_\text{NN}}} = 5.02$~\TeV. The dataset corresponds to an integrated luminosity of 1.93 nb$^{-1}$. Events are categorized into three regions using topologies with one of the \tleps decaying into a muon and two neutrinos and by the presence of one electron, or either one or three charged-particle track(s) from the second \tlep decay. Events without forward neutron emissions are used to suppress photonuclear backgrounds and to reduce contributions from dissociative processes which typically have larger initial-state photon virtuality. The data are corrected for detector-related effects through an iterative Bayesian unfolding procedure, obtaining differential fiducial cross-sections at particle-level for seven kinematic variables of the visible \tlep decay products for each of the three considered regions. Electromagnetic moments of the \tlep are extracted from maximum-likelihood fits to the muon transverse-momentum distributions in the three regions before unfolding is performed.


%
\section{ATLAS detector}

%
\newcommand{\AtlasCoordFootnote}{%
ATLAS uses a right-handed coordinate system with its origin at the nominal interaction point (IP)
in the center of the detector and the \(z\)-axis along the beam pipe.
The \(x\)-axis points from the IP to the center of the LHC ring,
and the \(y\)-axis points upwards.
Polar coordinates \((r,\phi)\) are used in the transverse plane,
\(\phi\) being the azimuthal angle around the \(z\)-axis.
The pseudorapidity is defined in terms of the polar angle \(\theta\) as \(\eta = -\ln \tan(\theta/2)\) and is equal to the rapidity
$ y = \frac{1}{2} \ln \left( \frac{E + p_z}{E - p_z} \right) $ in the relativistic limit.
Angular distance is measured in units of \(\Delta R \equiv \sqrt{(\Delta y)^{2} + (\Delta\phi)^{2}}\).}

The ATLAS experiment~\cite{PERF-2007-01,ATLAS-TDR-2010-19,PIX-2018-001} at the LHC is a multipurpose particle detector with a forward--backward symmetric cylindrical geometry.\footnote{\AtlasCoordFootnote} It consists of an inner tracking detector (ID) surrounded by a thin superconducting solenoid providing a \qty{2}{\tesla} axial magnetic field, electromagnetic and hadronic calorimeters, and a muon spectrometer (MS). The inner tracking detector covers the pseudorapidity range \(|\eta| < 2.5\). It consists of silicon pixel, silicon microstrip, and transition radiation tracking detectors. Lead/liquid-argon (LAr) sampling calorimeters provide electromagnetic (EM) energy measurements with high granularity within the region \(|\eta|< 3.2\). A steel/scintillator-tile hadronic calorimeter covers the central pseudorapidity range (\(|\eta| < 1.7\)). The endcap and forward regions are instrumented with LAr calorimeters for EM and hadronic energy measurements up to \(|\eta| = 4.9\). The muon spectrometer surrounds the calorimeters and is based on three large superconducting air-core toroidal magnets with eight coils each. The field integral of the toroids ranges between \num{2.0} and \qty{6.0}{\tesla\metre} across most of the detector. The muon spectrometer includes a system of precision tracking chambers up to \(|\eta| = 2.7\) and fast detectors for triggering up to \(|\eta| = 2.4\). For heavy-ion data taking, Zero Degree Calorimeters~(ZDCs)~\cite{ATLAS-TDR-ZDC} are installed at $z=\pm 140$~m from the interaction point. The ZDCs detect neutral particles in the forward region (approximately $|\eta|>8.3$), such as forward neutrons. Each ZDC detector consists of four modules, each with tungsten absorber read out by layers of vertical quartz rods. Minimum-bias Trigger Scintillators (MBTS) detectors situated at $z = \pm 3.56 $ m detect charged particles in the pseudorapidity range 2.07 $<|\eta|<$ 3.86. These detectors comprise two rings with azimuthal scintillator counters for the inner and outer rings.
The luminosity is measured mainly by the LUCID-2~\cite{LUCID2} detector, which is located close to the beam pipe, complemented with calorimeter and beam conditions systems for stability monitoring.

A two-level trigger system is used to select events~\cite{TRIG-2016-01,TRIG-2019-04}. The first-level trigger is implemented in hardware and uses a subset of the detector information to accept events. This is followed by a software-based trigger that reduces the accepted rate of complete events.

A software suite~\cite{ATL-SOFT-PUB-2021-001} is used in the simulation of collision events, in the reconstruction and analysis of real and simulated data, in detector operations, and in the trigger and data acquisition systems of the experiment.


%
\section{Data and Monte Carlo samples}
\label{sec:data-mc}

This analysis uses Pb+Pb data recorded at a center-of-mass energy of $\sqrt{s_{\text{NN}}} = 5.02$~\TeV during LHC Run 2. The data were recorded in 2015 and 2018, with an integrated luminosity of 0.49 nb$^{-1}$ and 1.44~nb$^{-1}$, respectively. The average number of hadronic interactions per bunch crossing was 0.0022 in 2015 and 0.003 in 2018. Only data that satisfy standard data-quality requirements~\cite{DAPR-2018-01} are included.

Monte Carlo~(MC) simulated signal events \ggtt were generated at leading-order in QED using the \Starlight~2.0~\cite{Klein:2016yzr} generator, interfaced with \PYTHIA[8.245]~\cite{Sjostrand:2014zea} to simulate final-state radiation (FSR) from the $\tau$-leptons, \textsc{Tauola}~\cite{Jadach:1993hs, Davidson:2010rw} to handle $\tau$-lepton decays and \textsc{Photos++} 3.61~\cite{Davidson:2010ew} to simulate FSR from the charged decay products of $\tau$-leptons (\varSampPF). The signal samples interfaced with \textsc{Tauola} are used as the nominal samples, and alternative signal samples employing \Pythia 8.245 itself, instead of \textsc{Tauola}, for \tlep decays (\varSampPFDecay) are used to estimate a systematic uncertainty from the modeling of \tlep decays. For the nominal and alternative signal samples, the di-$\tau$ invariant mass is required to be above 4 GeV and the \pt of the \tleps is required to be above 2 GeV at generator-level. A generator-level filter is also applied to select events with at least one charged particle with \pt above 3 GeV within $|\eta|<2.6$.

For the \ggtt process, it was shown in Ref.~\cite{Korchin:2025vzx} that the inclusion of full spin correlation effects in the \tlep pair production and decay can impact the kinematic distributions of the decay products at the level of a few percent, with the corresponding theoretical calculations being present in \textsc{TauSpinner}~2.1.2~\cite{Przedzinski:2018ett,Czyczula:2012ny}. The signal MC samples described above lack full spin correlation effects. The effect of full spin correlations in the \ggtt predictions was therefore evaluated using a generator-level \ggtt sample produced at leading-order in QED using \textsc{gamma-UPC} 1.0~\cite{gammaUPC} to simulate the photon flux with the charged form factor approach, combined with \MGNLO 3.5.1~\cite{Alwall:2014hca} to simulate the hard interaction process of \ggtt. This was interfaced with \textsc{Tauola} to handle the \tlep decays. The \textsc{TauSpinner} 2.1.2 package was then used to compute spin-weights on an event-by-event basis. Spin correlation bin-by-bin correction factors are then derived by taking the ratio of kinematic distributions of \tlep decay products with and without applying the spin-weights. The correction factors are computed for all fiducial regions and measured observables, and are found to be significant only for the one muon plus one track fiducial region $\mu$1T-FR defined in Section~\ref{sec:corrections}. The spin-weights applied for events in $\mu$1T-FR are normalized to the average spin-weight in that fiducial region, to remove the event selection bias on the spin-weight. The distribution of event-by-event spin-weights after normalization is approximately Gaussian, with a mean of 1.0 and standard deviation $\approx$ 0.2. The resulting bin-by-bin spin correlation correction factors are typically within 5\% of unity, with maximum deviations of up to 10\% observed only in a few bins.

One of the main sources of background is the \ggmmg process~\cite{HION-2016-02}. MC samples for this process are generated using the \Starlight generator, interfaced to \PYTHIA[8] to model FSR from the muons. If the FSR photon with the highest transverse momentum has $\pt > 2$~\GeV, a dedicated Next-to-Leading-Order (NLO) QED $\ggmm\gamma$ MC sample generated using \MGNLO2.9~\cite{Alwall:2014hca} interfaced to \PYTHIA[8] is used, instead of \Starlight interfaced to \PYTHIA[8]. This additional sample improves the modeling of high-\pt FSR photon emissions for cases where $\pt^\gamma \gtrsim \pt^\mu$. Since the version of \MGNLO used only generates initial-state photons from protons, the photon flux of this sample is reweighted to the one from \Starlight in \PbPb interactions. The reweighting is performed differentially in generator-level di-muon invariant mass and rapidity.

For the \ggtt and \ggmmg samples, the photon flux distribution is (further) reweighted to match that of ~\Superchic~3.05~\cite{Harland-Lang:2018iur}. In this case, \Superchic is indicated in the sample name, e.g., for the nominal signal sample as \nomSamp. This reweighting is motivated as \Starlight is known to underpredict the photon flux compared with previous measurements in Refs.~\cite{HION-2016-02,HION-2021-16}. The reweighting is performed differentially in generator-level di-$\tau$ or di-muon invariant mass and rapidity. In \Starlight, the photon flux is approximated using the Weizsäcker-Williams method~\cite{Fermi1, Fermi2, Weizsaecker, Williams, zolotorevMcDonald}, which approximates the electromagnetic field of a relativistic point charge as a cloud of virtual photons. To avoid interactions inside a nucleus that would lead to its breakup, the photon flux description in \Starlight is restricted to distances from the nucleus center larger than the geometric nuclear radius. To avoid physical collisions of the nuclei where hadronic interactions would dominate, the impact parameter of the collision is required to be larger than twice the geometric nuclear radius of the lead nuclei~\cite{Baltz:2009jk, Klein:2016yzr}. In contrast, \Superchic uses a nucleus charge form factor based on the Woods–Saxon distribution and a survival factor considering nucleon-nucleon interactions, where the survival factor is the probability of the large rapidity gaps in $\gamma\gamma\to X$ surviving the nucleus-nucleus interaction. Thus, for \Superchic the photon flux validity and impact parameter does not need to be restricted in this explicit way~\cite{Harland-Lang:2018iur}. For \Starlight and \Superchic MC samples, all possible configurations for Coulomb breakup of either nucleus (none, one or both), i.e., through additional electromagnetic interaction, is included.

Simulated samples for the \ggee process were generated using the \Starlight generator and \PYTHIA[8] to model FSR from the electrons. Dijet samples from the $\gamma\gamma\to q\bar{q}$ process with $q=u,d,s,c,b$ were generated using \PYTHIA[8]. Non-diffractive photonuclear events ($\gamma\text{A} \to X$) were generated using \Starlight interfaced with \textsc{Dpmjet-III}~\cite{Roesler:2000he}.

All MC samples described in this section which are not explicitly referred to as generator-level, were processed with a full detector simulation based on \GEANT~\cite{Agostinelli:2002hh, SOFT-2010-01}.


%
\section{Object definition}

Events that contain one \tlep decaying into a muon and two neutrinos and one \tlep decaying into a lepton or charged pion(s) or kaon(s) require the reconstruction of muons, electrons and charged-particle tracks from the detector signals. Additionally, photons and calorimeter topoclusters, defined below, are used to identify FSR and to reject hadronic backgrounds, respectively.

Muons are reconstructed from tracks in the ID and the MS\@. The muons must satisfy the dedicated ``LowPt''~\cite{MUON-2018-03} identification criteria that are optimized for increased efficiency and fake-rejection for muons with $\pt<$~5~\GeV. In addition, they are required to have transverse momentum $\pt>$~4~\GeV, $|\eta|<$ 2.4 and a transverse impact parameter relative to the measured beam-line position of $\left|d_0\right|<$~0.3~mm.

Electrons are reconstructed from tracks in the ID matched with EM clusters in the calorimeter. The EM clusters must be of good quality, meaning EM clusters with a large amount of energy from poorly functioning calorimeter cells are removed~\cite{EGAM-2018-01}, and a timing requirement is used to reject out-of-time candidates~\cite{HION-2019-08}. The electrons must satisfy the ``Loose''~\cite{EGAM-2018-01} likelihood-based identification criteria, based on shower-shape and track-quality variables. They must also satisfy $\pt>$ 4~\GeV, $|\eta|<$ 2.47, with the calorimeter transition region 1.37 $<|\eta|<$ 1.52 excluded, and $\left|d_0\right| < 0.5$~mm.

Photons are reconstructed from EM clusters in the calorimeter, and tracks in the ID to identify photon conversions to $e^+e^-$ pairs. The EM clusters and the reconstructed photons must both be of good quality. The photons must satisfy dedicated identification criteria defined in Ref.~\cite{HION-2019-08}, which are based on a neural network approach and are optimized for low-$E_T$ photons ($\et <$ 20~\GeV). They must also satisfy $\et>$ 1.5~\GeV and $|\eta|<2.37$ with the calorimeter transition region 1.37$<|\eta|<$1.52 excluded.

Charged-particle tracks are reconstructed from hit information in the ID~\cite{Akesson:2006sh, Cornelissen:2007aca}. The tracks must satisfy the ``Loose Primary'' criteria~\cite{ATL-PHYS-PUB-2015-051,PERF-2015-07,PERF-2015-08}, and must satisfy $\pt>$ 100~\MeV, $|\eta|<2.5$ and $\left|d_0\right|<$ 1.5~mm. When calculating track four-vectors and derived variables for tracks not associated with muons, electrons or photons, the particle is assumed to be a charged pion.

Three-dimensional collections of topologically connected calorimeter cells (``topoclusters'')~\cite{PERF-2014-07} are formed and must satisfy the energy significance criteria for individual cells described in Ref.~\cite{STDM-2011-01}. The topoclusters must satisfy $|\eta|<$ 4.9, $\pt>$ 1~\GeV for $|\eta|<$ 2.5 and $\pt>$ 0.1~\GeV for $2.5<|\eta|<4.9$. Additionally, topoclusters arising from noise bursts in the calorimeters or malfunctioning calorimeter cells are removed. The data-driven removal procedure involves studying two dimensional distributions of topocluster activity in $\eta$ and $\phi$ for each calorimeter layer. When calculating topocluster-related variables, the topoclusters are considered to be massless.


%
\section{Event selection}
\label{sec:event-selection}

In 2015, candidate events were recorded by triggers that require a muon track in the muon spectrometer (with no minimum \pt requirement) and a total transverse energy below 50~\GeV in the entire calorimeter. Additionally, in 2015 events were rejected if more than one hit was registered in the inner ring of the MBTS detectors on either side, and a well-reconstructed track with $\pt>$ 200~\MeV was required in the ID\@. In 2018, the trigger required at least one muon with $\pt>$ 4~\GeV, a total transverse energy below 50~\GeV in the entire calorimeter and a total transverse energy below 3~\GeV in the forward calorimeter on each side~\cite{ATL-DAQ-PUB-2019-001-customRef}. On average, muons lose roughly 3~\GeV of their energy while passing through the calorimeter system~\cite{MUON-2018-03}. To correct for differences between simulation and data, the reconstruction and identification efficiencies for electrons and muons and the trigger efficiencies for muons are measured using tag-and-probe methods that are similar to those described in Refs.~\cite{TRIG-2018-01,HION-2016-02,HION-2019-08}. The efficiency of the threshold of 50 GeV on the maximum total transverse energy accepted in the entire calorimeter is studied using a data-driven approach, and is found to be 100\%. Moreover, the trigger efficiency associated with the selection of the activity in the forward direction (MBTS in 2015 or the forward calorimeter in 2018) is evaluated in Ref.~\cite{HION-2019-08}, and is found to be above 98\%. The small inefficiency is accounted for in the analysis, along with its uncertainty. The triggers select $\sim6$\% of the signal events which pass the generator-level requirements including the filter requirements, described in Section~\ref{sec:data-mc}.

Three mutually exclusive \ggtt signal regions (SR) are defined. All SRs require exactly one muon, targeting events where one of the \tleps decays into a muon and two neutrinos, i.e., as \tmudecay. This requirement reduces dilepton and hadronic backgrounds. The events are then categorized according to the signature of the other \tlep decay. A summary of the SR selections is given in Table~\ref{tab:evt_sel}.

\begin{table}[tb]
\small
\centering
\renewcommand{\arraystretch}{1.1}
\begin{tabular}{llll}
\toprule
Region             & $\mu$1T-SR & $\mu$3T-SR & $\mu e$-SR  \\
\midrule\midrule
Trigger   & \multicolumn{3}{l}{Single-muon based}               \\
$0n0n$ topology              & \multicolumn{3}{l}{$E^{A,C}_{\text{ZDC}}<1~\TeV$ for data, $0n0n$ weights for MC}          \\
\midrule
$N_{\mu}^{\text{baseline}}$   & $= 1$ & $= 1$ & --- \\
$N_{\mu}$        & $= 1$ & $= 1$ & $= 1$ \\
$N_{e}$          & $= 0$ & $= 0$ & $= 1$ \\
$N_{\text{trk}}(\Delta R > 0.1$ from ${\mu})$   & $ = 1$ & $= 3$    & ---   \\
$N_{\text{trk}}(\Delta R > 0.1$ from ${\ell})$  & ---    & ---      & $= 0$   \\
$N_\text{topoclusters}^{\text{unmatched}}$                                         & $ = 0$  & $= 0$   & ---   \\
$\Sigma_{\mu,\text{trk(s)}\text{ or }e}$ charge             & $ = 0$  & $= 0 $  & $= 0$ \\
\pt of muon and track                           & $>$ 1~\GeV & ---   & ---  \\
\pt of muon, track and photon                    & $>$ 1~\GeV & ---   & ---  \\
\pt of muon, track and topocluster            & $>$ 1~\GeV & ---   & ---   \\
$m_{\text{trks}}$                                           & --- & $<$ 1.7~\GeV & ---  \\
$A_{\phi}^{\mu,\text{trk(s)}}$                              & $<$ 0.4 & $<$ 0.2 & ---  \\
\bottomrule
\end{tabular}
\caption{Summary of the signal region definitions. The $0n0n$ weights include the contribution from EM pileup. The notation trk(s) indicates a single track for the $\mu$1T-SR and the three track-system for the $\mu$3T-SR. The system \pt selections are not mutually exclusive, meaning an event may satisfy either one or both requirements.}
\label{tab:evt_sel}
\end{table}

The $\mu$1T-SR requires exactly one muon and exactly one track separated by $\Delta R> 0.1$ from the muon and with opposite charge to the muon. This SR targets cases where the second \tlep decays into one charged hadron, or leptonically, where the lepton has a \pt below the electron and muon \pt thresholds. To further reject $\ggmm/ee$ backgrounds, the event must contain zero electrons and no additional baseline muons that satisfy looser criteria (muons from matched ID and MS tracks with $\pt>$ 2~\GeV and $|\eta|<$ 2.5). The \pt of the muon-track system must be above 1~\GeV. This suppresses \pt-balanced backgrounds, for example \ggmm, while retaining signal events that are generally more \pt-imbalanced due to the presence of neutrinos. Additional system \pt selections are performed to reject backgrounds with FSR emissions, for example $\ggmmg$. Low-\pt photons or topoclusters are used to recover energy lost to FSR and compute a three-body system \pt. If a photon is found within $\Delta R< 1$ of the track, the \pt of the muon-track-photon system must be above 1~\GeV. If multiple photons meet these criteria, the highest-\pt photon is used. Furthermore, if a topocluster with \pt$> 2$~\GeV is found within $\Delta R< 1$ of the track but not directly associated with it, the \pt of the muon-track-topocluster system must be above 1~\GeV. A topocluster is considered to originate from the track if it lies within $\Delta R< 0.1$ of the track's extrapolation to the calorimeter for a track with $\pt>$ 0.7~\GeV, since lower-\pt tracks are unlikely to deposit significant energy in the calorimeter. If multiple topoclusters meet these criteria, the highest-\pt topocluster is used. These system \pt selections are not mutually exclusive, meaning an event may satisfy either one or both requirements.

The $\mu$3T-SR requires exactly one muon and exactly three tracks separated by $\Delta R> 0.1$ from the muon, targeting cases where the second \tlep decays into three charged hadrons. The total charge of the muon and tracks must sum to zero. The event must contain no electrons and there must be no additional baseline muons.  To ensure that the three tracks are compatible with a single 3-prong hadronic \tlep decay, while also suppressing background from exclusive $\rho$-meson production which coincides with \ggmm events, the mass of the three-track system $m_{\text{trks}}$ must be less than 1.7 ~\GeV.

The $\mu e$-SR requires exactly one muon and exactly one electron with opposite charge to the muon, targeting cases where both \tleps decay leptonically. The event must have no additional tracks with $\Delta R > 0.1$ from the electron and the muon.

To reject hadronic backgrounds, for example photonuclear processes, a topocluster veto is applied for $\mu$1T-SR and $\mu$3T-SR\@. This veto requires zero topoclusters located outside the proximity of the muon, with $\Delta R > 0.3$, and the track (for $\mu$1T-SR) or three-track system (for $\mu$3T-SR) with $\Delta R> 1.0$, referred to as `unmatched topoclusters' in the following. For the SRs that use the topocluster veto, a small correction of $\sim$2\% is applied to the MC to account for differences between the number of topoclusters in data and MC\@. Further selections are imposed on the acoplanarity, which is defined as $A_{\phi}^{\mu,\text{trk(s)}} = 1 - |\Delta\phi(\mu,\text{trk(s)})| / \pi$, to reduce photonuclear backgrounds. For the $\mu$1T-SR, the acoplanarity between the muon and the track must be less than 0.4, whereas for the $\mu$3T-SR, the acoplanarity between the muon and the track system is required to be less than 0.2. This requirement favors back-to-back production of the muon and the track(s).

For all signal regions, the energy recorded by the ZDC detectors must be less than 1~\TeV on both sides of the interaction point. This selects a $0n0n$ topology where no neutrons are detected in both ion directions. This suppresses dissociative processes which typically have larger initial-state photon virtuality, thus, this requirement helps to ensure that the initial-state photons have near-zero virtuality $q^2\sim 0~\GeV^2$. It also significantly reduces photonuclear backgrounds, which often involve nuclear dissociation leading to ion break up. Ion break up can, however, also arise in parallel to \ggtt or \ggmmg processes due to soft Coulomb exchanges. These exchanges typically excite one or both nuclei e.g., via a giant dipole resonance, inducing the emission of one or more neutrons. These neutrons, on average, carry the full per-nucleon beam energy~\cite{HION-2016-02}.

\begin{figure}
\begin{center}
\includegraphics[width=0.49\textwidth]{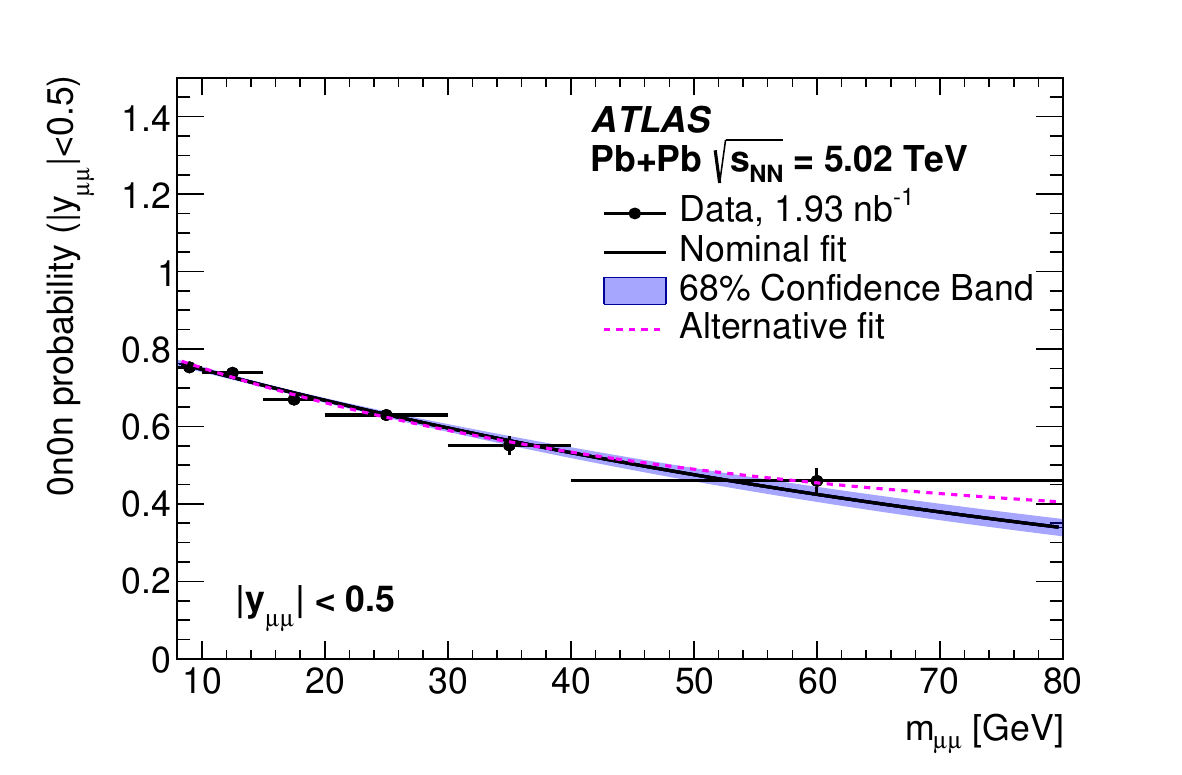}%
\includegraphics[width=0.49\textwidth]{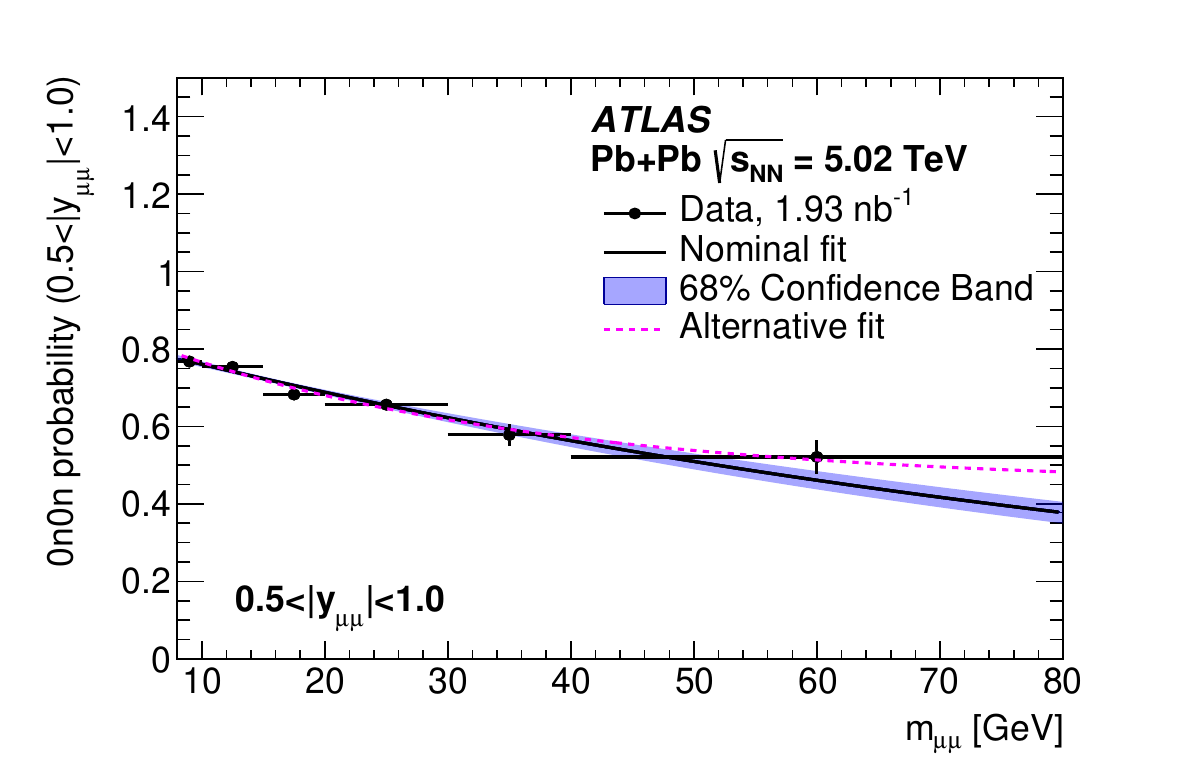}
\includegraphics[width=0.49\textwidth]{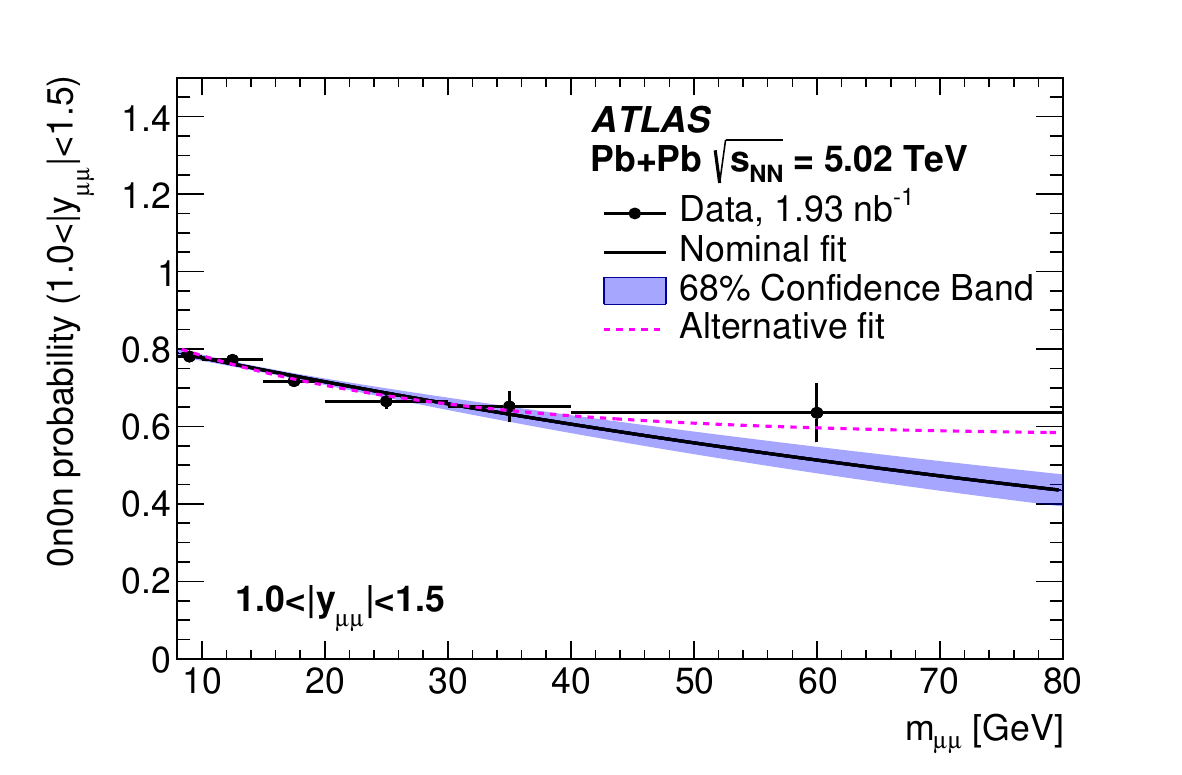}%
\includegraphics[width=0.49\textwidth]{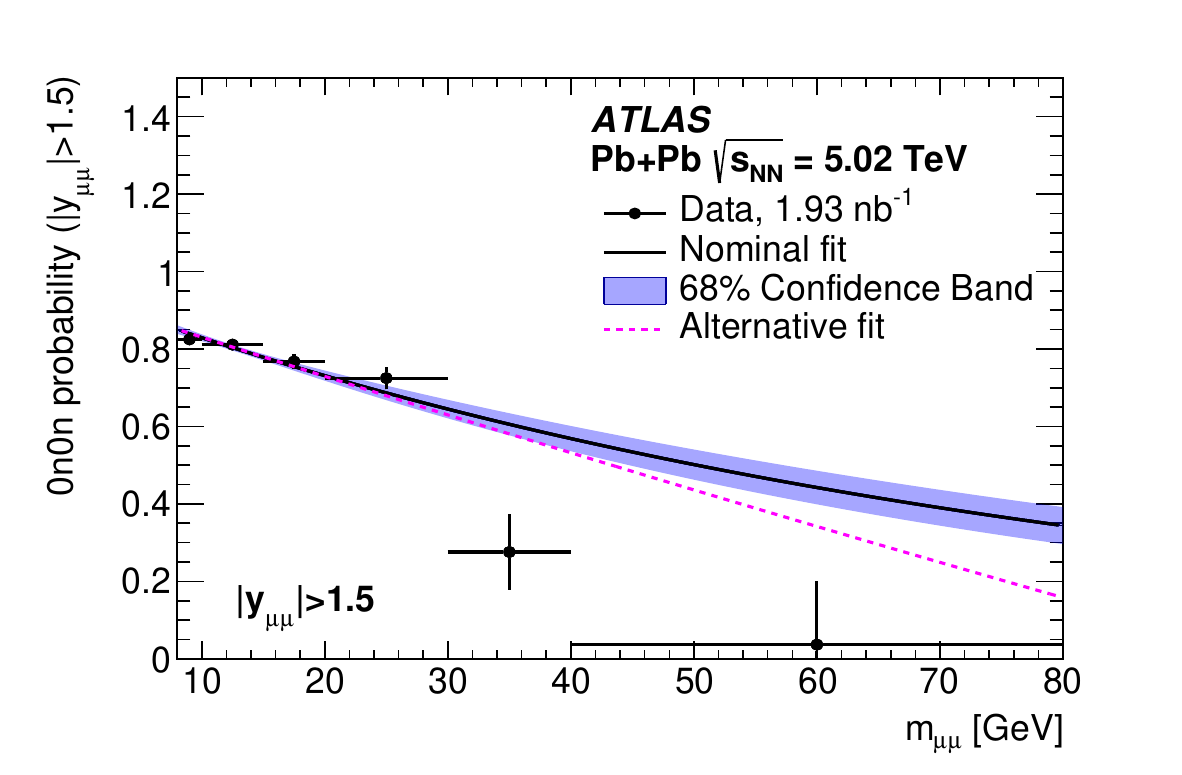}
\caption{Probabilities of $0n0n$ events extracted from 2015 and 2018 data using the $\ggmmg$ candidate events in the di-muon rapidity region (top-left) $|y_{\mu\mu}|<0.5$, (top-right) $0.5<|y_{\mu\mu}|<1.0$, (bottom-left) $1.0<|y_{\mu\mu}|<1.5$ and (bottom-right) $|y_{\mu\mu}|>1.5$.
Probabilities after correction for EM pileup (filled markers) are displayed. The error bars include statistical uncertainties and uncertainties in the exclusive single and double EM dissociation cross-sections measured by the ALICE Collaboration~\cite{ALICE:2012aa}, and their extrapolation to $\sqrt{s_{_\text{NN}}} = 5.02$~\TeV as described in Ref.~\cite{HION-2016-02}. To smooth statistical fluctuations, the corrected points are fitted using an exponential function (solid lines) with a shaded uncertainty band or an exponential function with an additional constant parameter (dashed lines).}
\label{fig:weights_0n0n_combined}
\end{center}
\end{figure}

The ZDC selection is applied only to data, as the ZDC response is not simulated in the MC samples, nor are forward neutrons typically produced by the generators. The \ggtt and \ggmmg MC samples utilized here do not distinguish among neutron topologies, requiring a data-driven correction to be applied to specifically account for the probability of having a $0n0n$ topology. The probability of an event being $0n0n$ is extracted from data using \ggmmg candidate events by measuring the ratio of $0n0n$ events to all events (without any ZDC requirements). The \ggmmg candidate events are selected by requiring exactly two muons and no additional charged-particle tracks, the di-muon system must have $\pT<$ 1~\GeV and acoplanarity $< 0.01$ %
to suppress backgrounds. Probabilities are extracted as a function of di-muon invariant mass in four rapidity intervals: $|y_{\mu\mu}|<0.5$, $0.5<|y_{\mu\mu}|<1.0$, $1.0<|y_{\mu\mu}|<1.5$, and $|y_{\mu\mu}|>1.5$. For the most forward rapidity bin, $|y_{\mu\mu}|>1.5$, there are very few data events at high di-muon mass, typically up to $\sim$5 events, and even fewer di-$\tau$ events so this region has a negligible impact. The $0n0n$ probabilities are then applied in the simulation as a weight depending on the generator-level di-muon or di-$\tau$ mass and rapidity for both \ggmmg and \ggtt events, which requires the measured probabilities to be corrected for migrations between measured bins and generator-level bins. Pileup from simultaneous electromagnetic interactions within the same LHC bunch crossing (EM pileup) can lead to additional neutrons generated, which are not associated with the observed di-muon pair. These neutrons can be detected in one or both ZDC arms, artificially lowering the $0n0n$ fraction. An additional correction is therefore determined to remove effects due to EM pileup, separately for 2015 and 2018, using a similar procedure to Refs.~\cite{HION-2016-02,HION-2021-16}. Following the EM pileup correction, the 2015 and 2018 $0n0n$ probabilities are combined. The combined $0n0n$ probabilities after EM pileup correction (filled markers) are shown in Figure~\ref{fig:weights_0n0n_combined}. The error bars include data statistical uncertainties and statistical and systematic uncertainties in the exclusive single and double EM dissociation cross-sections measured by the ALICE Collaboration~\cite{ALICE:2012aa}, and their extrapolation from $\sqrt{s_{_\text{NN}}} = 2.76$~\TeV to $\sqrt{s_{_\text{NN}}} = 5.02$~\TeV, as described in Ref.~\cite{HION-2016-02}.

To smooth statistical fluctuations, the $0n0n$ probabilities are fit with an exponential function $e^{p_0+p_1 \times m_{\mu\mu}}$, which determines the nominal $0n0n$ weights. The nominal fit to the combined EM-pileup-corrected points is displayed in Figure~\ref{fig:weights_0n0n_combined}, together with its 68\% Confidence Level (CL) uncertainty, which is considered as the statistical uncertainty in the $0n0n$ probability. A second fit with the functional form $e^{p_0+p_1 \times m_{\mu\mu}}+p_2$ is also shown and is referred to as the alternative fit. The difference between the alternative fit and the nominal one serves as a systematic uncertainty in the $0n0n$ probability. In addition, a systematic uncertainty arising from the choice of binning is also applied. In the alternative version of the binning, three rapidity bins are utilized instead of four and coarser mass bins are utilized. The same fitting procedure and uncertainty treatment is also performed for non-EM-pileup-corrected points separately for 2015 and 2018 data. The parameterized probabilities are subsequently used to reweight the MC samples as a function of the generator-level di-muon or di-$\tau$ mass and rapidity. For MC simulation compared directly to data, the fits to the non-EM-pileup-corrected points are used, i.e., the MC then includes contributions from EM pileup, in line with measured data. The application of the $0n0n$ weights reduces the nominal signal yield by $\sim$62\% relative to the after-trigger yield. For MC simulation compared with measured cross-sections, the fits to the EM-pileup-corrected points are used, i.e., the MC then excludes contributions from EM pileup. Using this approach for the MC-based unfolding inputs, the unfolding procedure described in Section~\ref{sec:corrections} automatically extrapolates the data from the $0n0n$ phase space including EM pileup to the $0n0n$ phase space with EM pileup subtracted. Additionally, statistical uncertainties and systematic uncertainties from using the alternative fit parameterization and from using the alternative binning for the nominal fit are also provided for use when applying the provided parameterized $0n0n$ probabilities.

In principle, the $0n0n$ probabilities should also be corrected for a potential ZDC inefficiency in detecting the emitted neutrons. As reported in Ref.~\cite{HION-2016-02}, the overall ZDC efficiency was, however, estimated to be over 99\%. The small inefficiency is therefore addressed in the systematic uncertainties, by varying the $0n0n$ probabilities applied to the MC simulation, when compared with measured data, by $\pm$1\%.

%
\section{Background estimate}
\label{sec:backgrounds}

After applying the event selection, there are two main sources of background, \ggmmg and photonuclear events with low activity in the ATLAS detector.

Photon-induced di-muon production with the emission of a FSR photon ($\gamma\gamma\to\mu\mu\gamma$) is especially likely to mimic the \ggtt signal as the FSR photon can cause a momentum imbalance between the muons, making the event more likely to satisfy the $\mu$1T-SR selection. It can also cause additional particle production, for example due to photon conversions that can produce two additional tracks, making the event more likely to satisfy the $\mu$3T-SR requirements, or the FSR photon could fake an electron which is pertinent for the $\mu e$-SR\@. The background from the \ggmmg processes is estimated by using the combination of MC samples discussed in Section~\ref{sec:data-mc}. The modeling of the \ggmmg background is checked and corrected using a dedicated control region 2$\mu$-CR\@. The 2$\mu$-CR requires exactly two muons with a di-muon invariant mass $m_{\mu\mu}$ greater than 11~\GeV to reject low mass di-muon resonances. Additionally, no extra tracks with $\Delta R>$ 0.1 from the muons are allowed. Similar to the SRs, the 2$\mu$-CR requires a $0n0n$ topology.

\begin{figure}
\begin{center}
\includegraphics[width=0.45\textwidth]{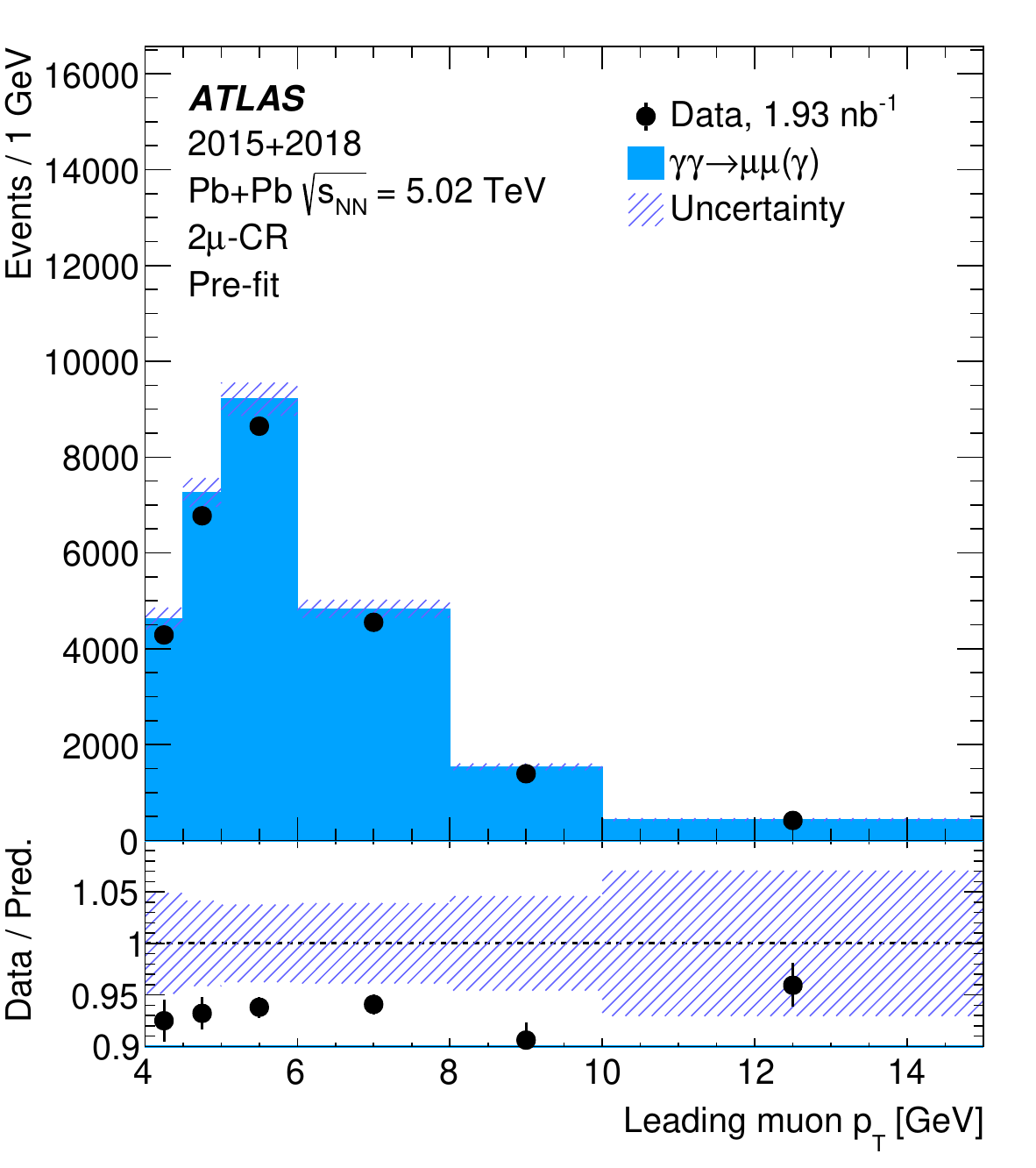}
\includegraphics[width=0.45\textwidth]{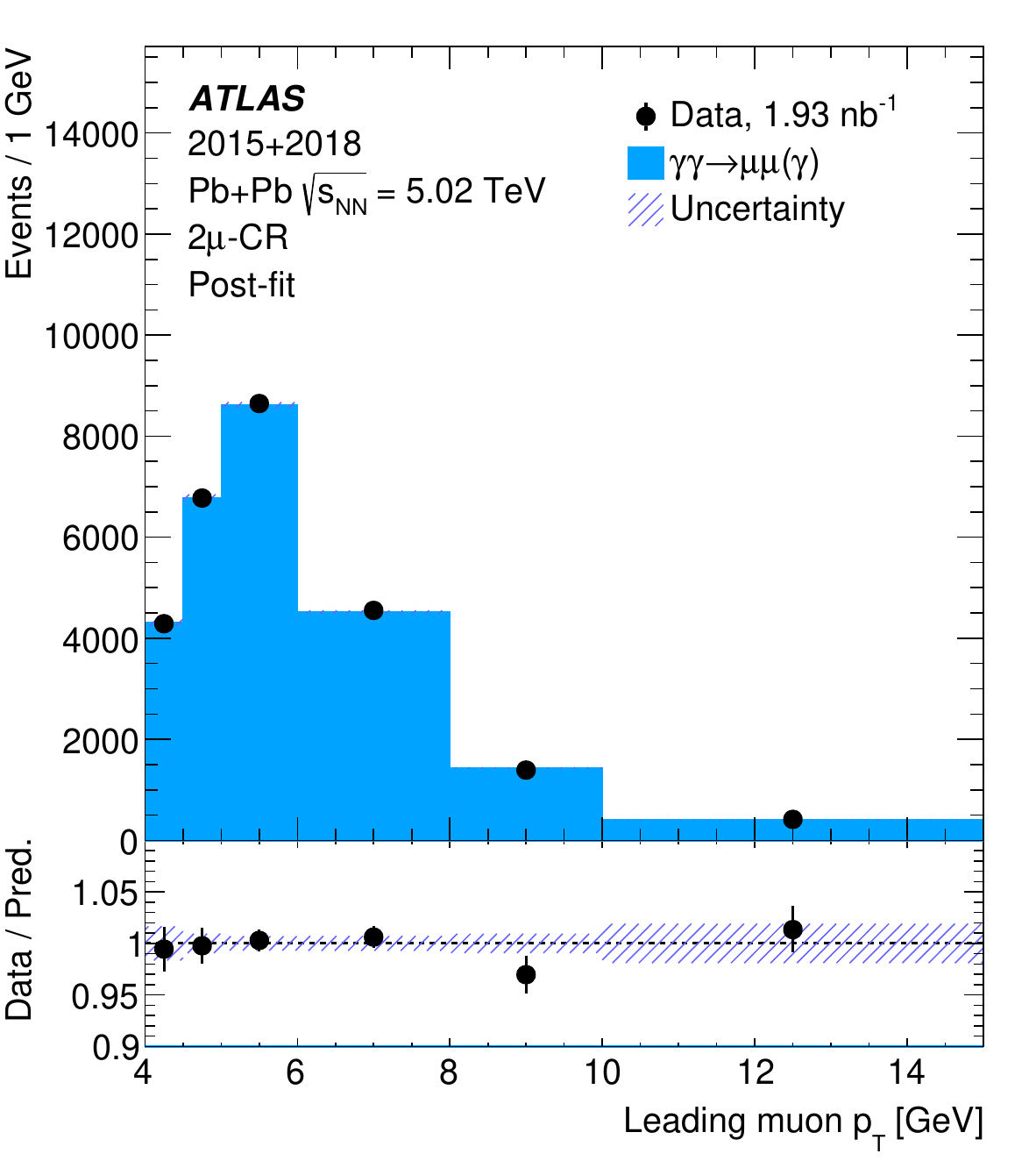}
\caption{Leading muon \pt distribution in the 2$\mu$-CR category for the combined 2015 and 2018 data, pre-fit (left) and post-fit (right). The data (filled markers) are compared with the \ggmmg background. The hatched band indicates the total uncertainties in the prediction. The error bars on the data points indicate the statistical uncertainty in the data. Event counts are shown normalized by bin width. The bottom panel displays the ratio of the data to the predictions.}
\label{fig:fits-GRFF_detlevel}
\end{center}
\end{figure}

A summary of the selection requirements is given in Table~\ref{tab:evt_sel_CRs}. The \ggtt signal contamination in the 2$\mu$-CR is below 0.1\% of the total signal plus background prediction for both 2015 and 2018 data, and is thus neglected. Figure~\ref{fig:fits-GRFF_detlevel} (left) shows the data-to-MC agreement for 2$\mu$-CR before any corrections are applied. Reasonable agreement in shape is found, however, there is an offset in the normalization which has also been observed in the ATLAS \ggee analysis in Ref.~\cite{HION-2021-16}. The residual offset between the \ggmmg prediction and the data is considered to result from inadequacies in the photon flux description~\cite{HION-2016-02}. A data-driven normalization correction factor, referred to as the Global Residual Flux Factor (GRFF), is thus extracted from the 2$\mu$-CR and is applied to the \ggmmg background and the nominal \ggtt signal predictions. The GRFF is extracted using a profile likelihood fit~\cite{trexfitter} to the highest-\pt (leading) muon \pt distribution in the 2$\mu$-CR\@. Nuisance parameters representing systematic uncertainties, described in Section~\ref{sec:systematics}, are included in the fit. The extracted value of the GRFF from the fit is $0.935^{+0.035}_{-0.033}$. After the GRFF is applied as an overall scale factor, the data-to-MC agreement is significantly improved, as shown in Figure~\ref{fig:fits-GRFF_detlevel} (right).

\begin{table}[tb]
\small
\centering
\renewcommand{\arraystretch}{1.1}
\begin{tabular}{lllll}
\toprule
Region             & $2\mu$-CR   & $\mu$2T-CR & $\mu$4T-CR \\
\midrule\midrule
0n0n topology      & As in SRs & \multicolumn{2}{c}{$E_{\text{ZDC}}<1~\TeV$ for side A and/or C} \\
$N_{\mu}^{\text{baseline}}$   & ---  & $=1$ & $=1$ \\
$N_{\mu}$        & $=2$ & $=1$ & $=1$ \\
$N_{\text{trk}}(\Delta R > 0.1$ from ${\mu})$   &  $= 0$ & $=2$ & $=4$  \\
\pt of lowest-\pt trk   &  --- & $<500$~\mev & $<500$~\mev  \\
$N_\text{topoclusters}^{\text{unmatched}}$  & ---  & $\leq8$ & $\leq8$ \\
$\Sigma_{\mu,\text{highest-\pt trk}}$ charge	& --- & $=0$ & ---  \\
$m_{\text{trks}}$ (or $A_{\phi}^{\mu,\text{highest-\pt trk}})$   & --- & $>1$~\gev (or $>0.2$)  & $<$ 1.7~\GeV  \\
$m_{\mu\mu}$  & $>$ 11~\GeV & --- & ---   \\
\bottomrule
\end{tabular}
\caption{Summary of the control region definitions. The $2\mu$-CR is used to improve the photon flux modeling and $\mu$2T-CR and $\mu$4T-CR are used to obtain templates for estimating the photonuclear background.
}
\label{tab:evt_sel_CRs}
\end{table}

Diffractive photonuclear processes also constitute a non-negligible background and can mimic the \ggtt signal due to their characteristic low particle activity. These interactions involve the exchange of a colorless object, known as a pomeron, which can result in one-sided rapidity gaps being partially filled by pomeron remnants~\cite{Helenius:2019gbd, Schafer:2020bnm}. The diffractive photonuclear background in the $\mu e$-SR is negligible in both 2015 and 2018 data, due to the very high purity of this region. The diffractive photonuclear background in $\mu$1T-SR and $\mu$3T-SR is estimated by using a fully data-driven method, exploiting the fact that the photonuclear events are often accompanied by extra particle production. Template distributions are built using control regions, which are similar to the signal regions but require an additional low-\pt track with $\pt<$~500~\MeV. The control regions are referred to as $\mu$2T-CR and $\mu$4T-CR, related to the $\mu$1T-SR and $\mu$3T-SR, respectively. Aside from modifying the requirement on the number of tracks, for $\mu$2T-CR and $\mu$4T-CR the $0n0n$ topology requirement is loosened to require that the energy recorded by the ZDC detectors is less than 1~\TeV on one or both sides of the interaction point. Loosening the $0n0n$ topology requirement enhances statistics, given that photonuclear events are often accompanied by break up of one of the ions. Additionally, the topocluster veto is relaxed to require less than or equal to eight unmatched topoclusters instead of zero. For the $\mu$2T-CR, the mass of the two-track-system is required to be above 1~\GeV or the acoplanarity between the muon and the highest-\pt track must be above 0.2, to reduce \ggtt signal contamination. For the $\mu$4T-CR, the requirement of zero net charge summing over the muon and tracks is removed. The other selections remain the same as those of $\mu$3T-SR\@. The contributions from \ggtt and \ggmmg were checked for the $\mu$2T-CR and $\mu$3T-CR using simulation and were found to be negligible as no event passes the criteria for either region. A summary of the selection criteria is given in Table~\ref{tab:evt_sel_CRs}. The measured kinematic distributions in the $\mu$2T-CR and $\mu$4T-CR are used as diffractive photonuclear template distributions. The normalization of the templates is performed using the ratio of the integrated data distribution and the integrated photonuclear background distribution, in a region where the SR selections are applied except for the topocluster veto, and instead $4\leq N_{\text{topoclusters}}^{\text{unmatched}}\leq 8$ is required. The fully reconstructed MC samples demonstrate that this region has no contribution from the \ggtt signal or the \ggmmg background. For 2015 data, the diffractive photonuclear background is negligible in the $\mu$3T-SR, compared with 3.2 $\pm$ 2.1 events for 2018 data.

All other considered background sources are negligible. The backgrounds from \ggee production and non-diffractive photonuclear processes were estimated by using simulation and were found to be negligible as no event passes any of the three SR criteria. The contribution from $\gamma\gamma \to q\bar{q}$ is estimated with simulation and found to be negligible. The same is found for resolved $\gamma\gamma$ interactions where at least one of the photons fluctuates into a hadronic state of a $q\bar{q}$ pair, which is modeled in \PYTHIA[8]~\cite{Helenius:2017aqz}. The contribution due to the simultaneous production of \ggmm or \ggtt and exclusive $\rho^0$-meson production $\gamma P \to\rho^0\to\pi\pi$ in the same ion-ion interaction~\cite{HION-2023-13} is estimated by using a fully data-driven method~\cite{STDM-2019-19} and is negligible.

\begin{figure}
\begin{center}
\includegraphics[width=0.45\textwidth]{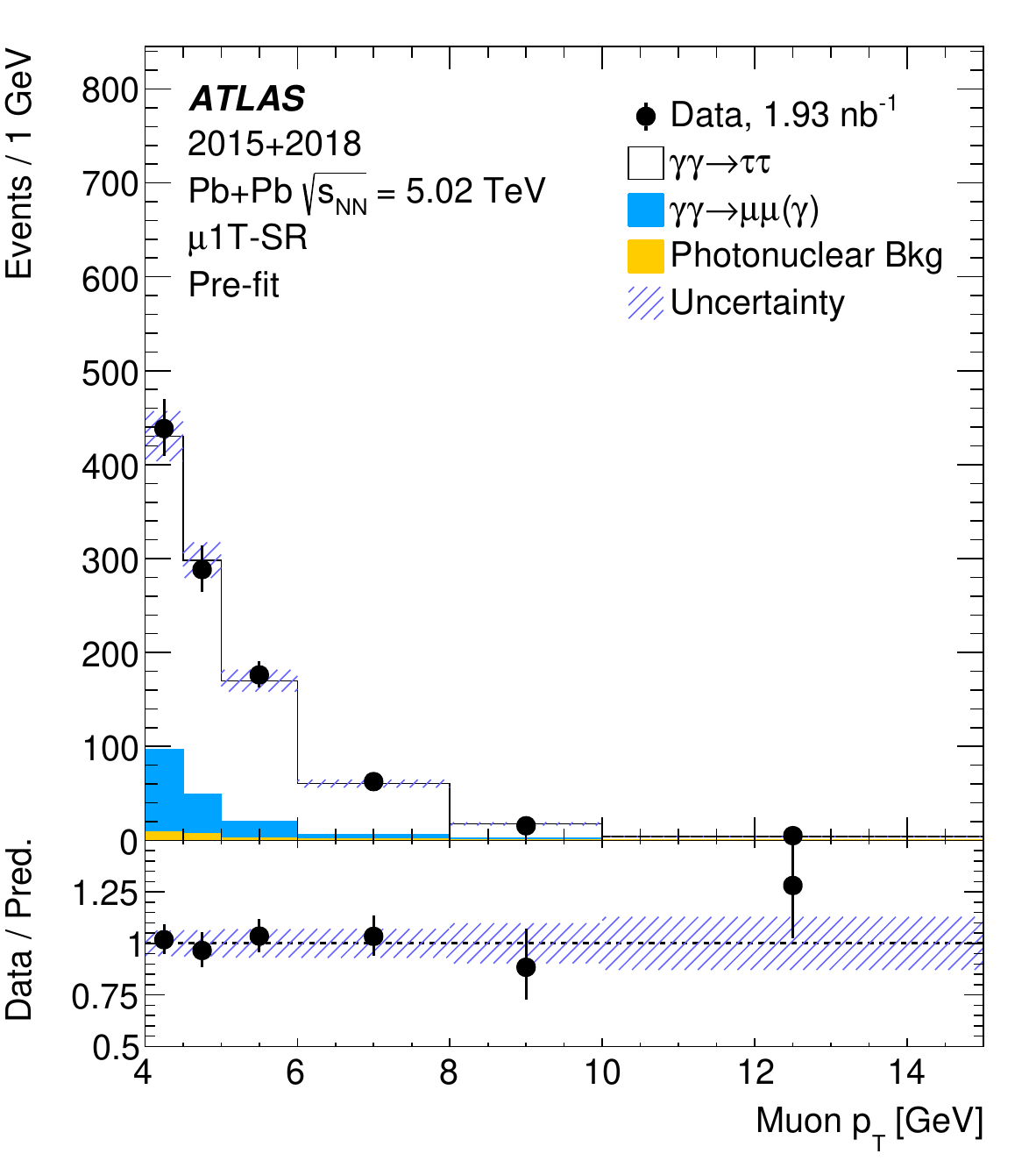}%
\includegraphics[width=0.45\textwidth]{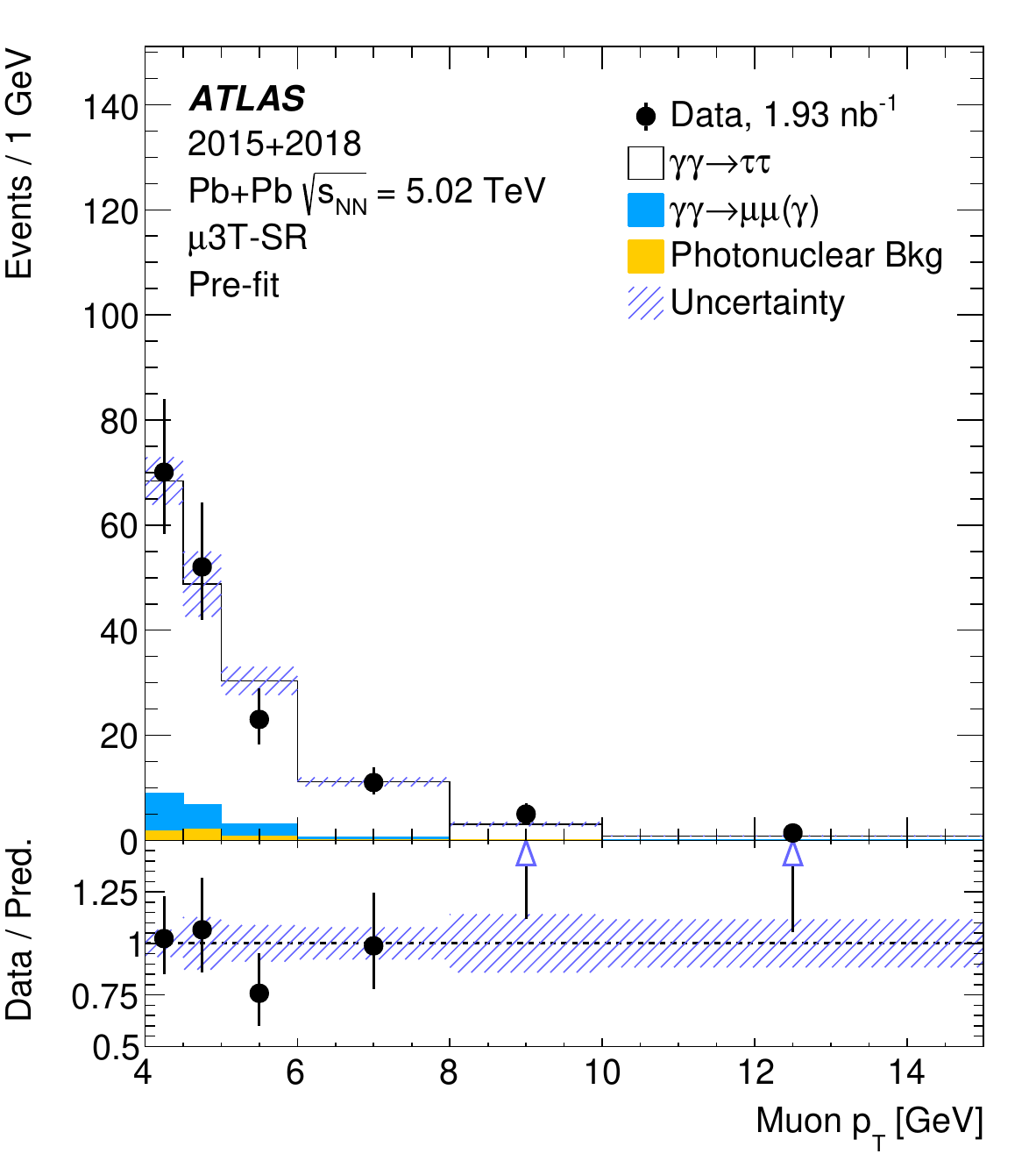}
\includegraphics[width=0.45\textwidth]{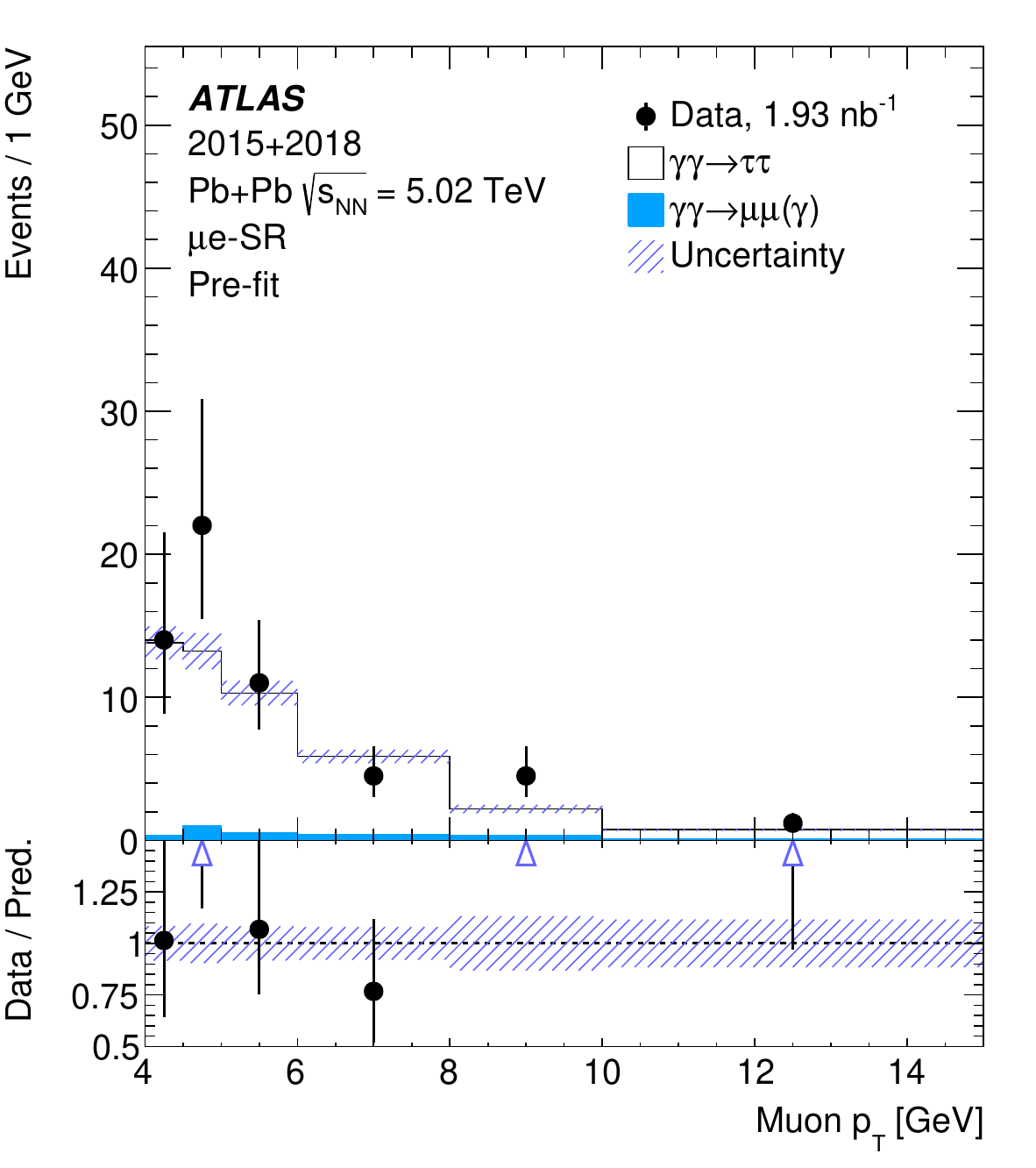}
\caption{Muon \pt distributions in the (top left) $\mu$1T-SR, (top right) $\mu$3T-SR, (bottom) $\mu e$-SR categories for the combined 2015 and 2018 data. The data (filled markers) are compared with the stacked histograms showing \ggtt signal with $\atau=\dtau=0$, the \ggmmg background and the photonuclear background. The hatched band indicates the total uncertainties in the prediction. The error bars on the data points indicate the statistical uncertainty in the data. Event counts are shown normalized by bin width. The bottom panel displays the ratio of the data to the predictions; arrows in this panel denote a data point that is outside the displayed range. }
\label{fig:fits-atau_detlevel-prefitdistrs}
\end{center}
\end{figure}

Figure~\ref{fig:fits-atau_detlevel-prefitdistrs} shows the detector-level muon \pt distributions for $\mu$1T-SR, $\mu$3T-SR and $\mu e$-SR for the combined 2015 and 2018 data. The observed data yields after all event selections for the combined 2015 and 2018 data, compared with the expected background events, are listed in Table~\ref{tab:sigbkg_yields}, demonstrating the very low background levels. The predicted signal-plus-background yields in Table~\ref{tab:sigbkg_yields} assume $\atau=\dtau=0$ for the signal and are compatible with the observed data yields.

\begin{table}[tb]
\small
\centering
\begin{tabular}{llll}
\toprule
Region  & $\mu$1T-SR & $\mu$3T-SR & $\mu e$-SR \\
\midrule
\ggtt signal & 598.4 $\pm$ 39.2 & 109.0 $\pm$ 9.1 & 40.6 $\pm$ 3.3\\
\ggtt contribution [\%] & 84.3$\%$ $\pm$ 5.5$\%$ & 89.6$\%$ $\pm$ 7.5$\%$ & 92.7$\%$ $\pm$ 7.5$\%$\\
\midrule
\ggmmg background & 95.8 $\pm$ 8.1 & 9.4 $\pm$ 1.0 & 3.2 $\pm$ 0.5\\
\ggmmg contribution [\%] & 13.5$\%$ $\pm$ 1.1$\%$ & 7.7$\%$ $\pm$ 0.8$\%$ & 7.3$\%$ $\pm$ 1.1$\%$\\
\midrule
Photonuclear background & 15.6 $\pm$ 11.4 & 3.2 $\pm$ 2.1 & --- \\
Photonuclear contribution [\%] & 2.2$\%$ $\pm$ 1.6$\%$ & 2.6$\%$ $\pm$ 1.7$\%$ & --- \\
\midrule
Combined background & 111.4 $\pm$ 14.0 & 12.6 $\pm$ 2.3 & 3.2 $\pm$ 0.5 \\
Combined background contribution [\%] & 15.7$\%$ $\pm$ 2.0$\%$ & 10.4$\%$ $\pm$ 1.9$\%$ & 7.3$\%$ $\pm$ 1.1$\%$\\
\midrule
Combined signal+background & 709.7 $\pm$ 46.2 & 121.6 $\pm$ 9.9 & 43.8 $\pm$ 3.6 \\
\midrule
Data & 723 & 124 & 53 \\
\bottomrule
\end{tabular}
\caption{Signal and background yields in the three SRs, together with the observed event counts in data. Negligible contributions are indicated by ---. Uncertainties in the predictions are the total uncertainties in the respective prediction. Percentages, including the uncertainty values, use the total signal-plus-background prediction as denominator.}
\label{tab:sigbkg_yields}
\end{table}

%
%
%
%
%
%
%
%
%
%
%
%
%


%
\section{Data unfolding}
\label{sec:corrections}

Differential cross-sections are measured for seven variables in each of the three fiducial regions. %
The fiducial regions (FRs) referred to as $\mu$1T-FR, $\mu$3T-FR, and $\mu$e-FR (defined at particle-level) correspond to $\mu$1T-SR, $\mu$3T-SR, and $\mu$e-SR (defined at detector-level), respectively. The FR selections are defined in Table~\ref{tab:fiducial_selection} and each FR is statistically independent to the others. The definition of the FRs only considers final-state particles with lifetimes greater than \SI{30}{ps}, which are considered 'stable' -- this is also referred to as "particle-level" in the simulation and for cross-section measurements. The definitions of the variables in each FR are given in Table~\ref{tab:fiducial_vars}.

\begin{table}[tb]
\small
\centering
\renewcommand{\arraystretch}{1.1}
\begin{tabular}{llll}
\toprule
Object                & Requirements       \\
\midrule
Leptons ($e$,$\mu$)   &  \multicolumn{3}{l}{$\pt >$ 4~\GeV and $|\eta| <$ 2.5} \\
Charged hadrons               &  \multicolumn{3}{l}{$\pt >$ 100~\MeV and $|\eta| <$ 2.5} \\
Low-\pt Leptons & \multicolumn{3}{l}{100~\MeV $< \pt <$ 4~\GeV and $|\eta| <$ 2.5} \\
Track particles (trk)   & \multicolumn{3}{l}{Charged hadron or low-\pt lepton}  \\
\midrule\midrule
Region           & $\mu$1T-FR & $\mu$3T-FR & $\mu$e-FR  \\
\midrule\midrule
$0n0n$ topology  & \multicolumn{3}{l}{$0n0n$ weights for MC} \\
\midrule
$N_\mu$ & $=1$ & $=1$ & $=1$ \\
$N_e$ & $=0$ & $=0$ & $=1$ \\
$N_\text{trk}(\Delta R > 0.1\text{ from }\mu\text{ or }e)$ & $=1$ & $=3$ & $=0$ \\
$\Sigma_{\mu,\text{trk(s)}\text{ or }e}$ charge & $=0$ & $=0$ & $=0$ \\
\pt of muon and track & $>1$~\GeV & --- & --- \\
$m_{\text{trks}}$  & --- & $< 1.7$~\GeV & --- \\
$A_{\phi}^{\mu,\text{trk}(s)}$  & $<0.4$ & $< 0.2$ & --- \\
\bottomrule
\end{tabular}
\caption{Summary of fiducial object requirements and fiducial region definitions. The $0n0n$ weights here exclude the contribution from EM pileup. The notation trk(s) indicates a single track for the $\mu$1T-FR and the three track-system for the $\mu$3T-FR.}
\label{tab:fiducial_selection}
\end{table}

\begin{table}[tb]
\small
\centering
\renewcommand{\arraystretch}{1.1}
\begin{tabular}{llll}
\toprule
Variable   & $\mu$1T-FR & $\mu$3T-FR & $\mu$e-FR  \\
\midrule
Muon \pt   & $\pt^\mu$ & $\pt^\mu$ & $\pt^\mu$ \\
Track(s)/Electron \pt & $\pt^\mathrm{trk}$ & $\pt(\textbf{p}_\mathrm{trks})$ & $\pt^e$ \\
$\pt\left(\mu,\mathrm{trk(s)}/e\right)$ & $\pt(\textbf{p}_\mu+\textbf{p}_\mathrm{trk})$ & $\pt(\textbf{p}_\mu+\textbf{p}_\mathrm{trks})$ & $\pt(\textbf{p}_\mu+\textbf{p}_e)$\\
$m_{\mathrm{inv}}(\mu,\mathrm{trk(s)}/e)$ & $m_{\mathrm{inv}}(\textbf{p}_\mu+\textbf{p}_\mathrm{trk})$ & $m_{\mathrm{inv}}(\textbf{p}_\mu+\textbf{p}_\mathrm{trks})$ & $m_{\mathrm{inv}}(\textbf{p}_\mu+\textbf{p}_e)$  \\
$\eta(\mu,\mathrm{trk(s)}/e)$ & $\eta(\textbf{p}_\mu+\textbf{p}_\mathrm{trk})$ & $\eta(\textbf{p}_\mu+\textbf{p}_\mathrm{trks})$ & $\eta(\textbf{p}_\mu+\textbf{p}_e)$ \\
$\deta{\mu}{\mathrm{trk(s)}/e}$ & $\eta_\mu-\eta_\mathrm{trk}$ & $\eta_\mu-\eta(\textbf{p}_\mathrm{trks})$ & $\eta_\mu-\eta_e$ \\
$A_{\phi}^{\mu,\mathrm{trk(s)}/e}$ & $1-\left|\phi_\mu-\phi_\mathrm{trk}\right|/\pi$ & $1-\left|\phi_\mu-\phi(\textbf{p}_\mathrm{trks})\right|/\pi$ & $1-\left|\phi_\mu-\phi_e\right|/\pi$\\
\bottomrule
\end{tabular}
\caption{Definition of variables measured in each FR\@. The notation $\textbf{p}$ indicates the four-vector of the related particles. The four-vector sum of the three selected tracks in the $\mu$3T-FR is indicated as $\textbf{p}_\mathrm{trks}$, i.e., $\textbf{p}_\mathrm{trks}=\sum_{i=1}^{3}\textbf{p}_{\mathrm{trk},i}$. If a variable is derived from a sum of four-vectors, the definition of that four-vector sum is provided in parentheses. Otherwise, the related particle is indicated as sub- or superscript.}
\label{tab:fiducial_vars}
\end{table}

To determine the yield of \ggtt events in data, the estimated background contributions described in Section~\ref{sec:backgrounds} are subtracted from the data in each bin of the studied variable. Following the background subtraction, the data are corrected for detector effects ("unfolded") using simulation, where the data and background estimates are considered to be at detector-level, and are unfolded to particle-level. The unfolding is performed using the iterative Bayesian unfolding (IBU) method~\cite{DAgostini:1994fjx,dagostini2010improvediterativebayesianunfolding,Biondi:2017pzs}, implemented using the D’Agostini iterative scheme within RooUnfold~\cite{Adye:2011gm} with the generated signal distribution used as the initial prior. %
The relationships between the generated signal distribution and the corresponding reconstructed one are obtained from the nominal \ggtt signal sample (\nomSamp), and are summarized in the migration matrix. Events used to populate the matrix are required to satisfy both the SR (detector-level) and FR (particle-level) selections. Migrations between detector-level and particle-level bins are considered via this matrix. Typical off-diagonal contributions to the matrix are in the range $\sim$5--10\%, up to a maximum of $\sim$20\% for some variables and fiducial regions. The fraction of events that fall into the SR and originate from the FR, referred to as the fiducial fraction, is 0.63 for $\mu$1T-FR, 0.91 for $\mu$3T-FR, and 0.88 for $\mu$e-FR\@. The fraction of events that originate from the FR and are reconstructed in the SR, referred to as efficiency, is 0.34 for $\mu$1T-FR, 0.33 for $\mu$3T-FR, and 0.36 for $\mu$e-FR\@. The fiducial fractions are corrected for before unfolding by multiplying the background-subtracted data in each bin with the fiducial fraction corresponding to that particular bin. The efficiencies are corrected for in the application of the IBU method. The number of iterations in the IBU method is optimized to correspond to the minimal number where the $\chi^2$-difference between unfolded data and the generated signal distribution does not change by more than 1\% with further iterations. The resulting numbers of iterations are between one and four iterations depending on the considered distribution. The effect of systematic uncertainties on the background estimate, detector-level and particle-level inputs to the unfolding are described in Section~\ref{sec:systematics} and the uncertainties related to the unfolding procedure itself are described in this section.

A bootstrap method~\cite{ATL-PHYS-PUB-2021-011} is used to propagate the statistical uncertainties of the data through the unfolding, allowing to determine the correlations between bins of the same unfolded distribution, and between different unfolded distributions in the same FR\@. Bootstrap replicas are constructed by assigning each data event a random weight based on a Poisson distribution with mean of one, and employing this to generate varied copies of the measured data distributions. The unfolding procedure is repeated for each bootstrap replica. The statistical covariance between two bins is defined relative to the mean bin content over all replicas. The statistical uncertainty in the unfolded data distribution is given by the square-root of the diagonal entries in the covariance matrix. To determine the statistical covariance matrix, \SI{10000} replicas were generated.

The statistical uncertainty in the background estimate is propagated with pseudo-experiments through the unfolding. For this purpose, the background predictions are varied randomly in each bin, assuming a Gaussian probability distribution centered on the nominal content with standard deviation corresponding to the background statistical uncertainty in this bin. The unfolding is repeated with the varied background estimates, and the covariance matrix is calculated as previously, with only correlations within the same distribution being accessible. To propagate the statistical uncertainty in the background prediction, \SI{10000} pseudo-experiments were used. The propagated statistical uncertainties in the background estimate are considered to be systematic uncertainties in the cross-section distributions.

The statistical uncertainty in the signal MC-based unfolding inputs is considered as further systematic uncertainty in the unfolding. It is determined with a simpler and faster unfolding method: a variant of bin-by-bin unfolding that accounts for correlations between detector-level and particle-level. The statistical uncertainties in the bin-by-bin weights, which are the ratios of the generated vs. reconstructed events per bin, are determined through error propagation~\cite{DeNardo2002Propagation}, properly taking the statistical overlap of events in the numerator and denominator of the weights into account.

A data-driven non-closure uncertainty is determined to cover potential biases in the unfolding procedure arising from the potential mismodeling of the data by the nominal signal MC simulation, employed to produce the MC-based unfolding inputs. The uncertainty is determined using pseudodata generated from MC distributions that were re-weighted based on comparisons with data~\cite{Malaescu:2009dm,Malaescu:2011yg}. More specifically, for each unfolded variable, a pseudodata generated signal distribution and -- through folding with the correspondingly adjusted migration matrix -- the related pseudodata signal reconstructed distribution is produced. %
The pseudodata reconstructed distribution is unfolded with the original (non-adjusted) signal MC unfolding inputs and the result compared with the pseudodata generated distribution. The difference between the two is considered as systematic uncertainty. The same numbers of iterations are used for the unfolding of the pseudodata reconstructed distribution as used in the unfolding of data. To reduce sensitivity to statistical fluctuations, six differently smoothed data-driven weights are tested to determine the pseudodata generated distribution, and the average over the unfolded results from these variations is used to provide the non-closure systematic uncertainty for each considered observable.


%
\section{Systematic uncertainties}
\label{sec:systematics}

The leading sources of uncertainty for this measurement are the statistical uncertainties of the data. In addition to these and the unfolding uncertainties, which are described in Section~\ref{sec:corrections}, several sources of systematic uncertainties also affect this measurement. These include those arising from the trigger efficiency, object reconstruction and calibration, luminosity and the modeling of signal and background.

The uncertainties associated with the muon trigger efficiency include both statistical and systematic contributions and are applied to the muon trigger data-to-MC correction factors used to correct the simulated samples. The uncertainties associated with muon reconstruction and identification efficiency are applied to the muon reconstruction and identification correction factors. The systematic uncertainties in the muon trigger, reconstruction and identification efficiency correction factors are derived by varying selection requirements in the \ggmm tag-and-probe analysis used to derive the correction factors. Systematic uncertainties associated with the muon momentum scale and resolution corrections are determined following Ref.~\cite{PERF-2015-10}. The uncertainties associated with the electron and photon reconstruction and identification efficiency and energy scale and resolution corrections are derived following Ref.~\cite{HION-2019-08}.

The systematic uncertainty associated with the track reconstruction efficiency is determined using alternative MC samples that vary the detector material budget or the physics model for particle interactions in the detector material employed in the \GEANT simulation. The tracking efficiency is studied as a function of the charged particle \pt and $\eta$, separately for charged pions, electrons, and muons. The impact of systematic variations in the tracking efficiency is propagated by randomly removing tracks in the nominal MC samples, according to the varied track efficiency in bins of \pt and $|\eta|$.

The uncertainty in the topocluster reconstruction efficiency and energy calibration is determined using \ggee events in which one of the electrons emits a hard bremsstrahlung photon due to its interaction with the ID material. No data-to-MC correction factors are applied for the topocluster reconstruction efficiency but a global systematic uncertainty of 2\% is assigned to cover small data-to-MC discrepancies. The uncertainty in topocluster energy calibration is assessed by applying additional corrections (shifts) to the reconstructed topocluster \pt, based on small data-to-MC discrepancies~\cite{HION-2019-08}.

The estimate of the diffractive photonuclear background is impacted by both statistical and systematic uncertainties. The statistical uncertainty arises due to the limited number of events in the photonuclear background normalization region $4\leq N_{\text{topoclusters}}^{\text{unmatched}}\leq$ 8. The systematic uncertainty is estimated by using alternative control regions to construct diffractive photonuclear background templates. The control regions have the same event selections as the $\mu 2$T-CR and $\mu 4$T-CR, except that the net electric charge of the muon and the track/three-track system must be non-zero. The uncertainty is given by the difference between the diffractive photonuclear background estimate determined using the nominal control regions and that obtained using the alternative control regions.

To estimate the uncertainty arising from the modeling of the photon flux, parameters describing the proton and neutron densities are varied by their experimental uncertainties using the \textsc{SuperChic3} generator. In particular, the parameters $R_{p} =6.68\pm0.02$~fm, $R_{n}=6.69\pm0.03$~fm, $a_{p}=0.447\pm0.01$~fm and $a_{n}=0.560\pm0.03$~fm are varied by their uncertainties, where $R_{p,n}$ and $a_{p,n}$ are the radius and skin depth of the proton and neutron densities, respectively~\cite{Loizides:2017ack,Klos:2007is,Tarbert:2013jze}. The average of the largest absolute up and down deviations in the $\gamma\gamma\rightarrow\mu\mu$ cross-section, inclusive in forward neutron topologies, in bins of $m_{\mu\mu}$ and $\left|y_{\mu\mu}\right|$ is taken and symmetrized. Since the photon flux depends only on the invariant mass and rapidity of the centrally produced system (when $s \gg m_{X}^{2}$), the uncertainty computed using $\gamma\gamma\rightarrow\mu\mu$ events is also valid for $\gamma\gamma\rightarrow\tau\tau$ events.
The uncertainty is of order 0.5\% in bins with the largest number of observed $\gamma\gamma\rightarrow\tau\tau$ events i.e., $m_{\mu\mu}\lesssim 50$~\GeV and $\left|y_{\mu\mu}\right|\lesssim 2.0$. In addition, a flat 2.5\% uncertainty is assigned on the theoretical modeling of the survival factor~\cite{Harland-Lang:2021ysd}. The photon flux uncertainties are applied to both the \ggtt and \ggmmg samples. The uncertainty in the GRFF is described in Section~\ref{sec:backgrounds}. Potential correlations between the nuisance parameters in the fit to extract the GRFF and in the fits performed to extract the \tlep electromagnetic dipole moments, described in Section~\ref{sec:dipoles}, are not retained as the fits are performed independently to ensure that the GRFF extraction is free from any influence of signal. The uncertainties related to the modeling of \tlep decays are evaluated using alternative signal samples, which use \PYTHIA[8] instead of \textsc{Tauola} to model \tlep decays. The uncertainty in the modeling of spin correlation effects is defined as the size of the bin-by-bin correction factor. The uncertainty in the $0n0n$ reweighting and the efficiency of the ZDC detectors are described in Section~\ref{sec:event-selection}.

The luminosity uncertainty for the combined 2015 and 2018 data is 1.5\%. It is derived using beam-separation scans, following a similar method to Ref.~\cite{DAPR-2021-01}, and using LUCID-2 for the baseline luminosity measurements.

\begin{figure}[t]
\begin{center}
\includegraphics[width=0.33\textwidth]{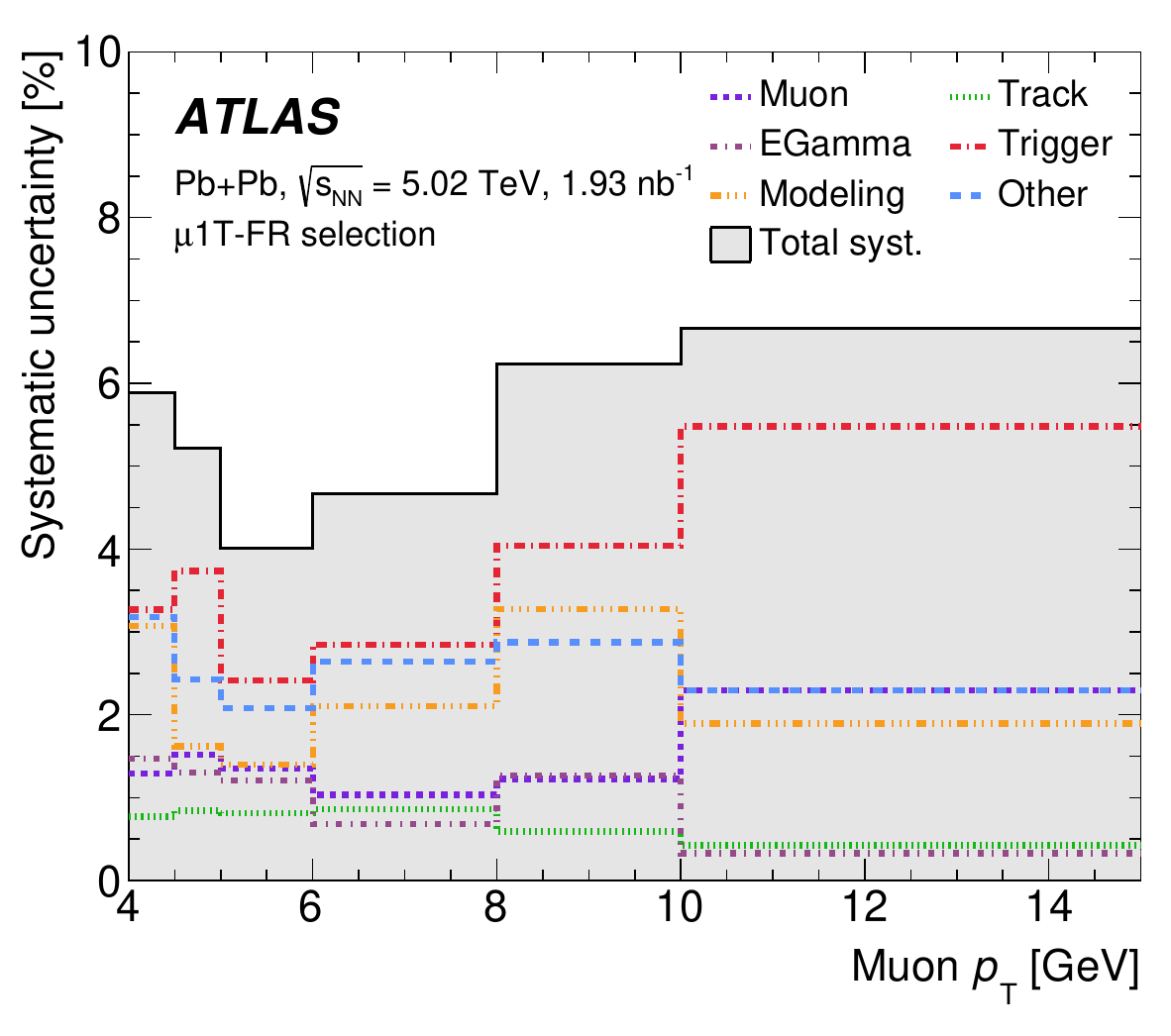}%
\includegraphics[width=0.33\textwidth]{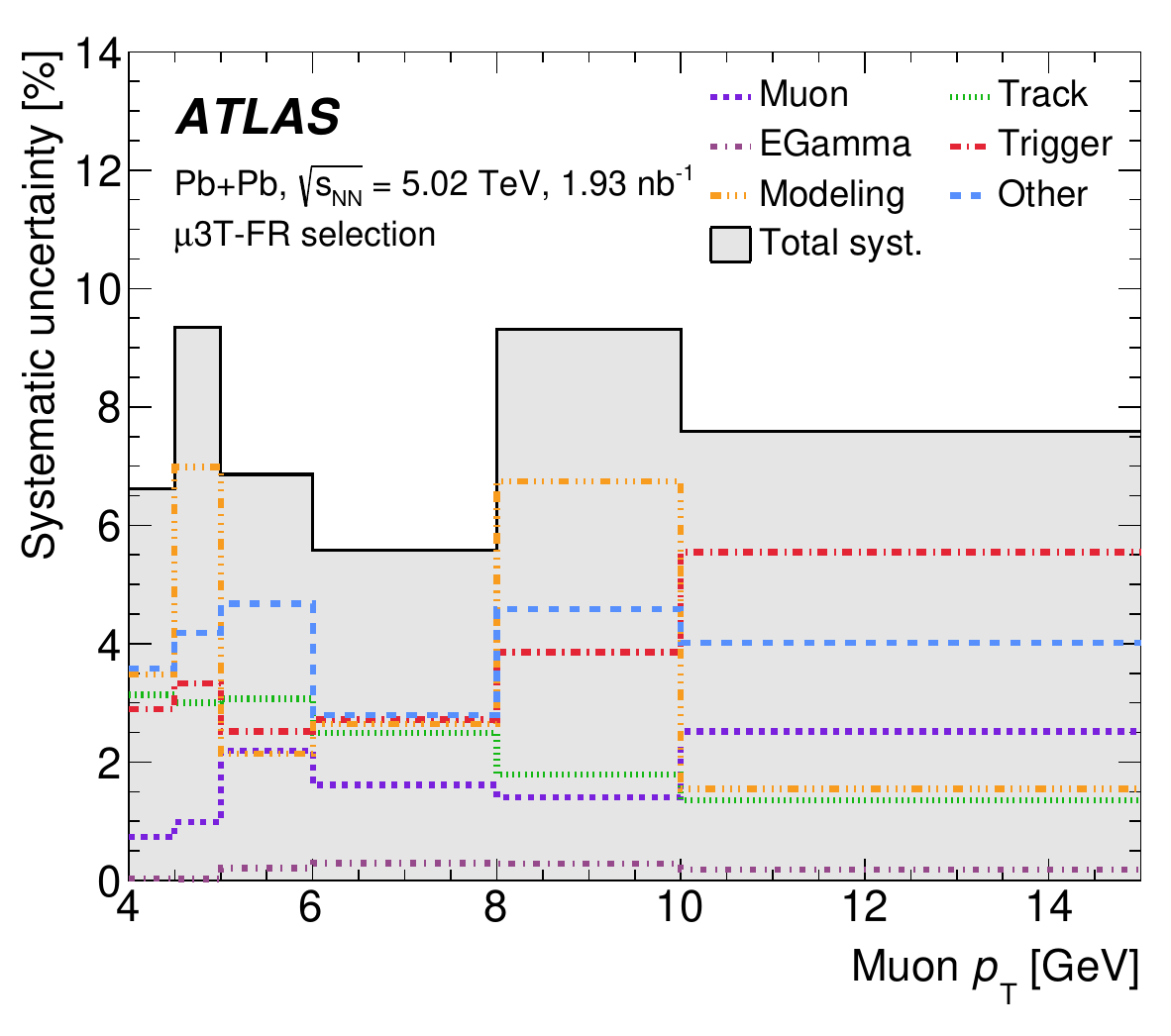}%
\includegraphics[width=0.33\textwidth]{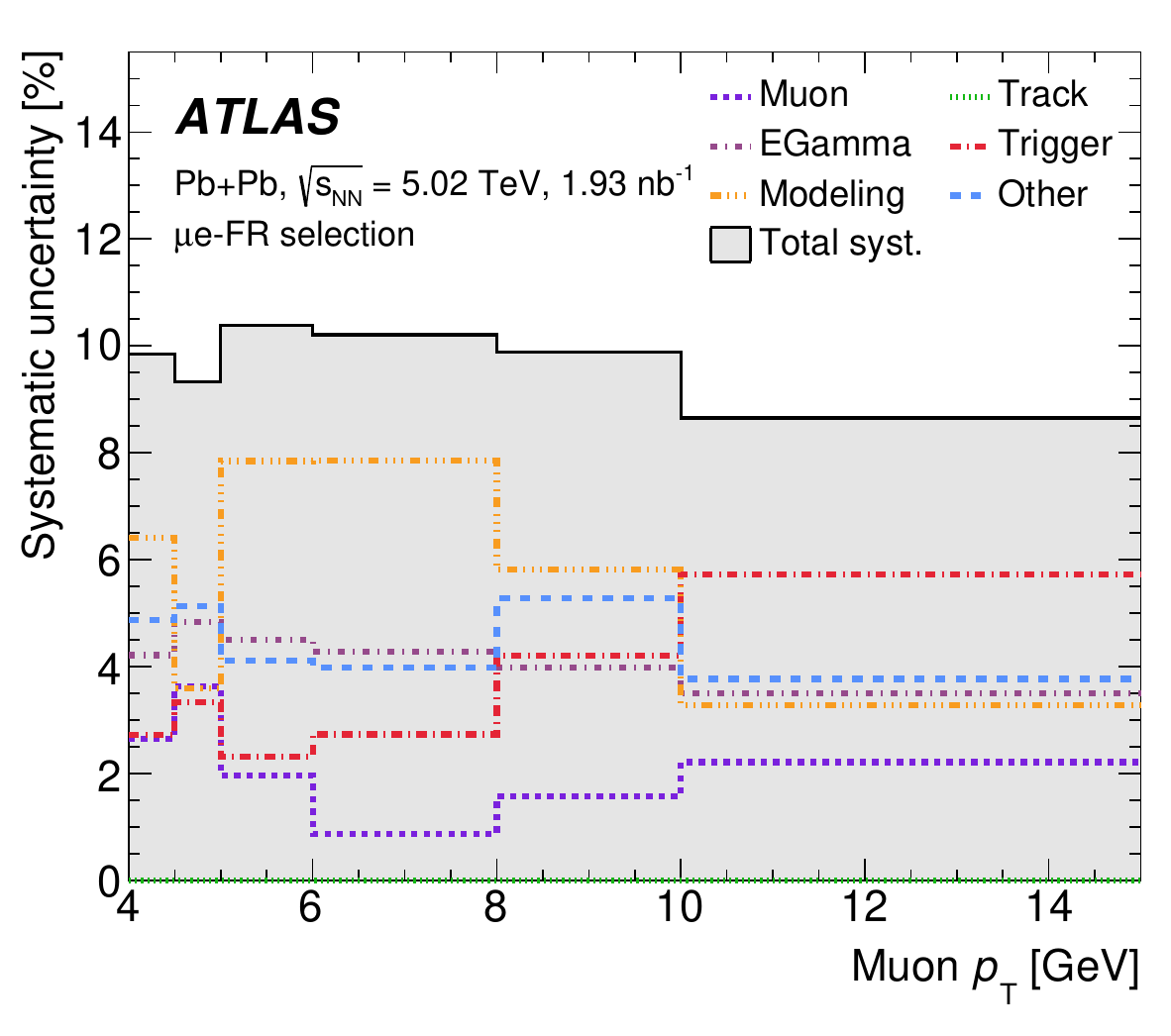}
\caption{Grouped systematic uncertainties in the cross-sections per bin of leading muon $\pt$ for $\mu$1T-FR (left), $\mu$3T-FR (middle) and $\mu e$-FR (right) for the combined 2015 and 2018 data. The contents of the various categories are explained in the text.}
\label{fig:groupwise-total-syst}
\end{center}
\end{figure}

For presentation purposes, the systematic uncertainties in the cross-sections are grouped into categories. The impact of each group of systematic uncertainties for the unfolded muon \pt distribution in $\mu$1T-FR, $\mu$3T-FR and $\mu e$-FR for the combined 2015 and 2018 data are shown in~\Cref{fig:groupwise-total-syst}. ``Muon'' includes all systematic uncertainties related to muon reconstruction, identification and calibration, ``Trigger'' includes muon trigger scale factor systematic uncertainties, ``EGamma'' includes electron and photon reconstruction, identification, energy scale and resolution systematic uncertainties, and topocluster systematic uncertainties. ``Track'' includes uncertainties in the track reconstruction efficiency, ``Modeling'' includes the $0n0n$ systematic uncertainties, photon flux, GRFF, spin correlation and \tlep decay systematic uncertainties. Finally, ``Other'' includes luminosity, background and unfolding uncertainties. Data statistical uncertainties are not shown as they are much larger than the systematic uncertainties. The largest uncertainty contributions in many bins come from the ``Trigger'', ``Modeling'', and ``Other'' categories.

%
\section{Differential fiducial cross-sections}

\begin{figure}[htbp]
\centering
\subfloat[]{\includegraphics[width=0.33\textwidth]{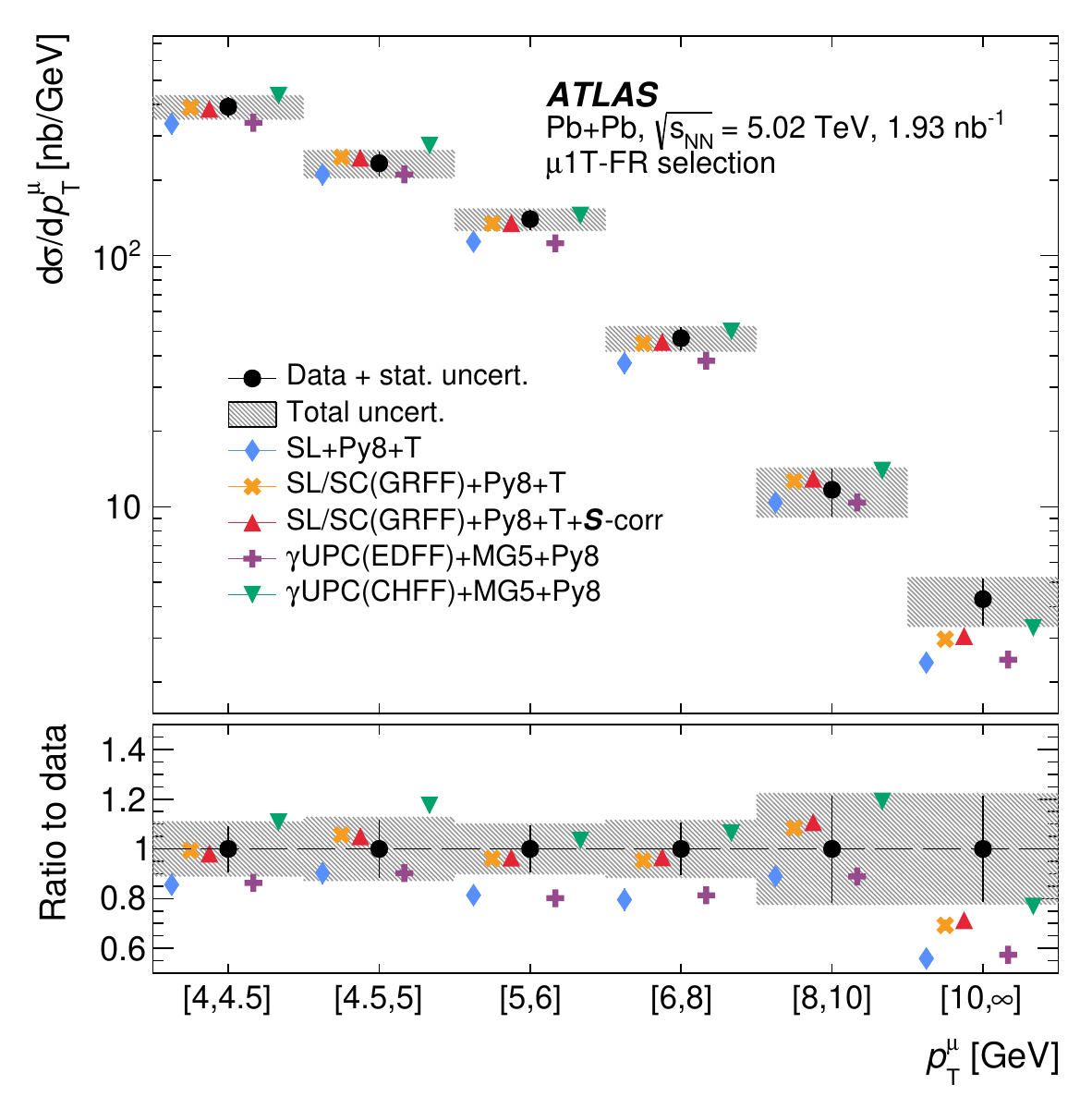}}
\subfloat[]{\includegraphics[width=0.33\textwidth]{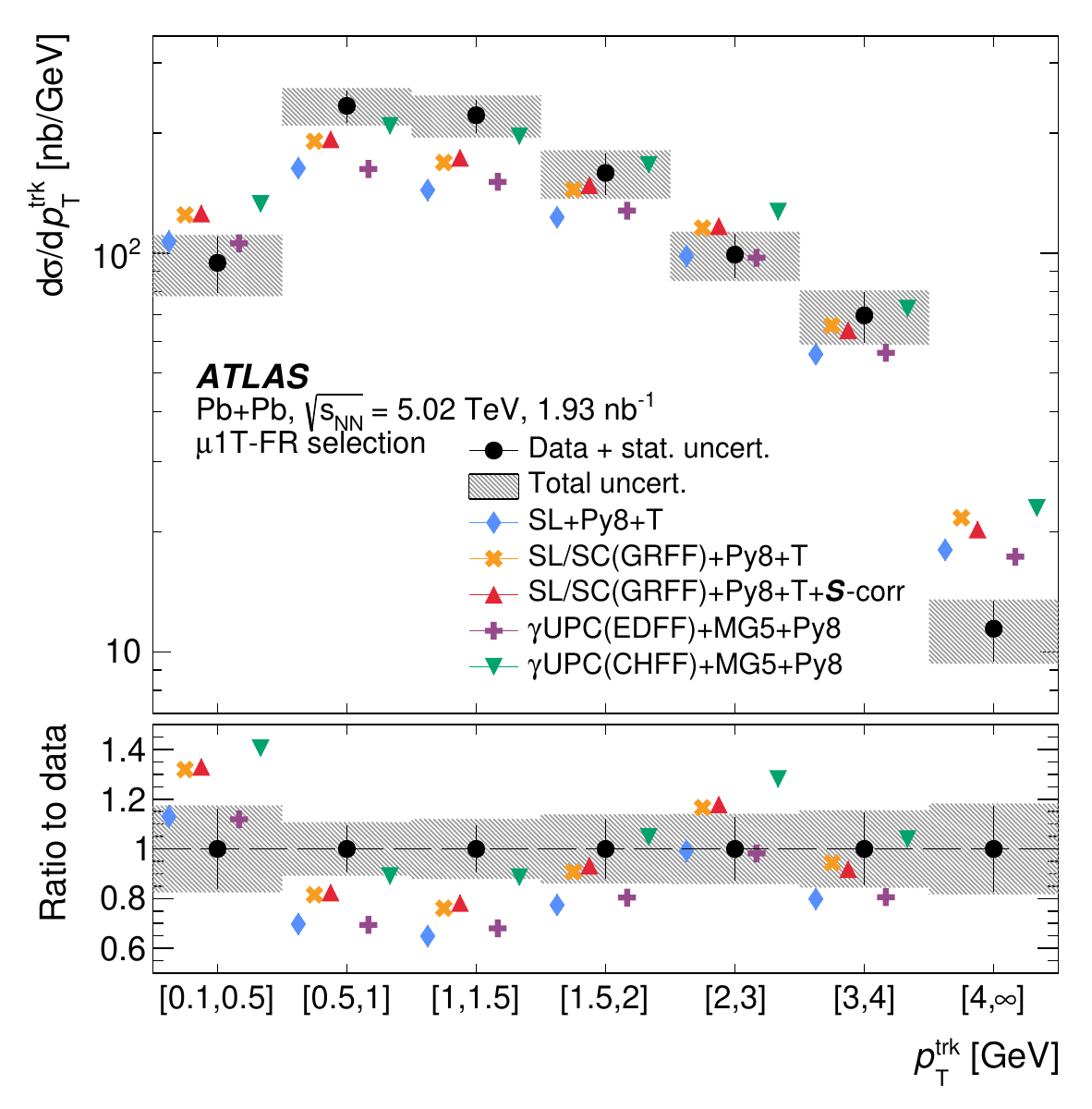}}\\
\subfloat[]{\includegraphics[width=0.33\textwidth]{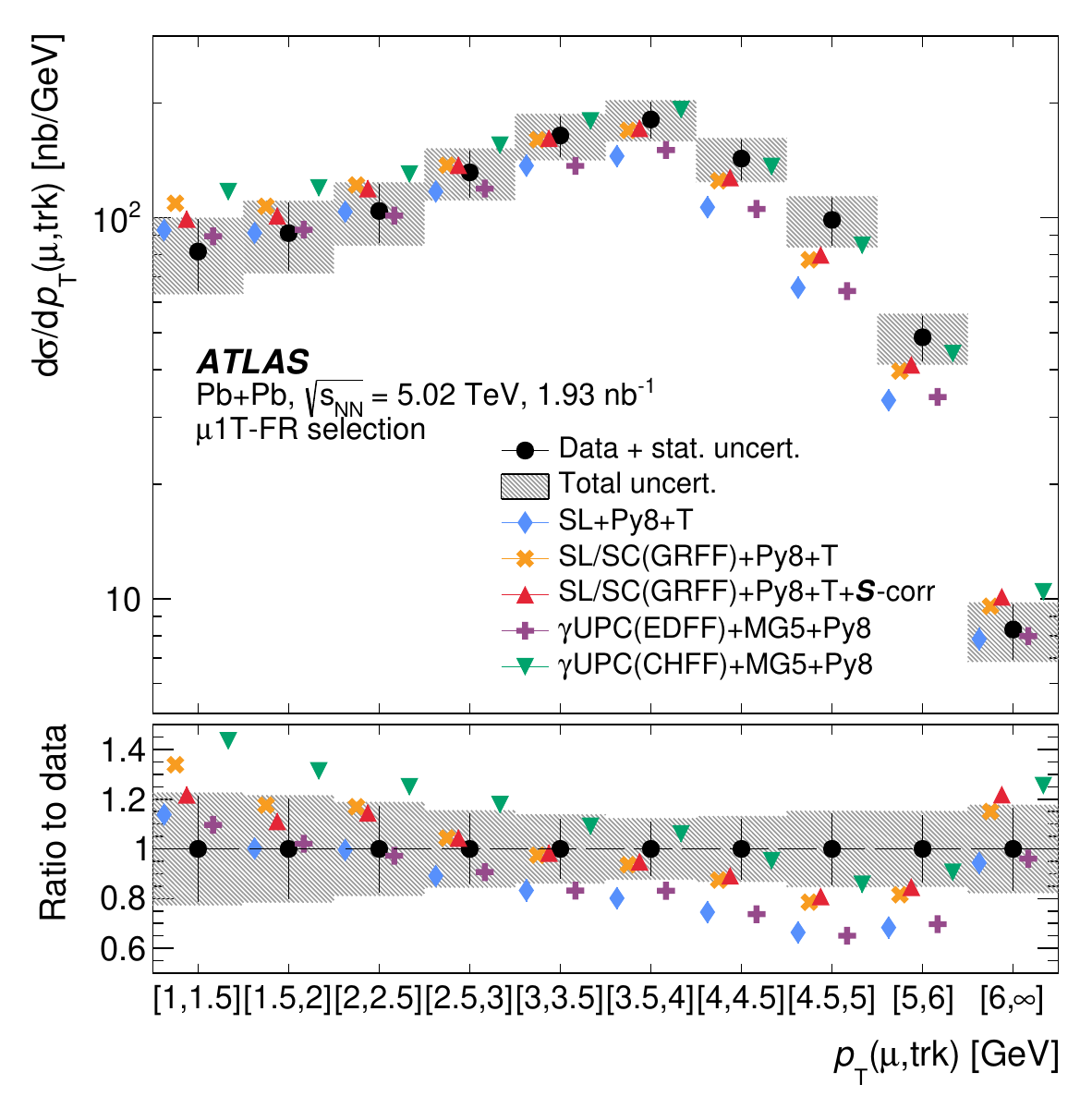}}
\subfloat[]{\includegraphics[width=0.33\textwidth]{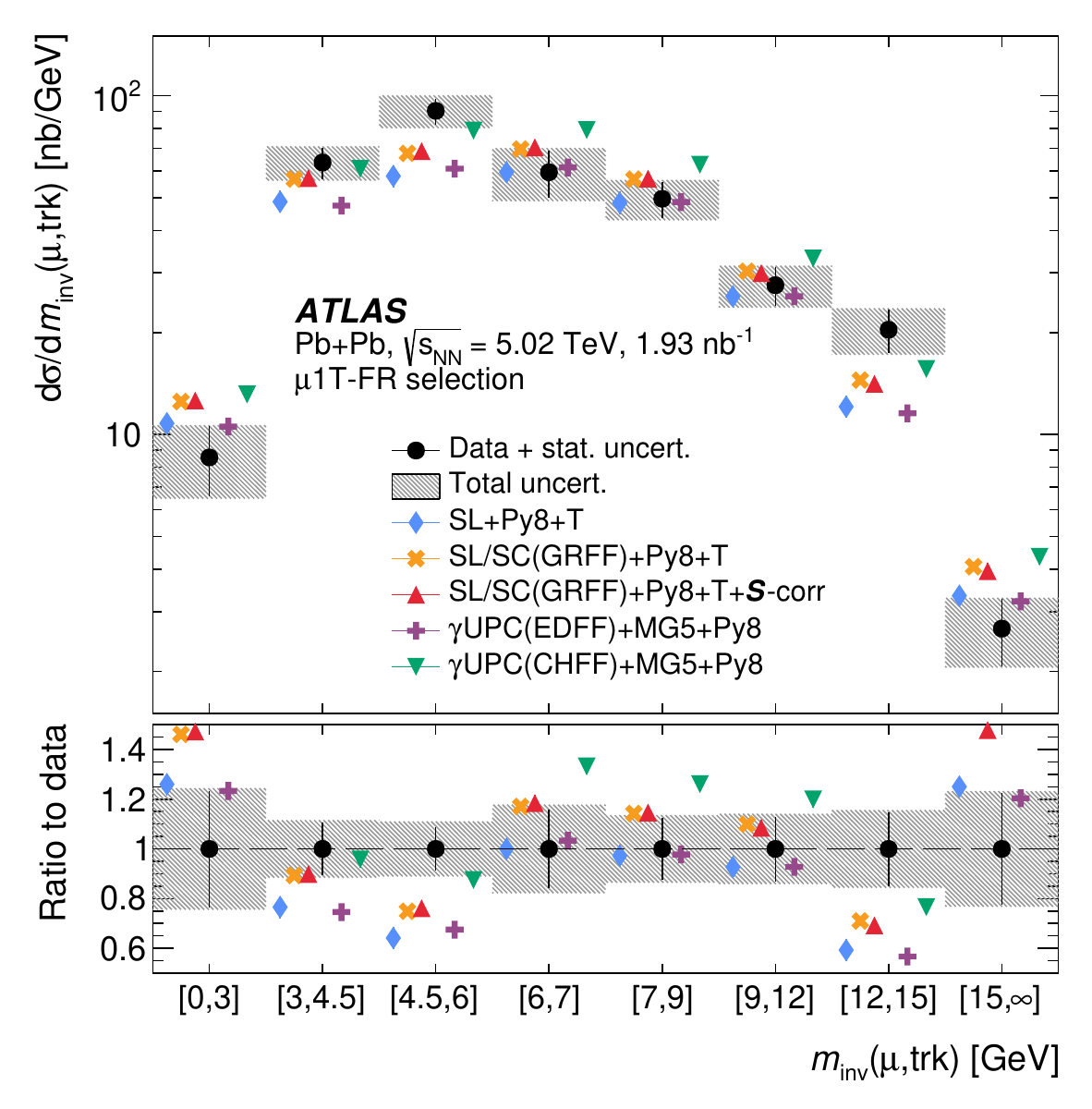}}
\subfloat[]{\includegraphics[width=0.33\textwidth]{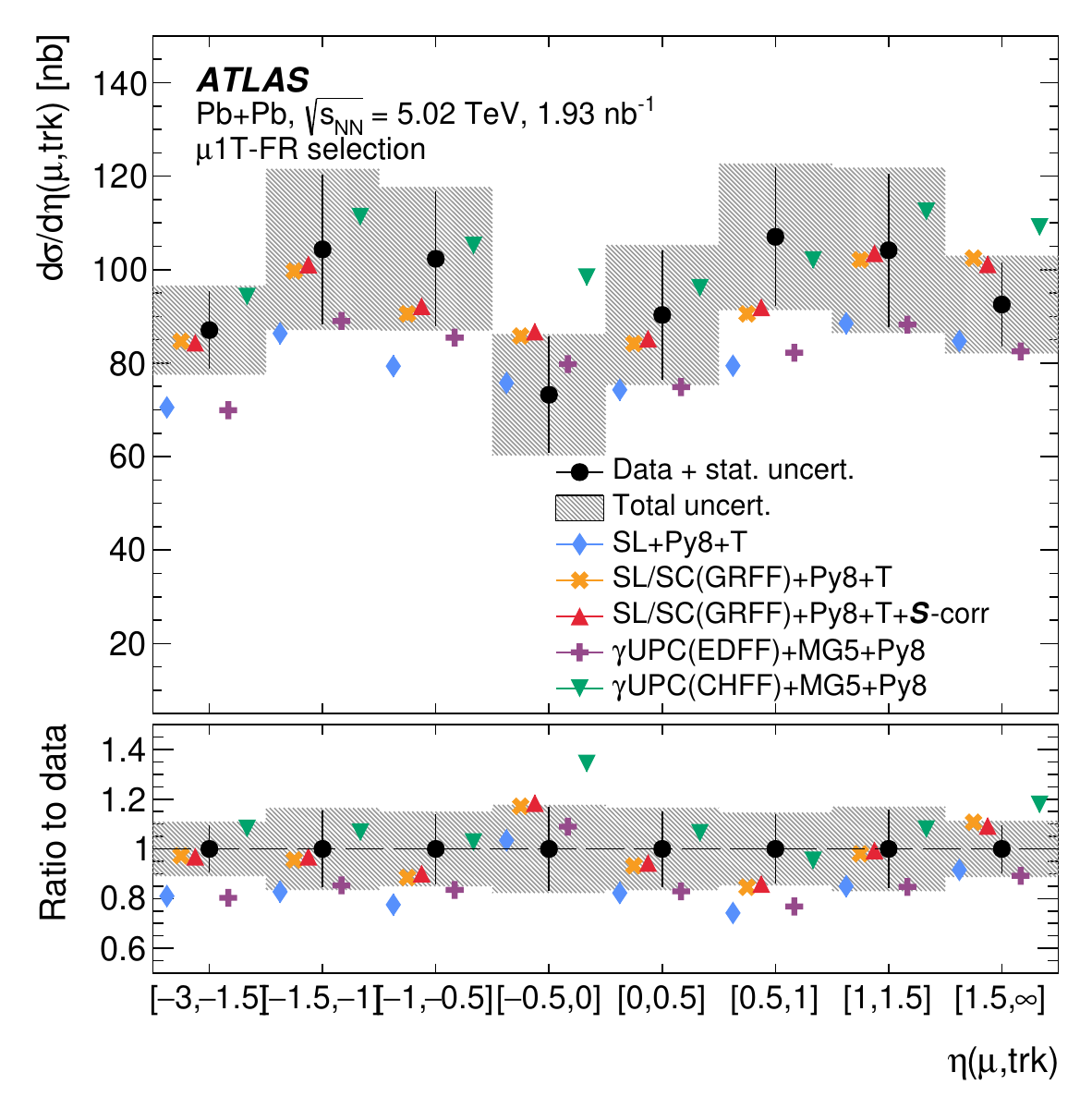}}\\
\subfloat[]{\includegraphics[width=0.33\textwidth]{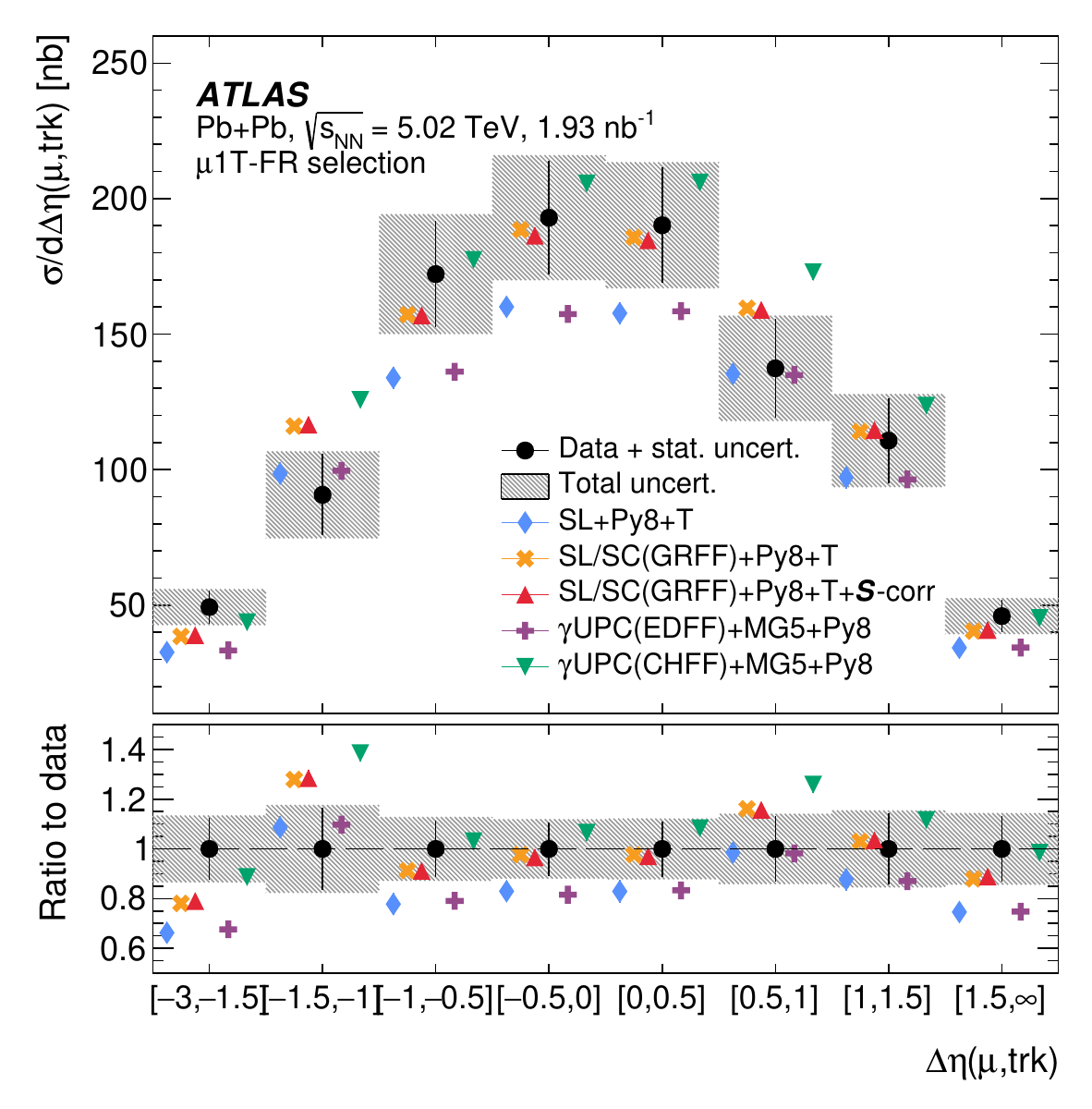}}
\subfloat[]{\includegraphics[width=0.33\textwidth]{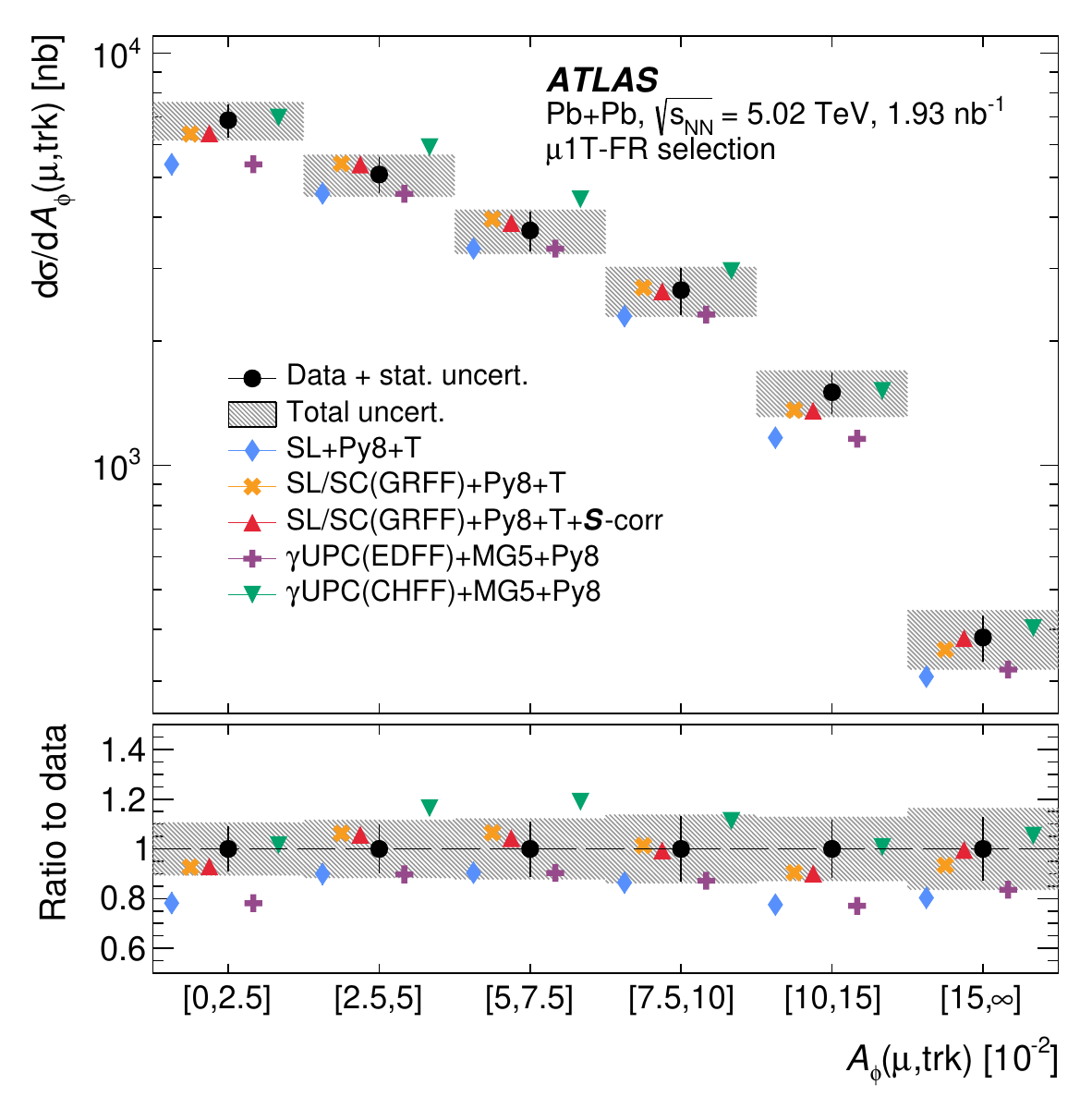}}
\caption{Differential fiducial cross-sections at particle-level for the variables (a) muon $\pt$, (b) track $\pt$, (c) $\pt\left(\mu,\trk\right)$, (d) $m(\mu,\trk)$, (e) $\eta(\mu,\trk)$, (f) $\deta{\mu}{\trk}$, and (g) $A_{\phi}^{\mu,\trk}$, in the $\mu$1T-FR, using combined 2015 and 2018 data. The statistical uncertainty in the data (error bars) and the total uncertainty (hatched band) are shown. The cross-sections are compared with several SM \ggtt predictions: nominal with GRFF (\nomSampGRFFShort), nominal with GRFF and spin correlations (\nomSampGRFFspinShort), varied photon-flux modeling (\varSampPFShort), \gammaUPC predictions with the electric dipole form factor photon-flux (\gammaUPCSampEDFFShort) and the charged form factor photon-flux (\gammaUPCSampCHFFShort). The error bars on the leading-order predictions are the statistical uncertainty. The lower panel displays the ratio of these predictions to data. The respective bin sizes are indicated in square brackets on the $x$-axis. Infinity is used to represent the upper boundary of the fiducial region, within the definitions given in Table~\ref{tab:fiducial_vars}.}
\label{fig:unfolded-crosssection-1M1T-Run2}
\end{figure}

\begin{figure}[htbp]
\centering
\subfloat[]{\includegraphics[width=0.33\textwidth]{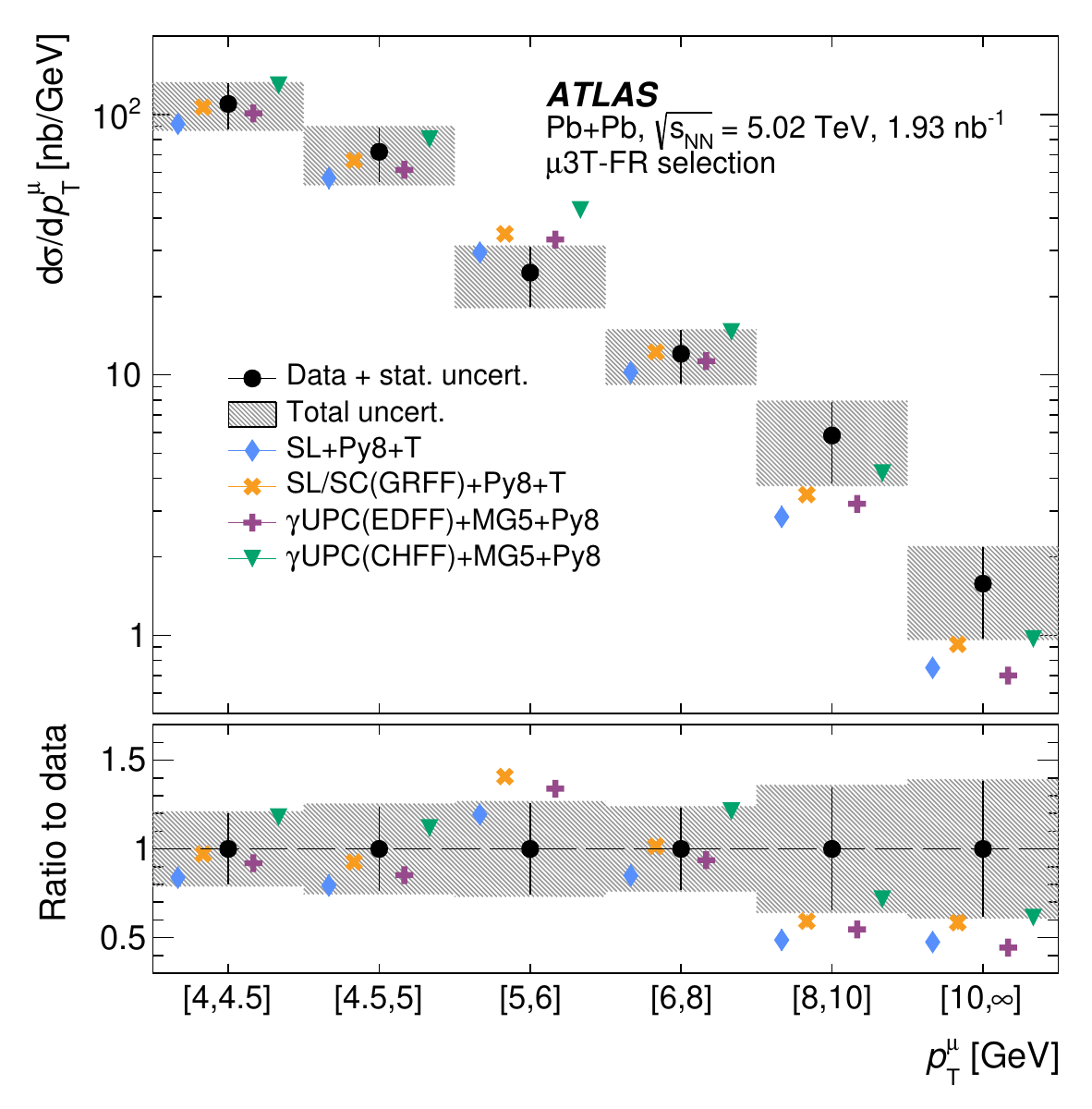}}
\subfloat[]{\includegraphics[width=0.33\textwidth]{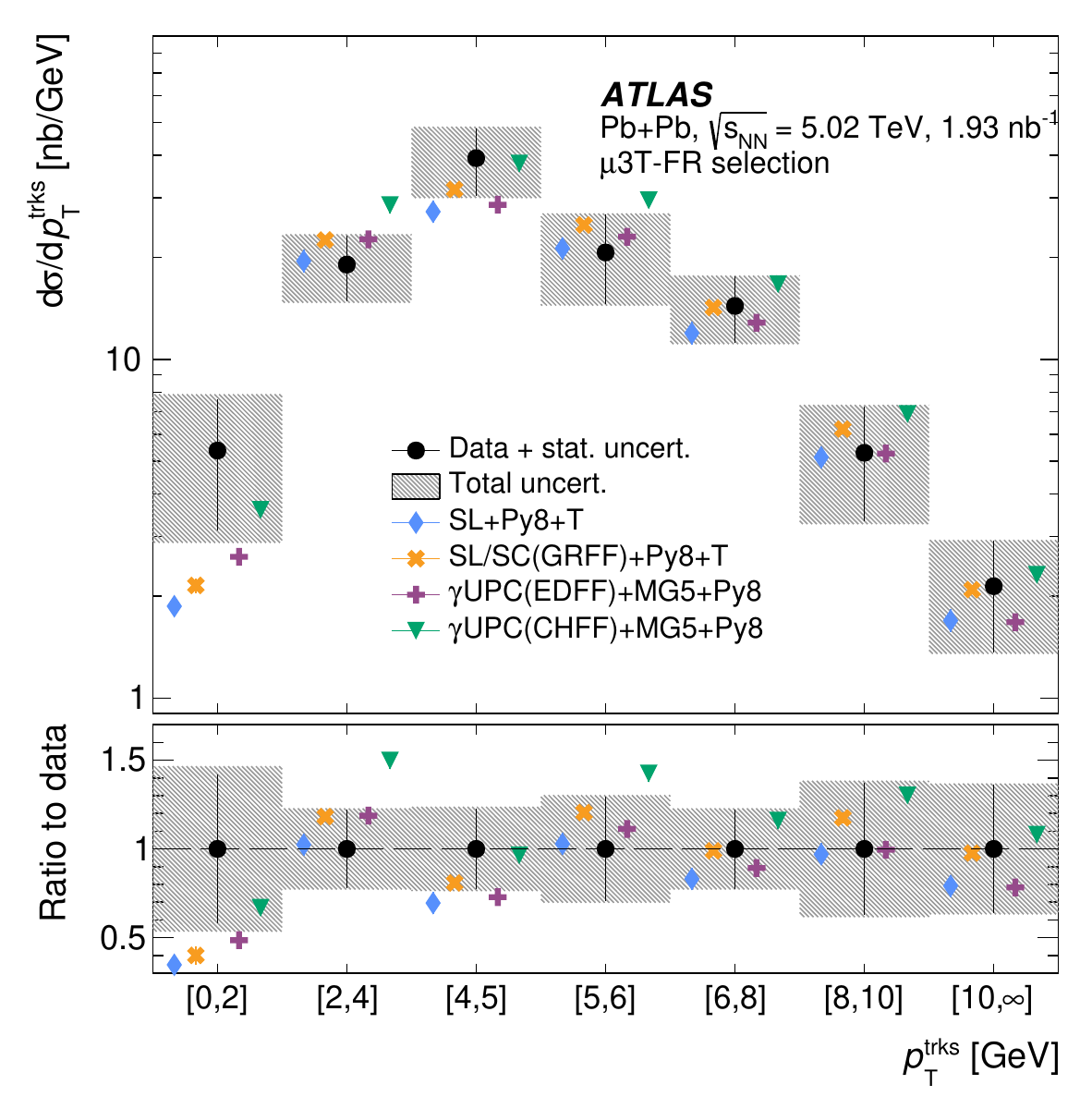}}\\
\subfloat[]{\includegraphics[width=0.33\textwidth]{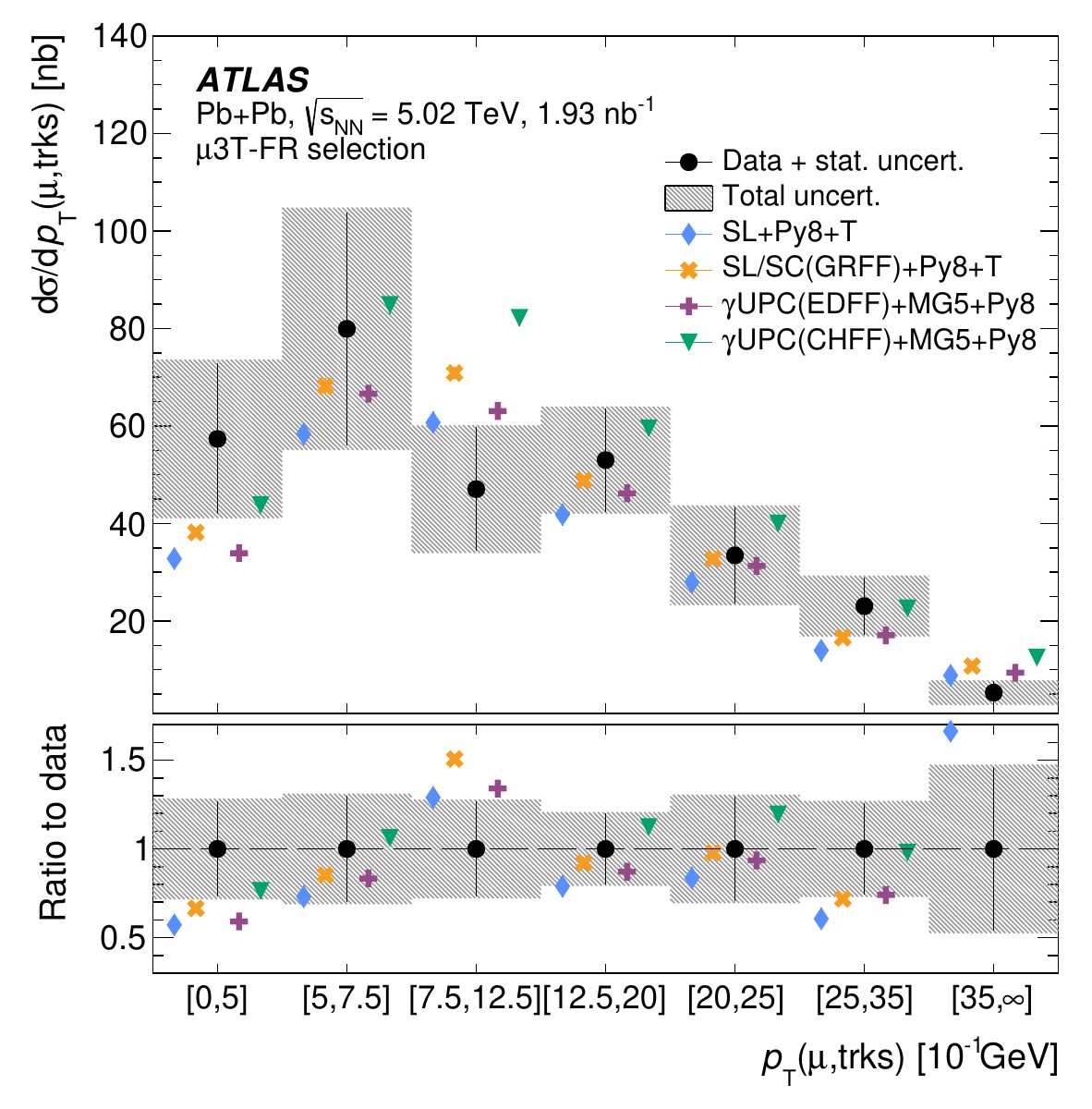}}
\subfloat[]{\includegraphics[width=0.33\textwidth]{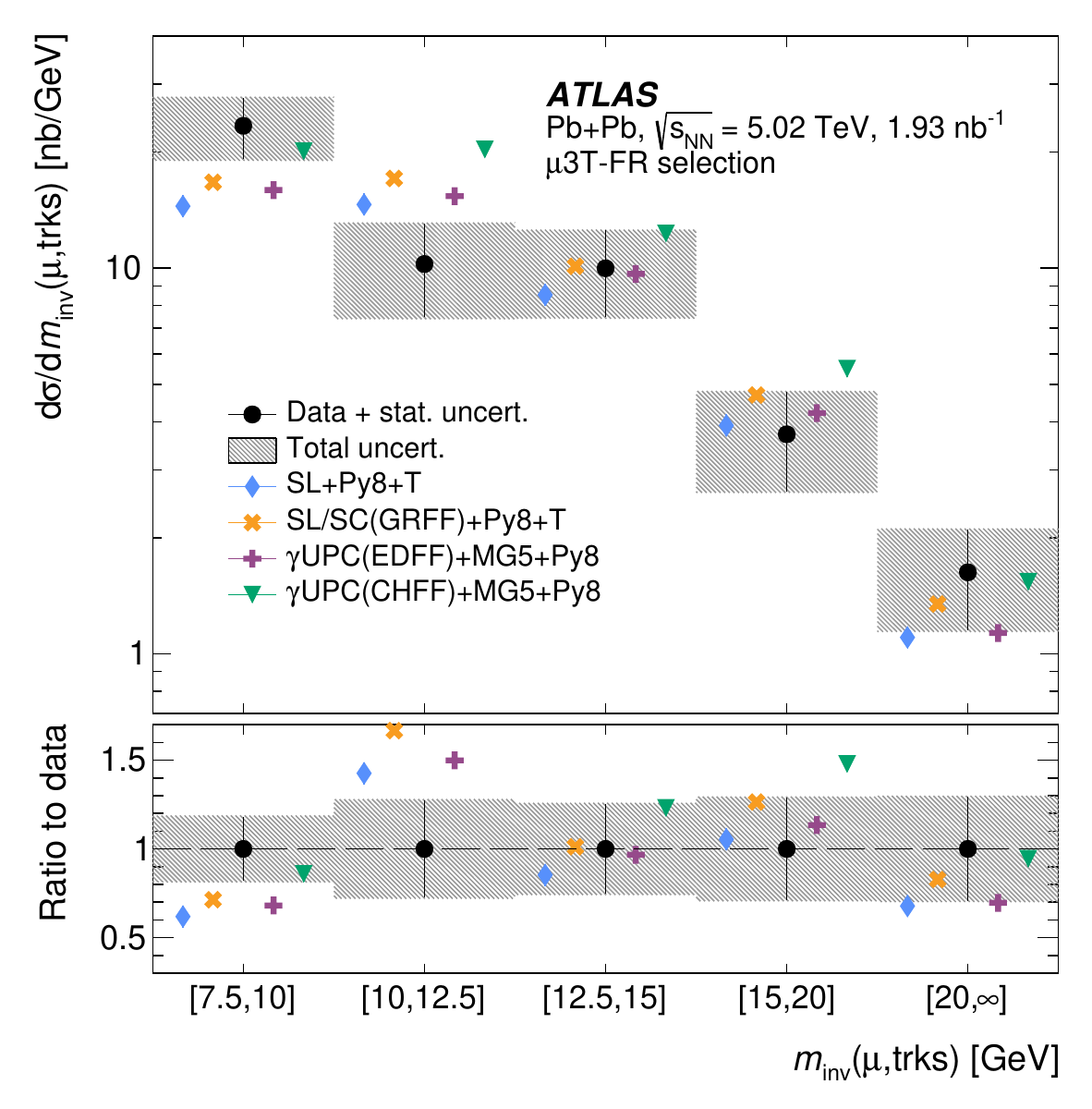}}
\subfloat[]{\includegraphics[width=0.33\textwidth]{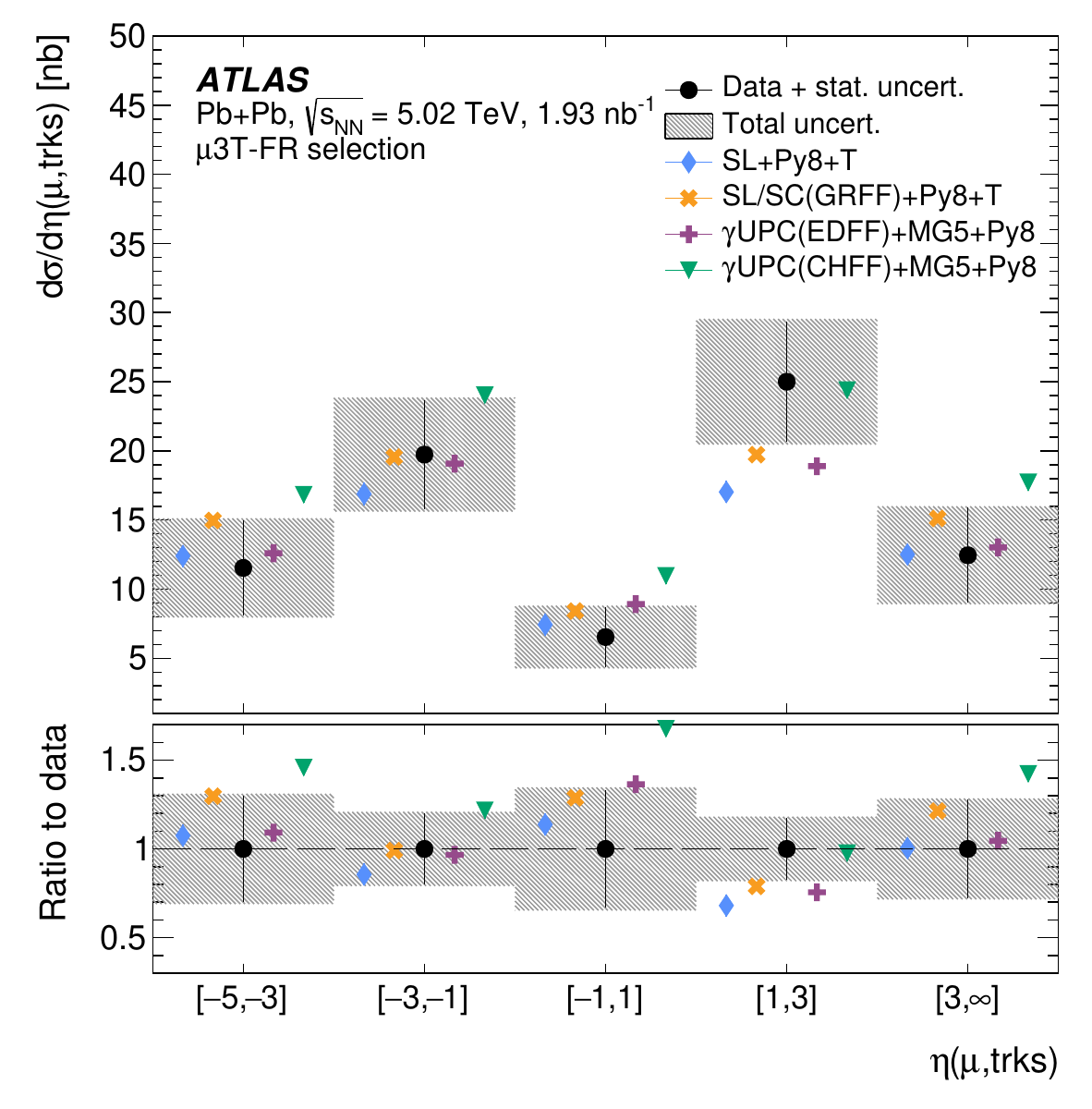}}\\
\subfloat[]{\includegraphics[width=0.33\textwidth]{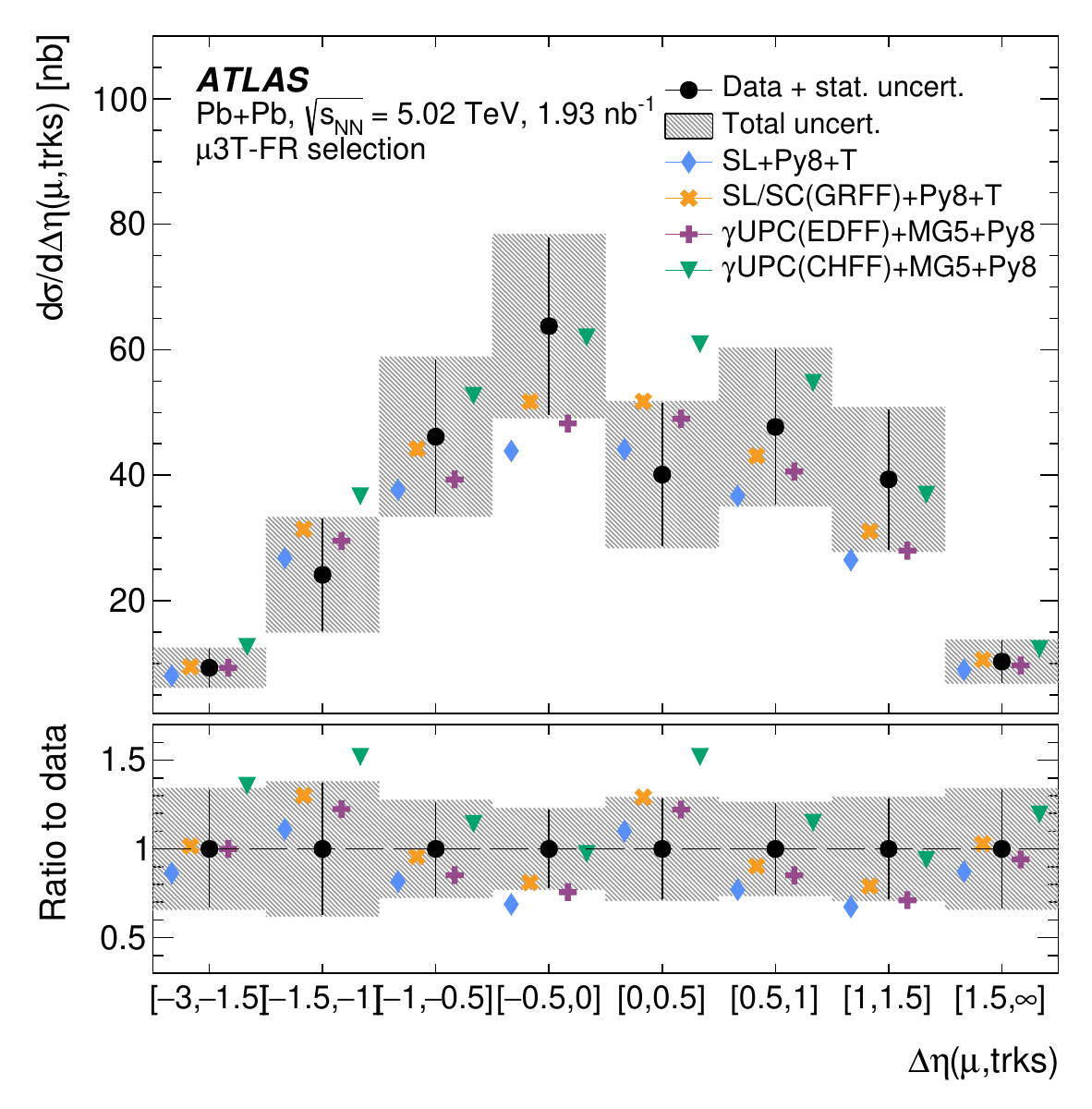}}
\subfloat[]{\includegraphics[width=0.33\textwidth]{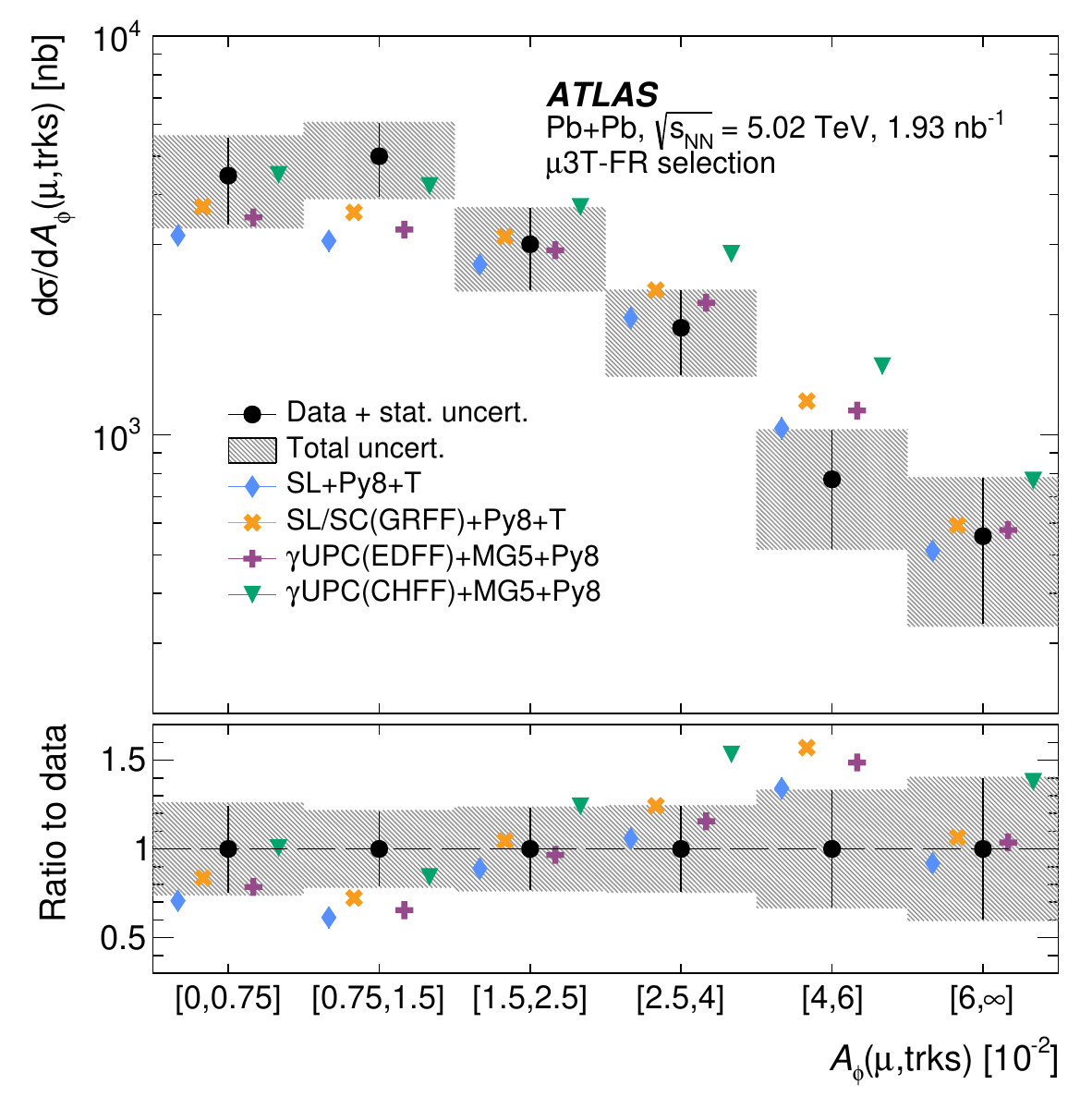}}
\caption{Differential fiducial cross-sections at particle-level for the variables (a) muon $\pt$, (b) $\pt\left(\trks\right)$, (c) $\pt\left(\mu,\trks\right)$, (d) $m(\mu,\trks)$, (e) $\eta(\mu,\trks)$, (f) $\deta{\mu}{\trks}$, and (g) $A_{\phi}^{\mu,\trks}$, in the $\mu$3T-FR, using combined 2015 and 2018 data. The statistical uncertainty in the data (error bars) and the total uncertainty (hatched band) are shown. The cross-sections are compared with several SM \ggtt predictions: nominal with GRFF (\nomSampGRFFShort),
varied photon-flux modeling (\varSampPFShort), \gammaUPC predictions with the electric dipole form factor photon-flux (\gammaUPCSampEDFFShort) and the charged form factor photon-flux (\gammaUPCSampCHFFShort). The error bars on the leading-order predictions are the statistical uncertainty. The lower panel displays the ratio of these predictions to data. The respective bin sizes are indicated in square brackets on the $x$-axis. Infinity is used to represent the upper boundary of the fiducial region, within the definitions given in Table~\ref{tab:fiducial_vars}.}
\label{fig:unfolded-crosssection-1M3T-Run2}
\end{figure}

\begin{figure}[htbp]
\centering
\subfloat[]{\includegraphics[width=0.33\textwidth]{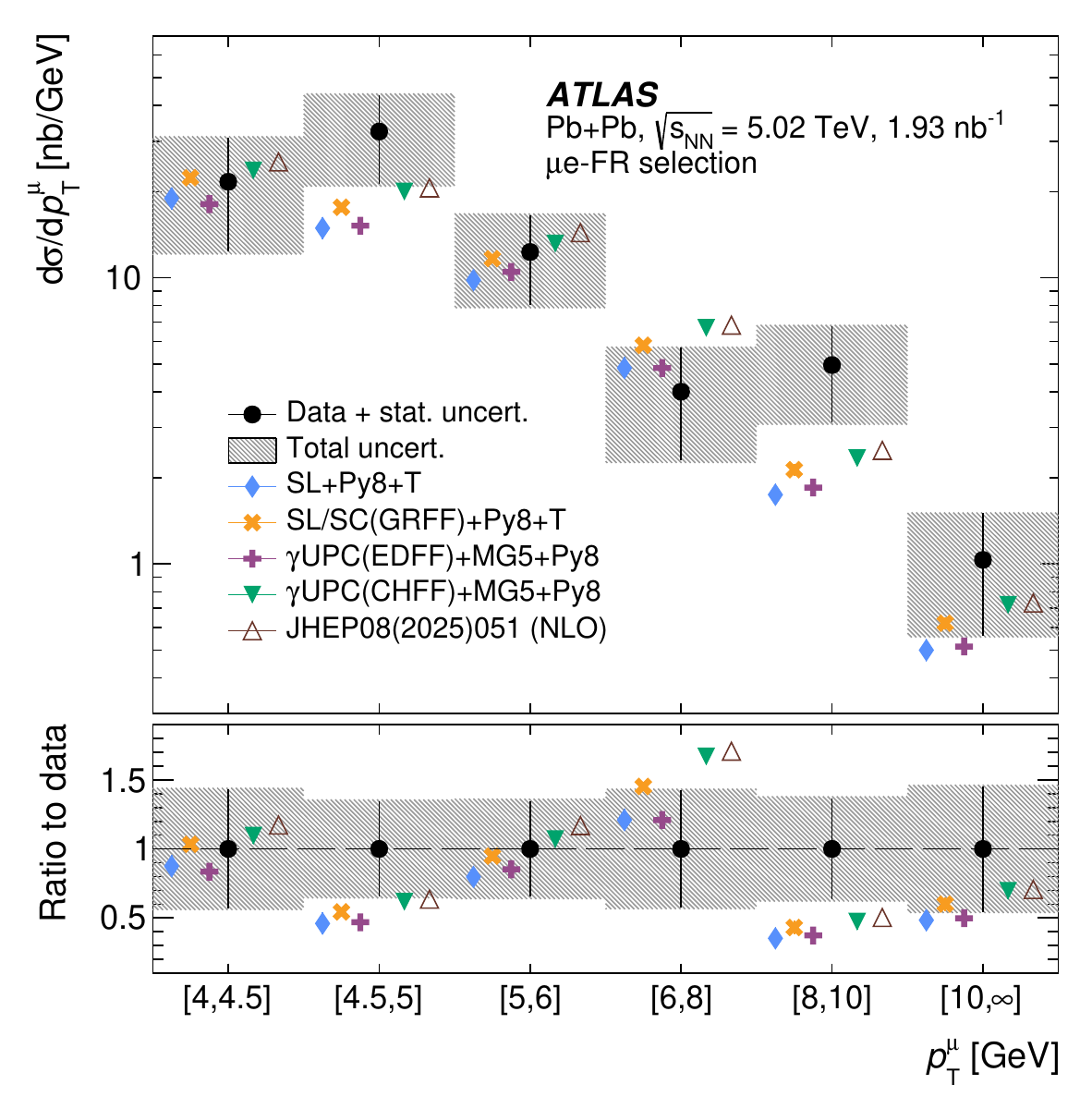}}
\subfloat[]{\includegraphics[width=0.33\textwidth]{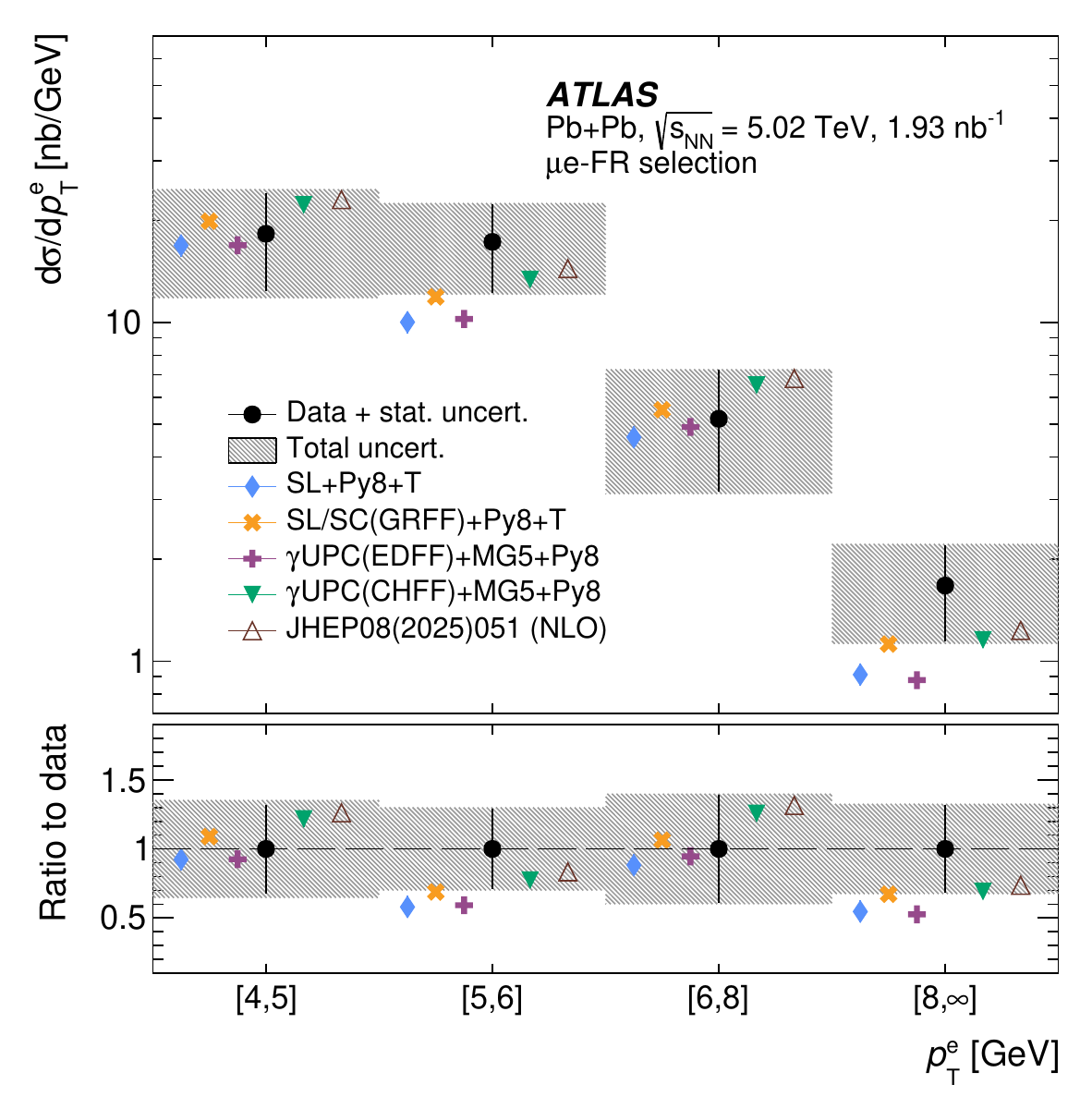}}\\
\subfloat[]{\includegraphics[width=0.33\textwidth]{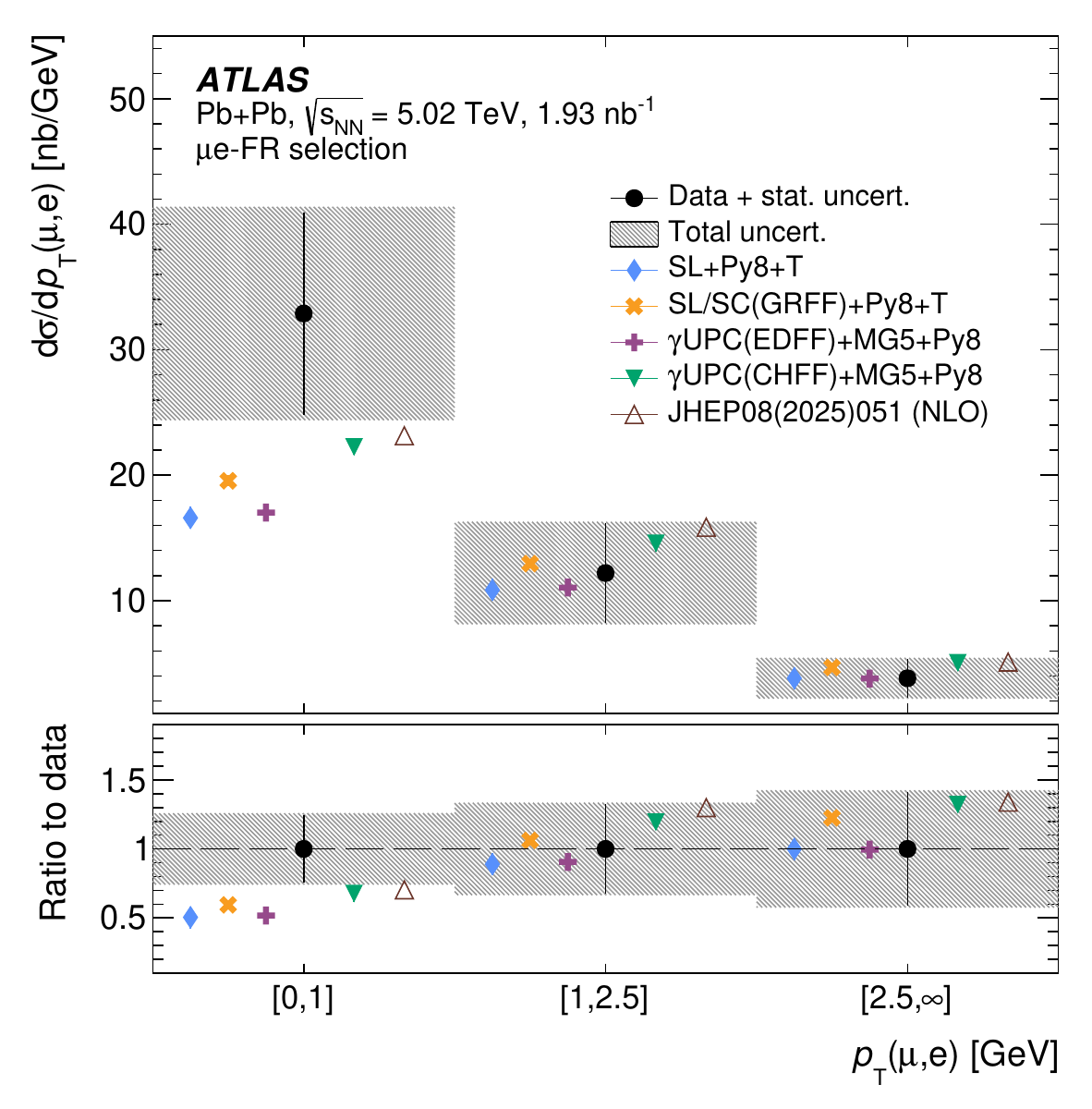}}
\subfloat[]{\includegraphics[width=0.33\textwidth]{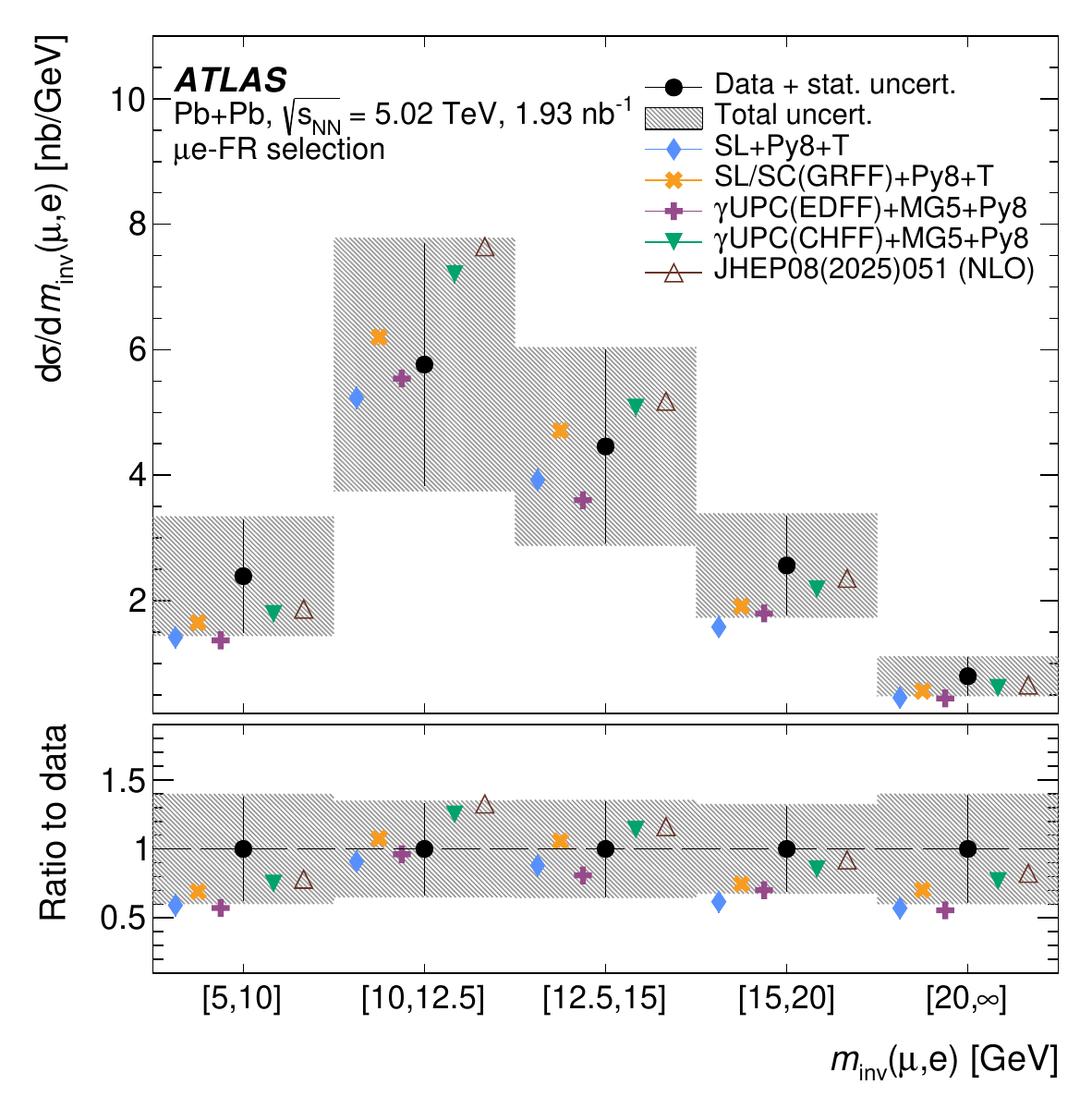}}
\subfloat[]{\includegraphics[width=0.33\textwidth]{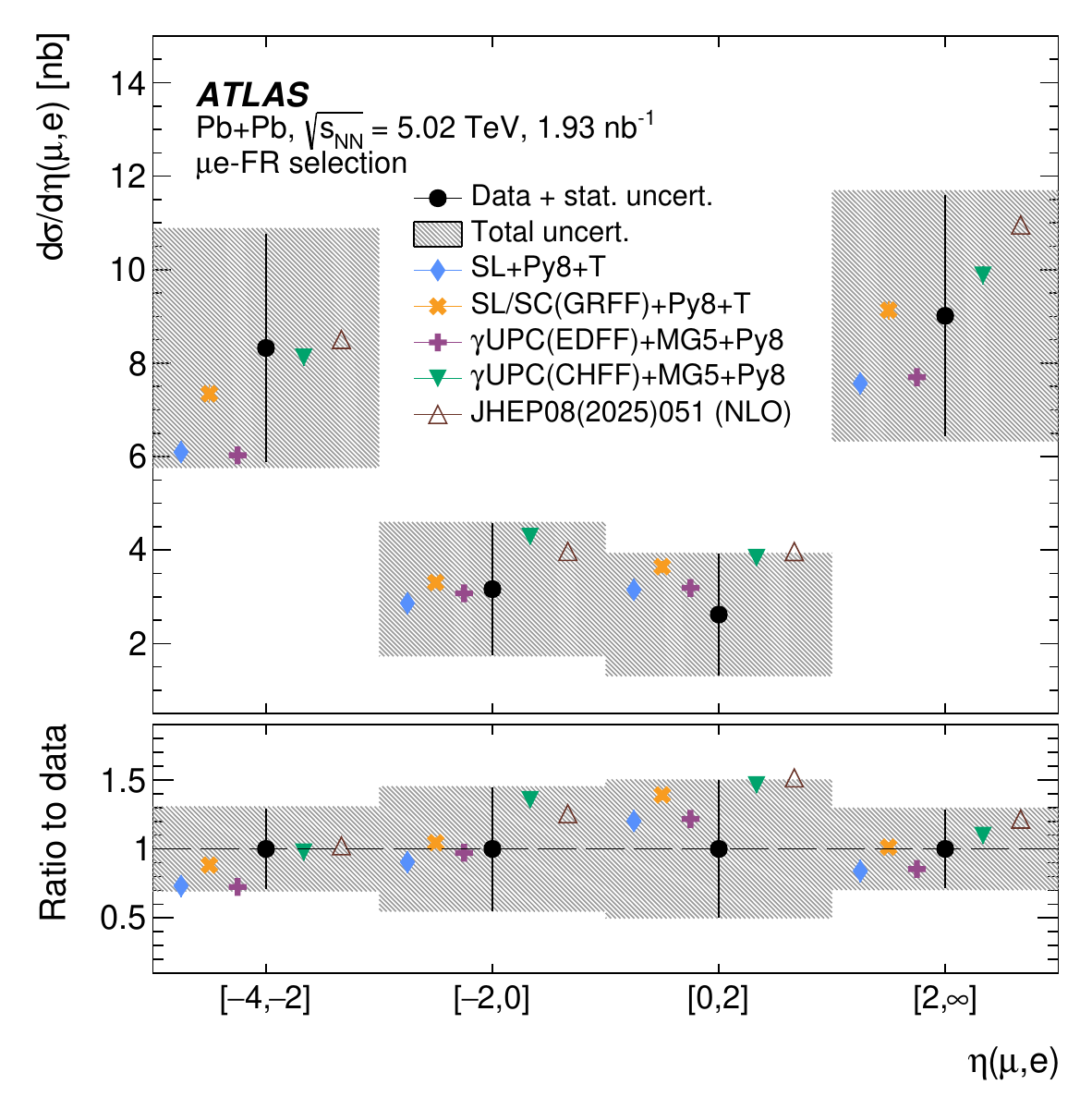}}\\
\subfloat[]{\includegraphics[width=0.33\textwidth]{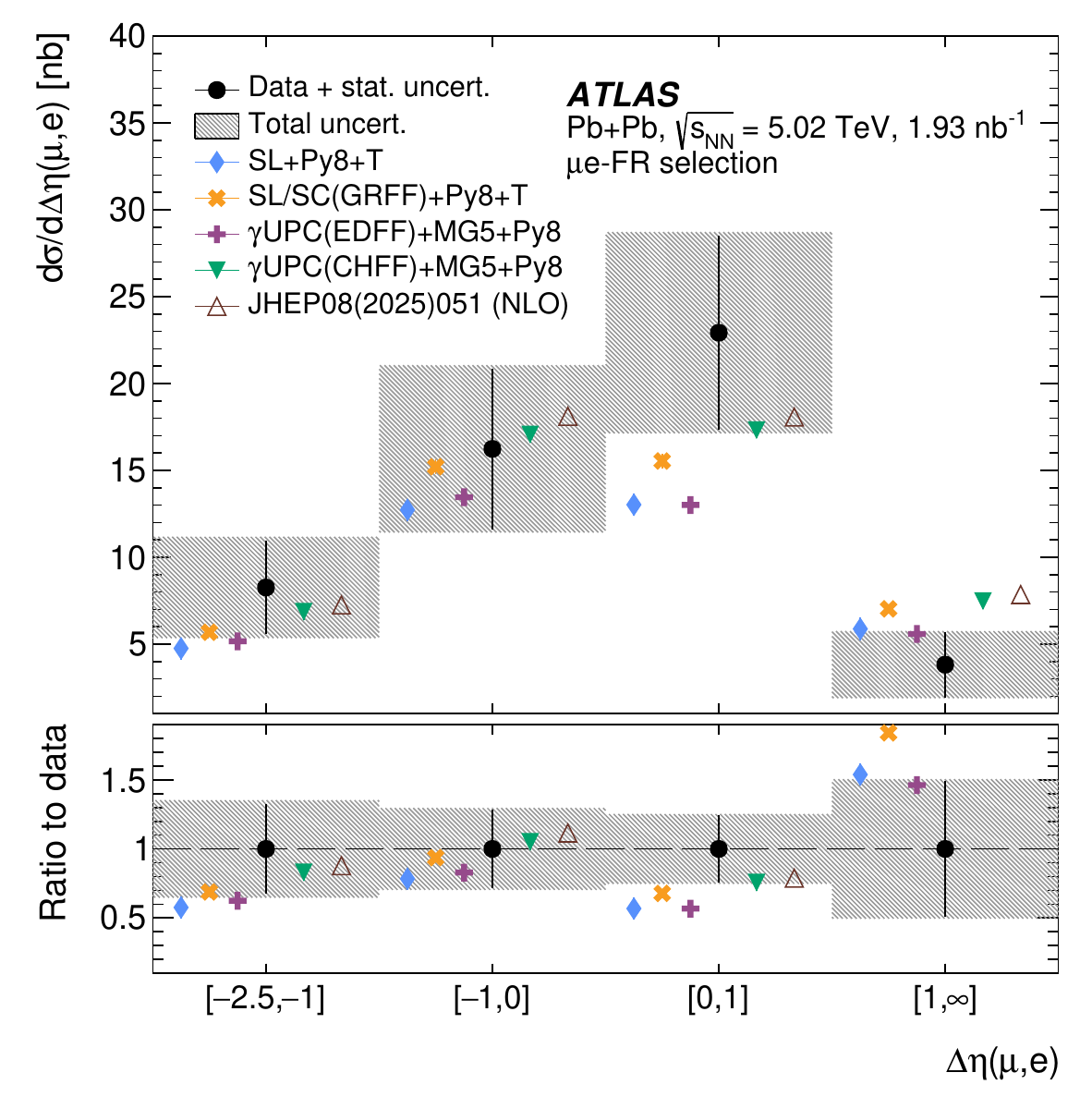}}
\subfloat[]{\includegraphics[width=0.33\textwidth]{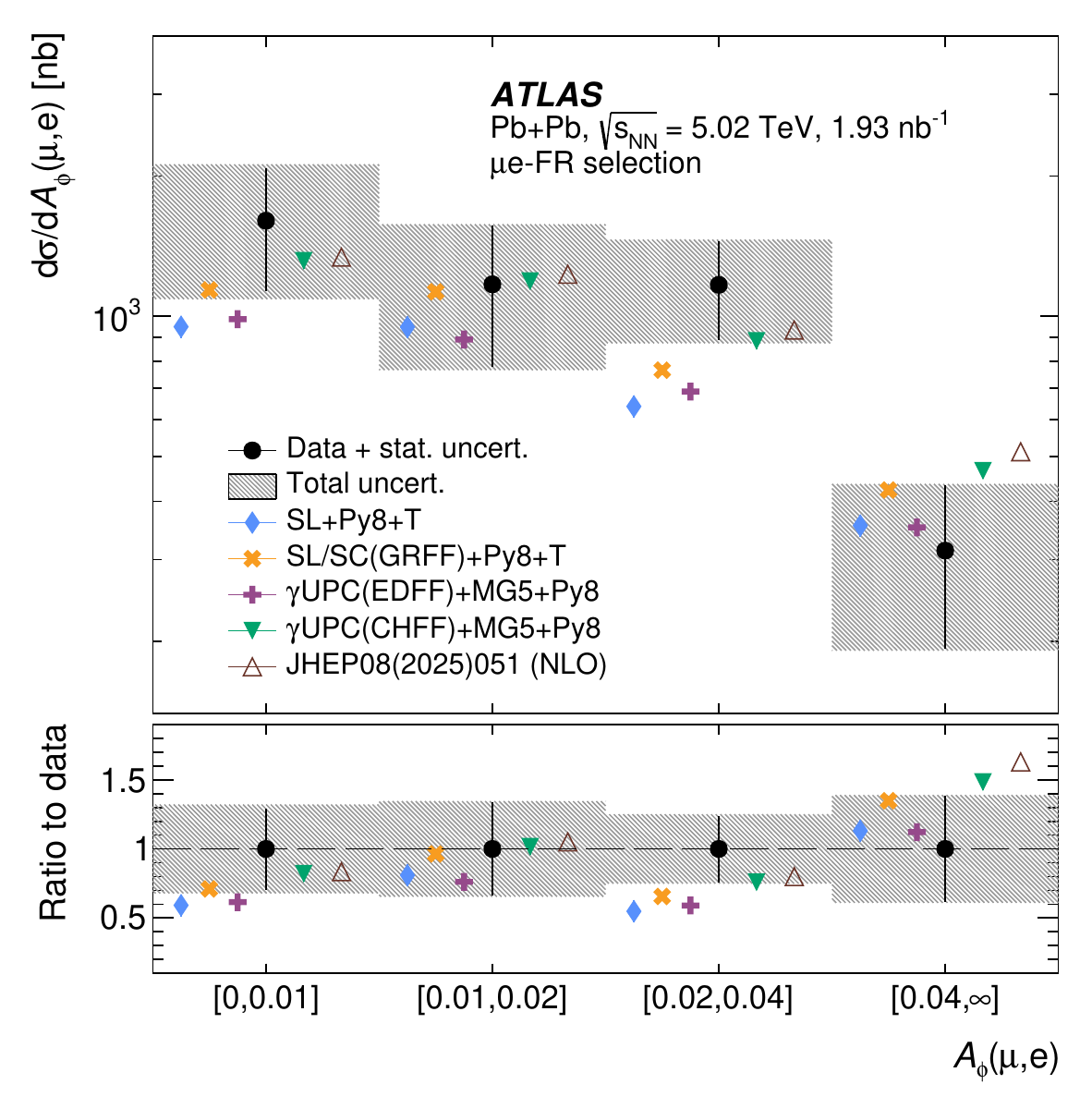}}
\caption{Differential fiducial cross-sections at particle-level for the variables (a) muon $\pt$, (b) electron $\pt$, (c) $\pt\left(e,\mu\right)$, (d) $m(e,\mu)$, (e) $\eta(e,\mu)$, (f) $\deta{e}{\mu}$, and (g) $A_{\phi}^{e,\mu}$, in the $\mu e$-FR, using combined 2015 and 2018 data. The statistical uncertainty in the data (error bars) and the total uncertainty (hatched band) are shown. The cross-sections are compared with several SM \ggtt predictions: nominal with GRFF (\nomSampGRFFShort),
varied photon-flux modeling (\varSampPFShort), \gammaUPC predictions with the electric dipole form factor photon-flux (\gammaUPCSampEDFFShort) and the charged form factor photon-flux (\gammaUPCSampCHFFShort), and NLO predictions (\NLOShort). The error bars on the leading-order predictions are the statistical uncertainty. The lower panel displays the ratio of these predictions to data. The respective bin sizes are indicated in square brackets on the $x$-axis. Infinity is used to represent the upper boundary of the fiducial region, within the definitions given in Table~\ref{tab:fiducial_vars}.}
\label{fig:unfolded-crosssection-1M1E-Run2}
\end{figure}

The differential fiducial cross-sections are measured in the FRs combining the 2015 and 2018 data before unfolding, and treating systematic uncertainties as correlated between the two data samples. The consistency of the 2015 and 2018 data was verified beforehand, and the two datasets were found to agree within uncertainties. The cross-sections per-bin are estimated by dividing the unfolded number of events by the bin width and the central value of the 2015 and 2018 luminosity. The last bin of the measured cross-sections is normalized by the bin width of the last detector-level bin. Figures~\ref{fig:unfolded-crosssection-1M1T-Run2}, \ref{fig:unfolded-crosssection-1M3T-Run2} and \ref{fig:unfolded-crosssection-1M1E-Run2} show the measured differential fiducial cross-sections for the $\mu$1T-FR, $\mu$3T-FR, and $\mu e$-FR, respectively. Beyond enabling constraints on \tlep electromagnetic dipole moments \atau and \dtau, these measurements probe initial-state and final-state dynamics including photon-flux modeling and the underlying nuclear form factors, as well as spin-correlation effects and higher-order electroweak corrections.

The measured differential cross-sections are compared with several SM signal predictions. The nominal \ggtt signal sample with the application of the GRFF, \nomSampGRFF (short form \nomSampGRFFShort) is shown, and with the application of the GRFF and spin-correlation corrections, \nomSampGRFFspin (short form \nomSampGRFFspinShort) for $\mu$1T-FR\@. In addition, predictions removing the reweighting to the \Superchic photon flux, i.e., utilizing the original \Starlight photon flux \varSampPF (short form \varSampPFShort) are shown. For the above listed predictions \textsc{Photos++} was used to simulate FSR from the charged decay products of the $\tau$-leptons.

Predictions using the \gammaUPC 1.0 generator integrated with \Madgraph 3.5.1 and using \PYTHIA[8.245] to handle \tlep decays and FSR from the $\tau$-leptons and \textsc{Photos++} 3.61~\cite{Davidson:2010ew} to simulate FSR from the charged decay products of $\tau$-leptons are also displayed. Two different assumptions on the modeling of the ion nucleus form factor, and hence the photon flux, are shown. The electric dipole form factor approach (short form \gammaUPCSampEDFFShort) assumes a point-like distribution for the Pb charge, but restricts the impact parameter integration range to distances from the nucleus center larger than the geometric nuclear radius, similarly to \Starlight. The charged form factor approach (short form \gammaUPCSampCHFFShort) assumes a 3-parameter Woods-Saxon distribution for the Pb charge, with no restriction on the impact parameter integration range, similarly to \Superchic.

For the $\mu e$-FR, NLO electroweak (EW) predictions based on the results of Ref.~\cite{Dittmaier:2025ikh} are shown (short form \NLOShort). The authors do the full matrix element calculation for the $\gamma\gamma\rightarrow \tau^+\tau^- \rightarrow e^+\mu^- 4\nu$ process instead of splitting the calculation between the \tlep pair production and decay, thereby retaining full spin correlation effects. A mixed EW input parameter scheme is utilized and the \gammaUPC charged form factor photon flux is used for the initial state, so the predictions are at a similar scale as the \gammaUPCSampCHFFShort predictions. The NLO predictions shown in Figure~\ref{fig:unfolded-crosssection-1M1E-Run2} were prepared by the authors of Ref.~\cite{Dittmaier:2025ikh} including applying an additional weighting factor for the combined EM-pileup-corrected $0n0n$ weights, since the results of Ref.~\cite{Dittmaier:2025ikh} are nominally presented for the case where no requirement on forward neutron activity is imposed.

Overall, reasonable agreement is observed between the measured differential fiducial cross-sections and the SM predictions in most of the bins. The \pt of the visible di-$\tau$ system ($\pt(\mu,\text{trk})$, $\pt(\mu,\text{trks})$, $\pt(\mu,e)$ in $\mu$1T-FR, $\mu$3T-FR and $\mu e$-FR, respectively) shows a slope in the ratio of the nominal prediction to data, though in opposite directions in $\mu$1T-FR vs. $\mu$3T-FR and $\mu e$-FR\@. For the nominal sample, the GRFF and spin-correlation corrections improve the agreement with the unfolded data. The predictions using the \Starlight photon flux (\varSampPF) and the \gammaUPC electric dipole form factor photon flux (\gammaUPCSampEDFFShort) shows poorer agreement with the data for most bins across the various distributions in all FRs.

The difference between the photon flux description in \Superchic and \Starlight has the largest impact on the predictions where predictions with the photon flux from \Starlight are typically below the ones from \textsc{Superchic.} The difference between the two predictions is typically in the range of 10\% and 30\%. The NLO EW predictions shown for the $\mu e$-FR are typically $\sim$1--2\% lower than the corresponding leading-order prediction. The overall measurement precision ranges typically between 12\% and 45\%, depending on the bin, the observable and the FR.

%
\section{Determination of the \tlep's electromagnetic moments}
\label{sec:dipoles}

The values of the electromagnetic moments of the \tlep (\atau and \dtau) can influence the interaction probabilities for \ggtt production and modify the expected event yields and the shape of differential distributions. To extract the values of \atau and \dtau best describing the data, predictions for a range of \atau and \dtau values are required. An event-level reweighting procedure, using generator-level \tlep kinematics and SM tree-level (i.e., \atau=\dtau=0) QED matrix element calculations of the \ggtt process, was employed to obtain these templates. In this procedure, the $\gamma\tau\tau$ vertex function is parameterized, depending on the momentum transfer, $q$, as follows:

\begin{equation}
i\Gamma^{(\gamma \tau\tau)}_{\mu}(q) =
-ie\left[ \gamma_{\mu} F_{1}(q^{2})+ \frac{i}{2 m_{\tau}}
\sigma_{\mu \nu} q^{\nu} F_{2}(q^{2}) + \frac{1}{2 m_{\tau}} \gamma^5
\sigma_{\mu \nu} q^{\nu}  F_{3}(q^{2})
\right]
\,,
\label{eq:gttvertex}
\end{equation}

where $F_1(q^2)$ and $F_2(q^2)$ are the Dirac and Pauli form factors, $F_3(q^2)$ is the electric dipole form factor and $\sigma_{\mu \nu} =i[\gamma_{\mu},\gamma_{\nu}]/2$ denotes the spin tensor, proportional to the commutator of the gamma matrices. In the limit \( q^2 \to 0~\GeV^2\), the asymptotic values of the form factors describe the electromagnetic properties of the $\tau$-lepton, with:
\begin{equation}
F_1(0) = 1\,,\;F_2(0) = a_{\tau}\,,\;\text{ and }F_3(0) = d_{\tau} \frac{2m_\tau}{e}\,.
\end{equation}
Since the photon virtualities in UPC Pb+Pb collisions at the LHC are near-zero, this asymptotic condition is well satisfied. A similar parameterization was employed in previous ATLAS and LEP results~\cite{STDM-2019-19,Ackerstaff:1998mt,Acciarri:1998iv,Abdallah:2003xd}. More details on the event-level reweighting procedure are provided in Appendix~\ref{app:reweight}. To extract the best-fit \atau and \dtau values, 76 templates were produced for each \atau and \fthrz with a spacing of 0.002 for $|\atau~\mathrm{or}~\fthrz|<0.06$ and a spacing of 0.005 for $0.06<|\atau~\mathrm{or}~\fthrz|<0.10$. The weights are applied to the nominal \nomSamp signal MC simulation. Furthermore, the GRFF is also applied for both the \ggtt MC simulation, and the \ggmmg background simulation. Effects of spin correlations between the two produced \tleps in \ggtt production, and their propagation to the \tlep decay products are included for $\mu$1T-SR via bin-by-bin correction factors.

The values of \atau and \dtau best describing the data are extracted using a maximum-likelihood fit~\cite{trexfitter,Cowan:2010js,Cranmer:1456844} of the detector-level muon \pt distributions in the three SRs, $\mu$1T-SR, $\mu$3T-SR, and $\mu e$-SR, as muon \pt was found to be the most sensitive variable from a set of tested variables. The fits are performed independently, assuming the tree-level SM value for \atau, i.e., zero, in the \dtau fit and vice versa. The likelihood $\mathcal{L}$ for the parameter of interest (POI) \atau or \dtau is given as the product of the Poisson probability distributions of the observed event count in data $n_i$ given the expected values $\nu_i$ for each bin $i$ of these distributions:
\begin{equation}
\mathcal{L}(\atau) = \prod_{i\,\in\text{ bins}} P\left(n_i|\nu_i\left(\atau~\mathrm{or}~\dtau,\boldsymbol{\theta}\right)\right) \times  \prod_{j\,\in\text{ NPs}} G(\theta_j)
\end{equation}
The expected values $\nu_i$ include the \ggtt signal prediction as a function of \atau or \dtau and the background estimates.
Systematic uncertainties $\boldsymbol{\theta}$ are added as nuisance parameters (NPs) in the fit, including additional terms with Gaussian probability distributions $G(\theta_j)$ centered at zero with standard deviations corresponding to the respective systematic uncertainty $\theta_j$. In this way, large shifts in the nuisance parameters are penalized through a reduction of the overall likelihood, while small shifts are possible if the agreement with data is improved in this way. Statistical uncertainties in the background predictions are included as separate nuisance parameters for each bin. Statistical uncertainties in the signal MC prediction largely fall below a minimal-size threshold of 0.3\% and those that do fall below this threshold are neglected in the fit to improve numerical stability. %
In the fit, the value of \atau or \dtau that minimizes the negative log-likelihood (NLL) is extracted.

\begin{figure}[htbp]
\centering
\includegraphics[width=0.37\textwidth]{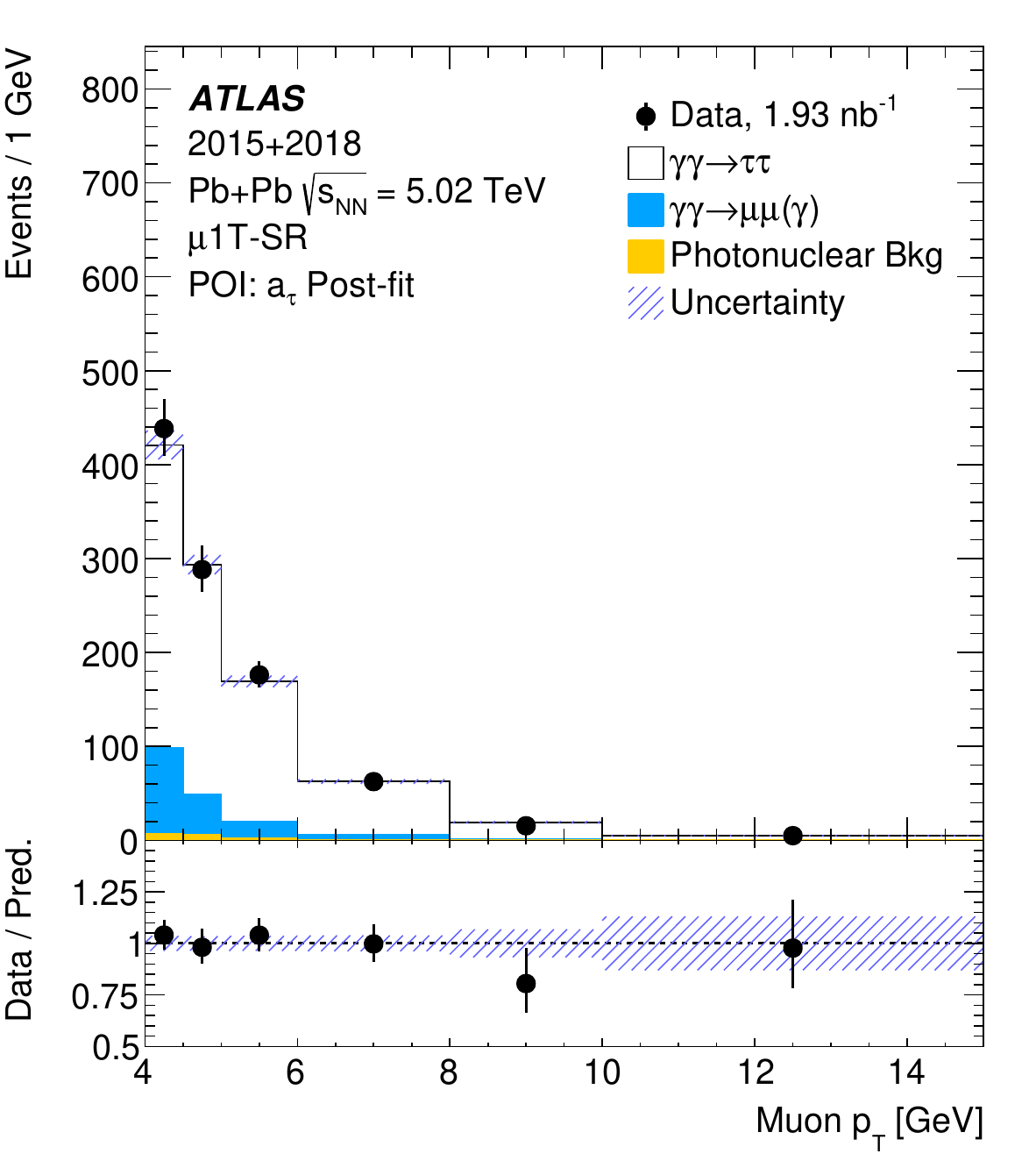}%
\includegraphics[width=0.37\textwidth]{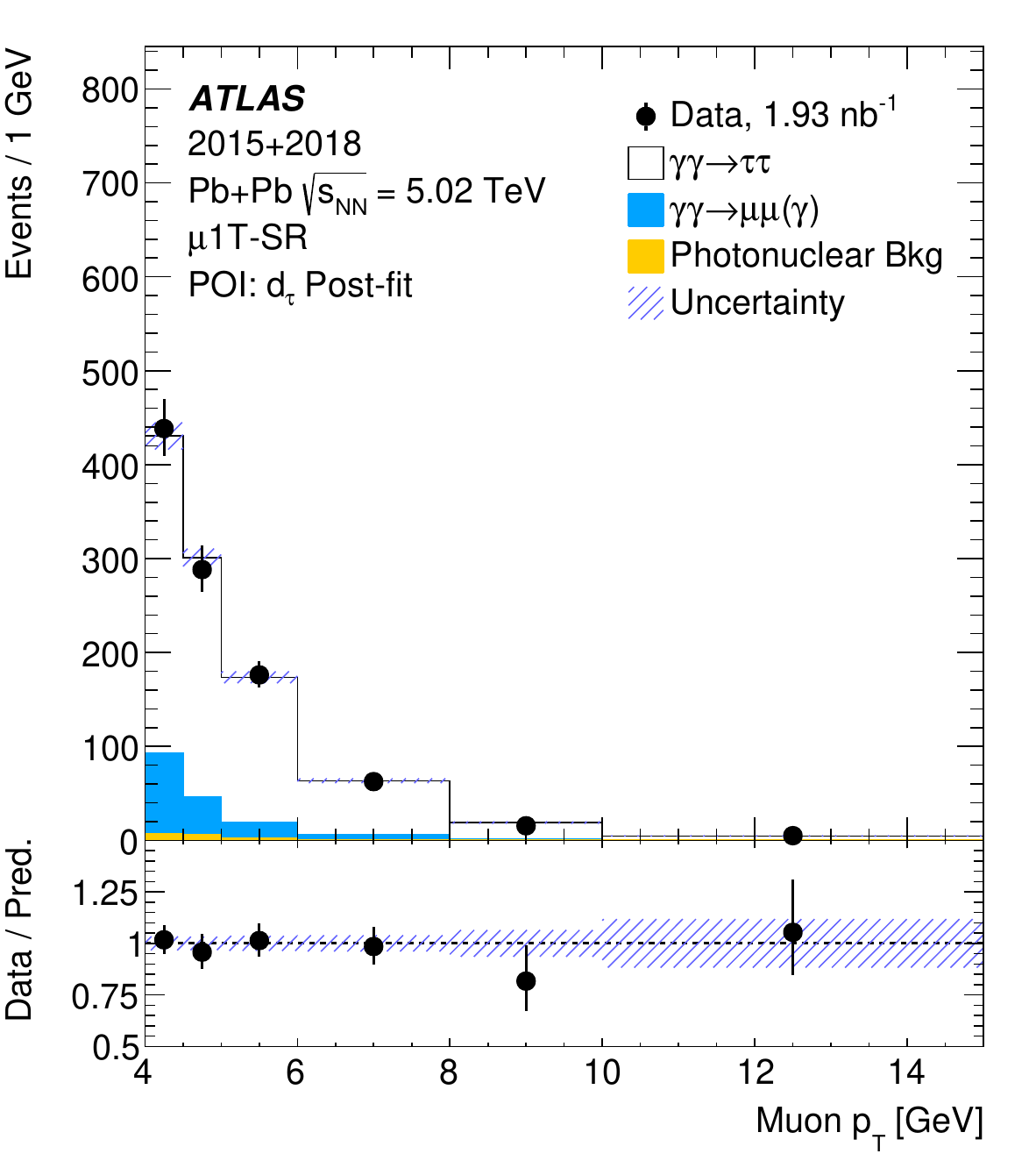}
\includegraphics[width=0.37\textwidth]{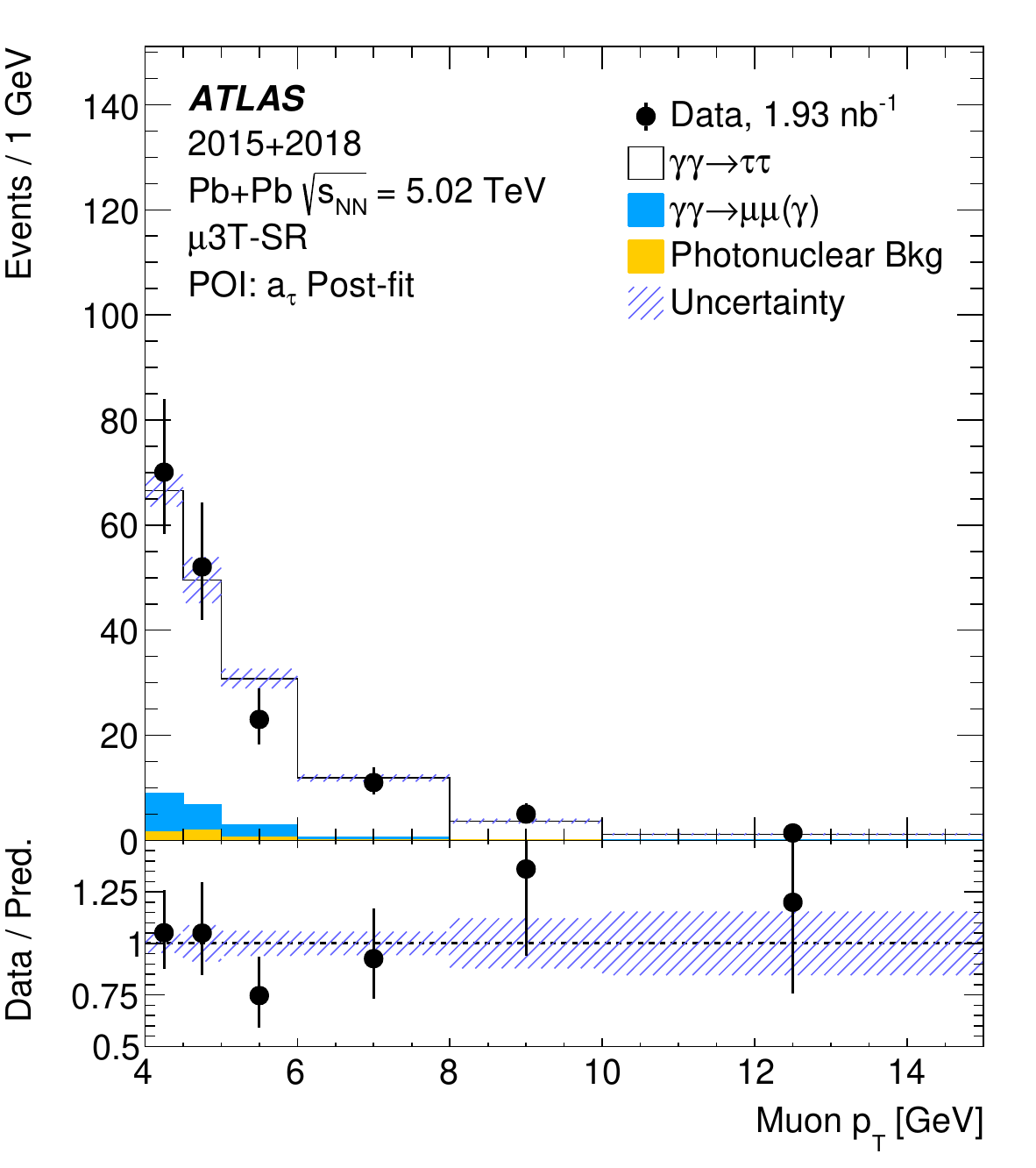}%
\includegraphics[width=0.37\textwidth]{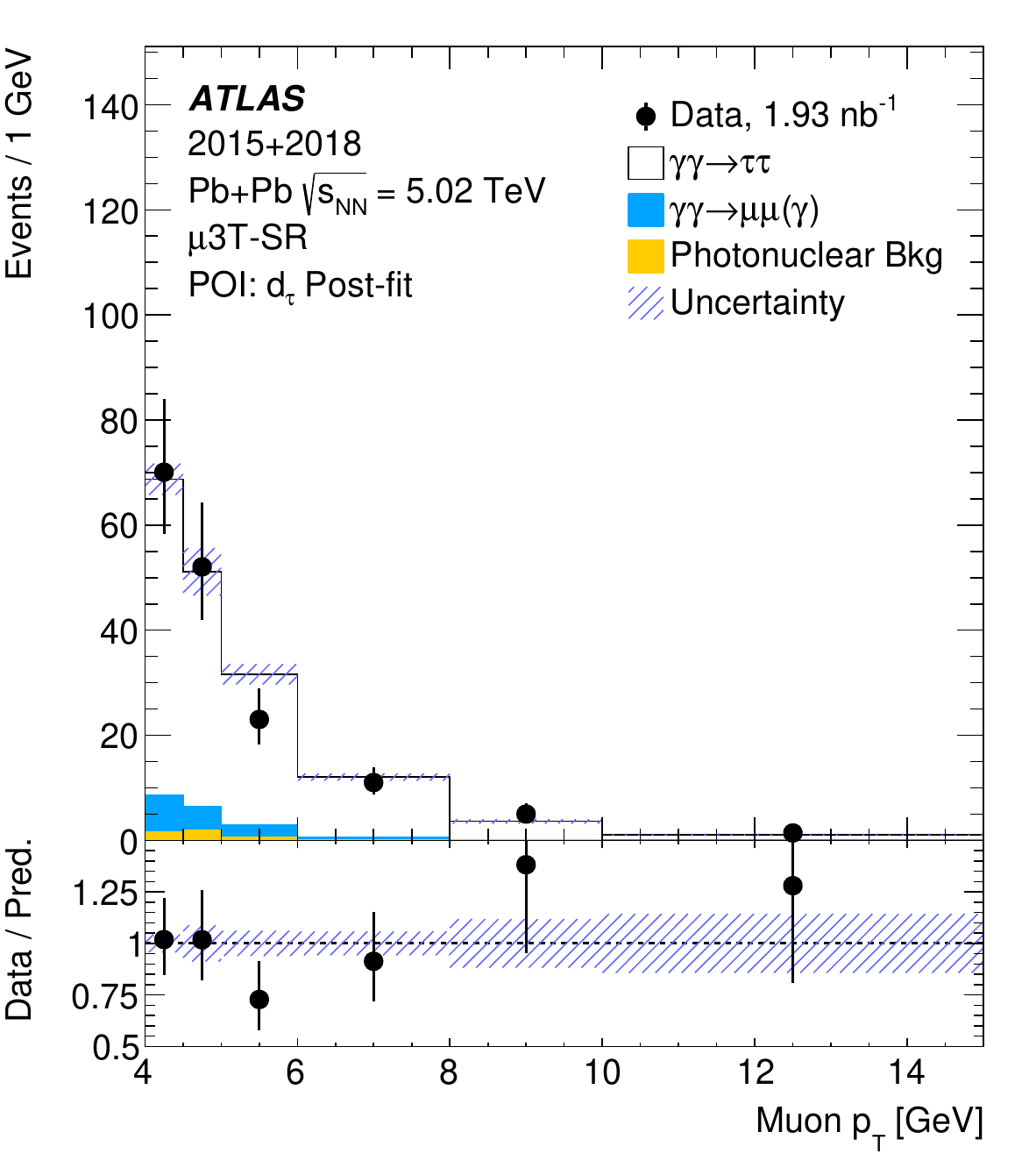}
\includegraphics[width=0.37\textwidth]{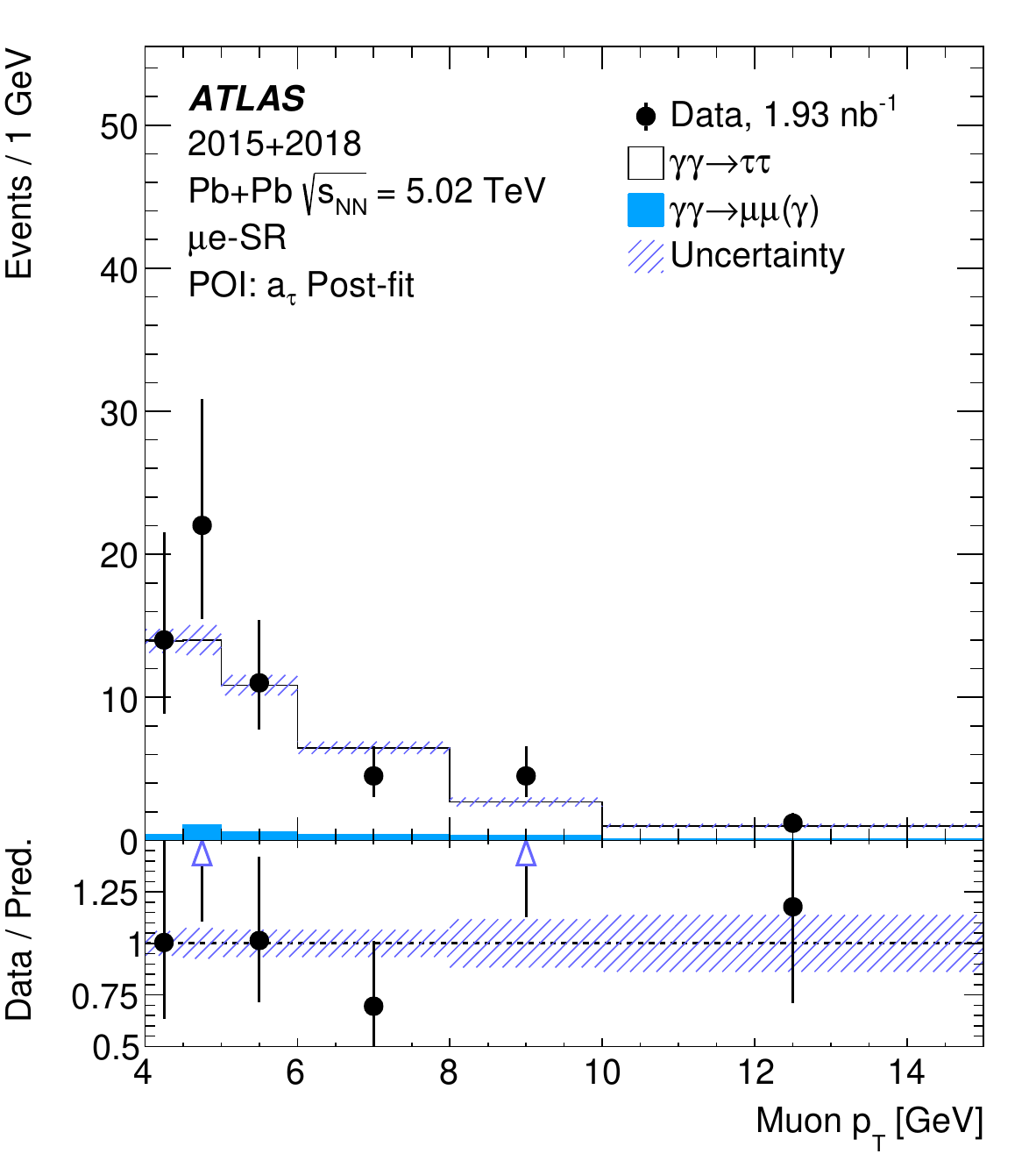}%
\includegraphics[width=0.37\textwidth]{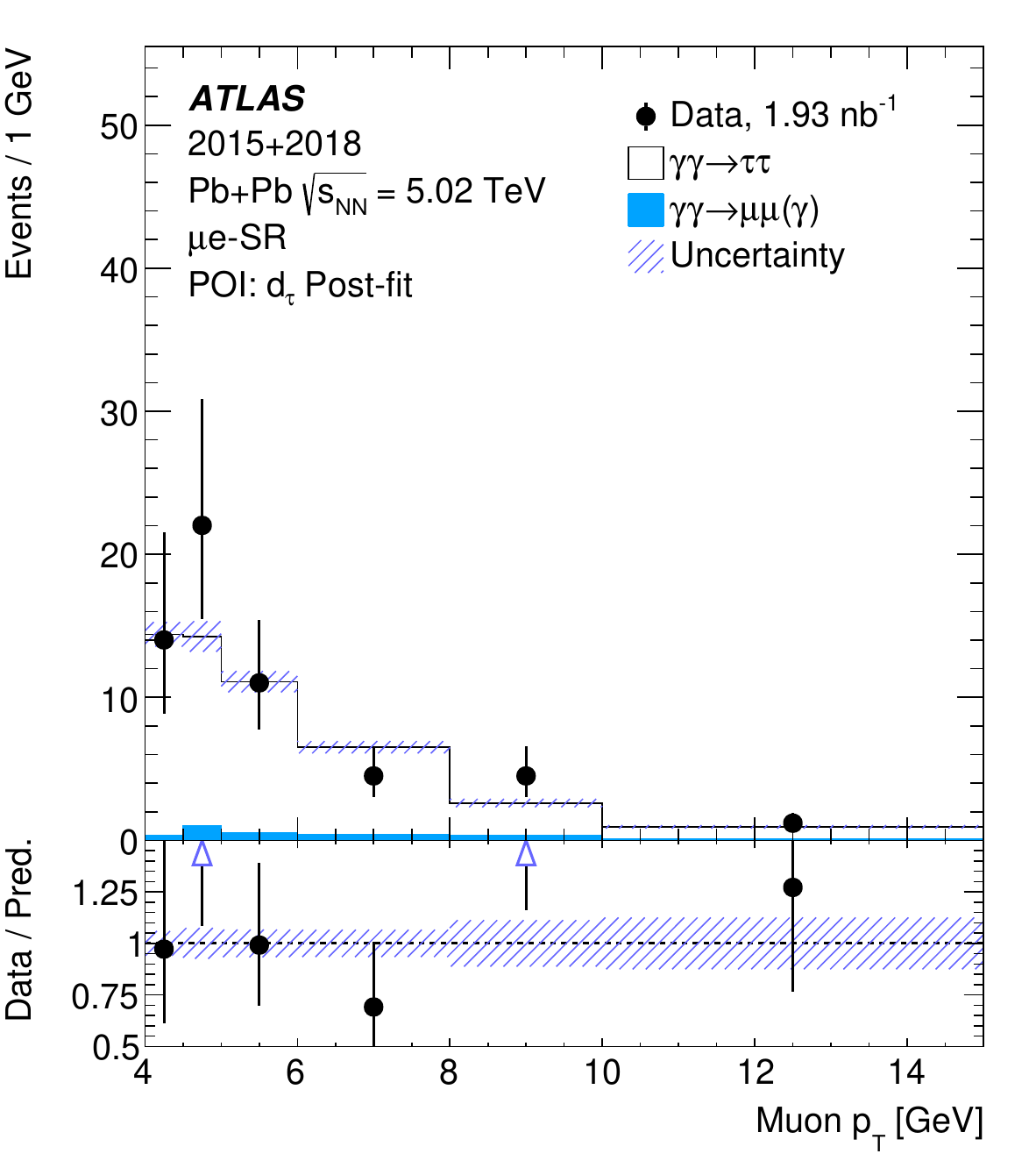}
\caption{Muon \pt distributions in the (top) $\mu$1T-SR, (middle) $\mu$3T-SR, (bottom) $\mu e$-SR\@. Data (filled markers) are compared with the post-fit predictions for the \atau fit (left) and \dtau fit (right) summing signal and background contributions (stacked histograms). Event counts are shown normalized by bin width. The bottom panel displays the ratio of the data to post-fit predictions. Uncertainties from the finite number of data events (filled markers) and $\pm1\sigma$ systematic uncertainties of the prediction with the constraints from the fit applied (hatched bands) are indicated to better judge variations between data and predictions. Arrows in the ratio panel denote a data point outside the displayed range.}
\label{fig:fits-atau_dtau_detlevel-postfitdistrs}
\end{figure}

The data are compared with the post-fit expectations for the \atau and \dtau fits, shown in Figure~\ref{fig:fits-atau_dtau_detlevel-postfitdistrs}. The best-fit predictions agree better with data, compared with the pre-fit predictions shown in Figure~\ref{fig:fits-atau_detlevel-prefitdistrs}, as expected. In particular, at high muon \pt the small pre-fit data excess is reduced.

\begin{figure}[htbp]
\centering
\includegraphics[width=0.49\textwidth]{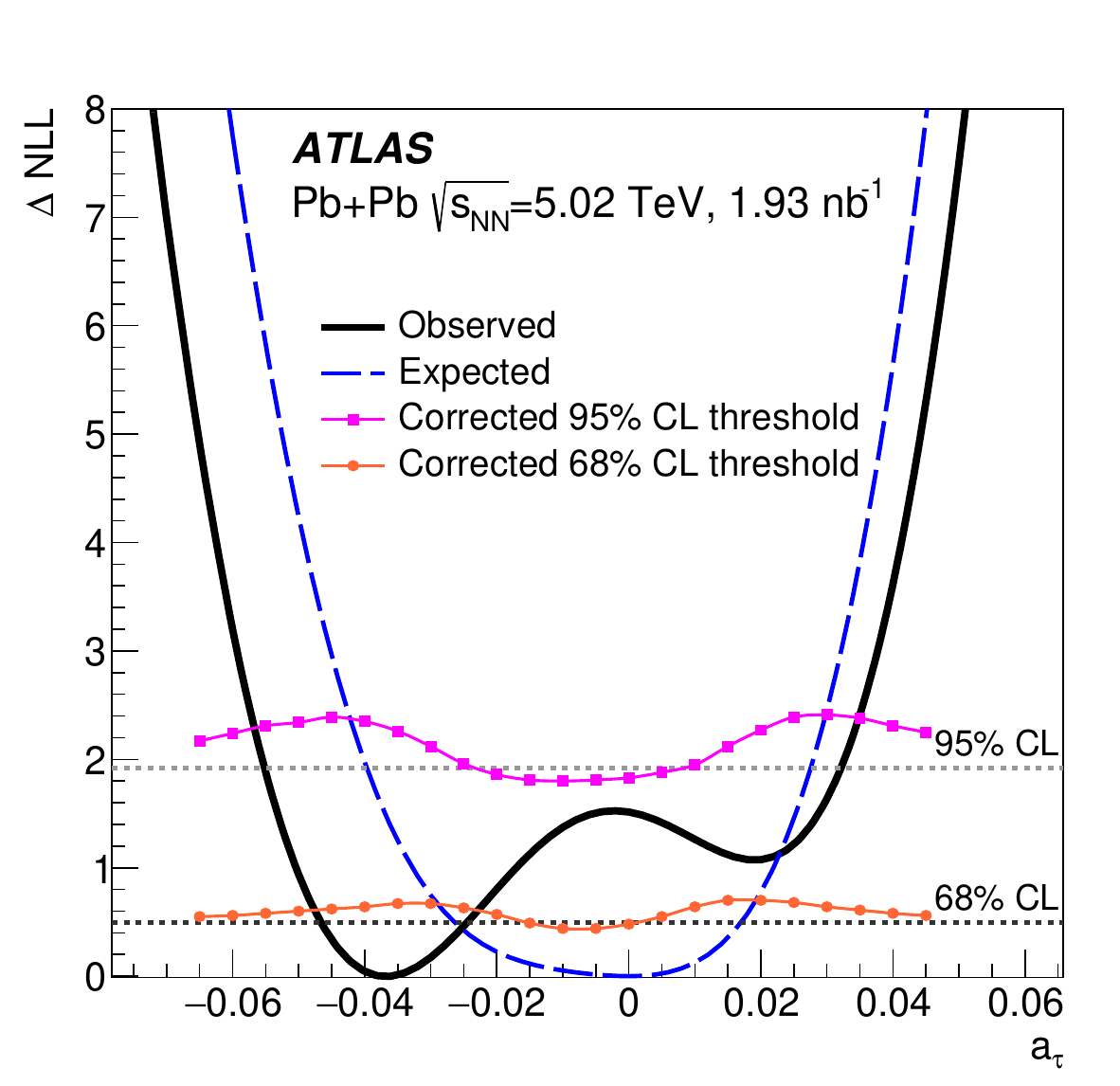}%
\includegraphics[width=0.49\textwidth]{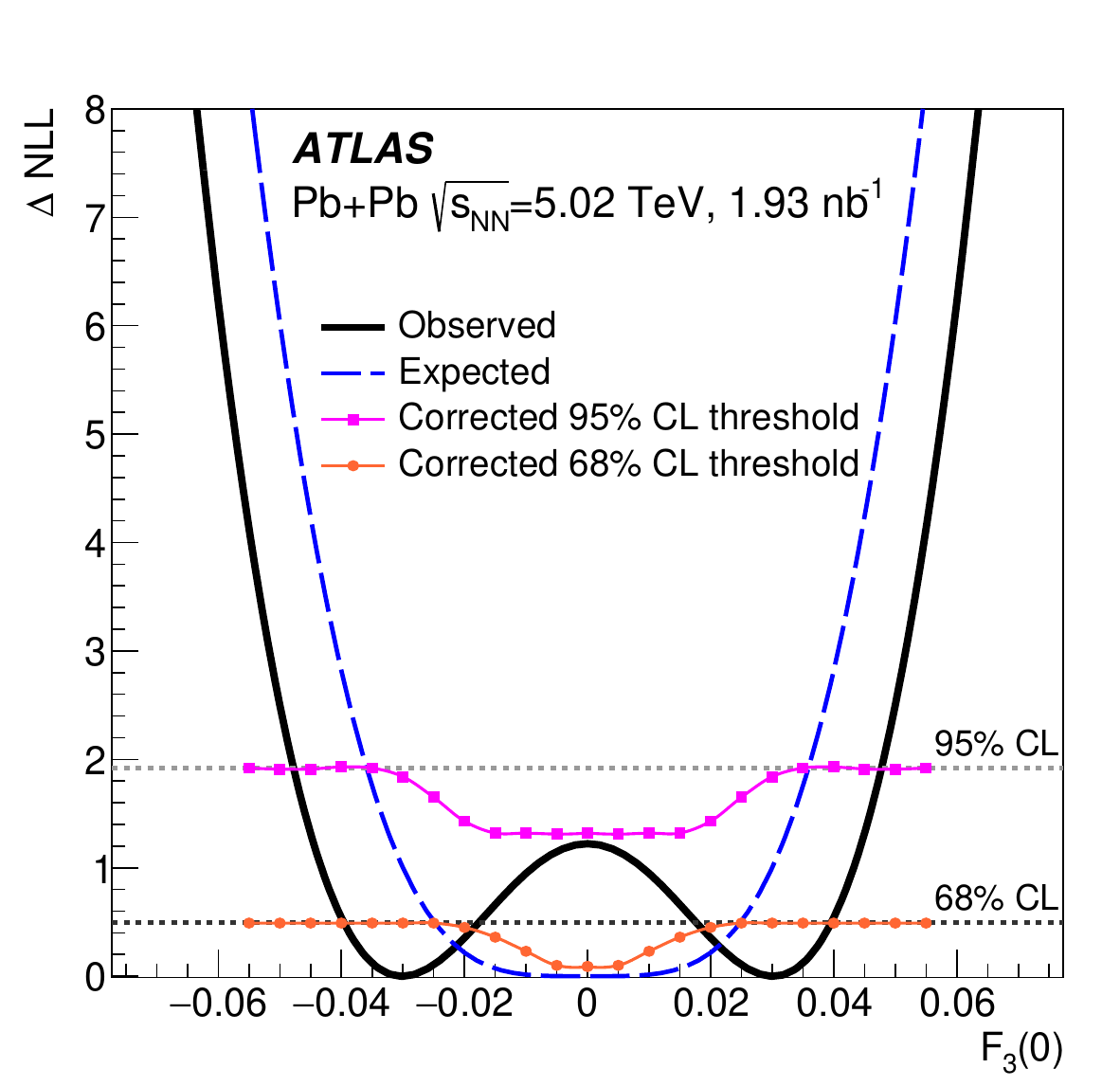}
\caption{Observed and expected NLL curves for the combined fit to all SRs, normalized to its minimum $\Delta$NLL, as a function of \atau (left) and $F_3(0)$ (right). The coverage-corrected threshold points, calculated via pseudoexperiments~\cite{ATLAS:2025yww}, and interpolated lines for 68\% and 95\% coverage are shown.}
\label{fig:det_level_coverage_corrected}
\end{figure}

A pseudodata-derived correction is used to obtain the Confidence Intervals (CIs) for $\atau$ and $\dtau$ since the typically used asymptotic assumption relies on Wilk’s Theorem~\cite{10.1214/aoms/1177732360}. Wilk's Theorem may not be valid for cases where the parameters of interest involve a quadratic cross-section dependence~\cite{Bernlochner:2022oiw}, which is the case for both \atau and \dtau. Observed and expected NLL scan curves for \atau and \dtau combined fits to all SRs (the latter in units of $F_3(0)$), normalized to their minimum $\Delta$NLL, are shown in Figure~\ref{fig:det_level_coverage_corrected}. The 68\% and 95\% coverage lines, under the asymptotic assumption, are shown by the dotted lines. The 68\% and 95\% coverage points, calculated using pseudodata, are also shown. These points are calculated following the approach presented in Ref.~\cite{ATLAS:2025yww}. Rather than assuming a $\chi^2$ distribution of the profile-likelihood-ratio (PLR), pseudodata are used to directly generate the PLR distribution of many pseudoexperiments, and NLL thresholds with correct coverage are calculated. Coverage lines are then obtained by interpolating between the points. CIs are calculated by determining the intersections between the NLL-scan curves and the coverage-corrected lines.

\begin{figure}[htbp]
\centering
\includegraphics[width=0.49\textwidth]{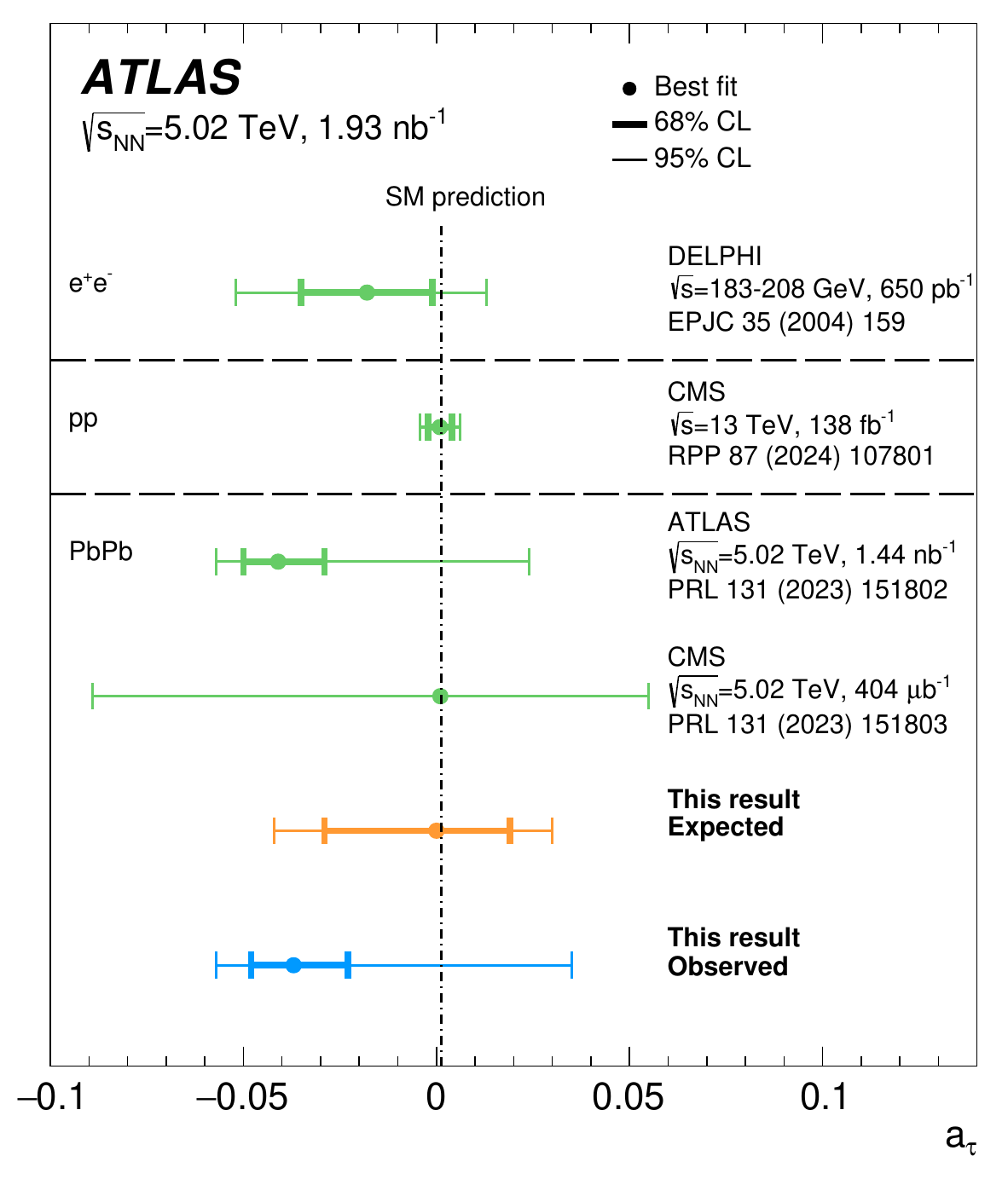}%
\includegraphics[width=0.49\textwidth]{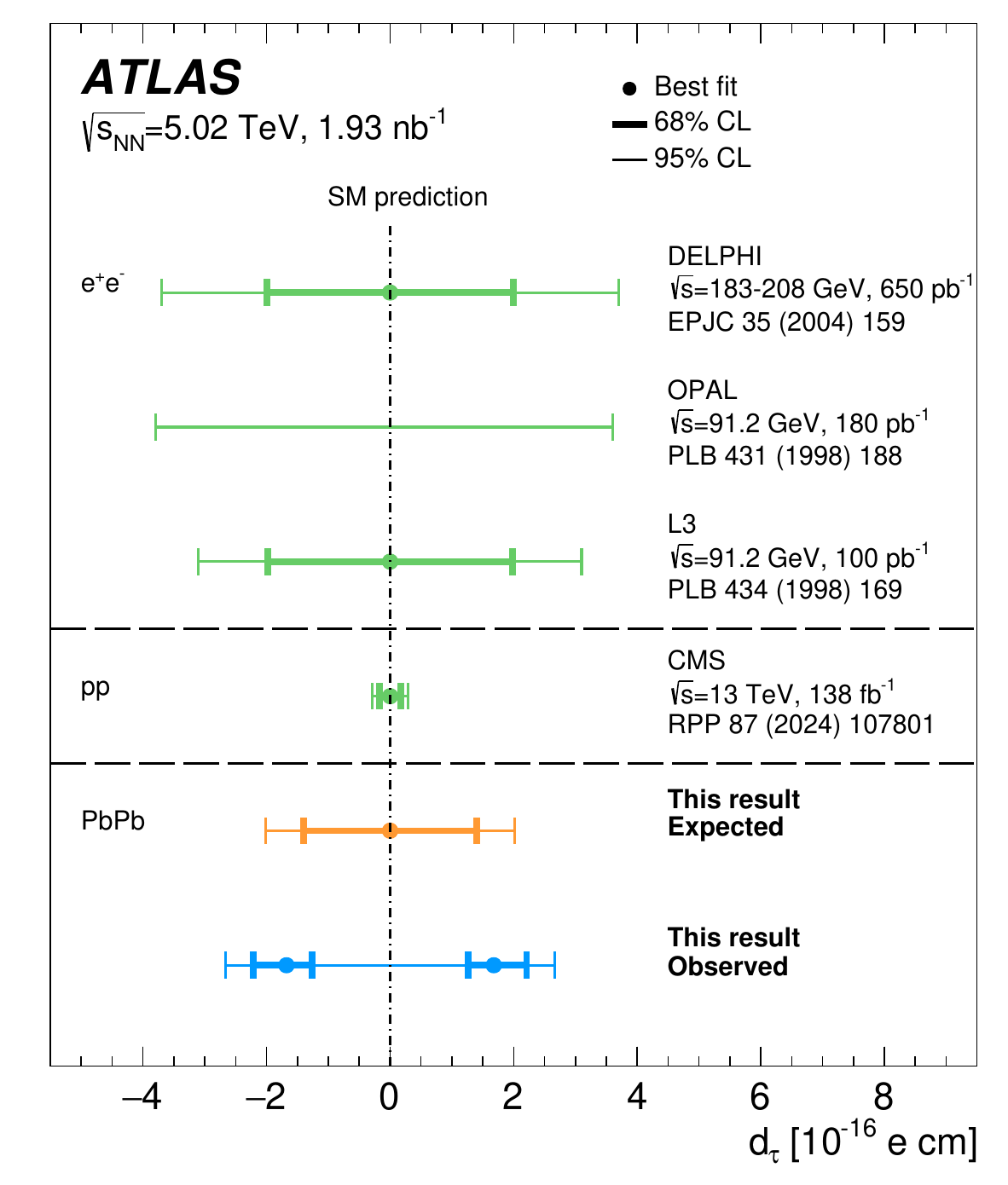}
\caption{Constraints on \atau (left) and \dtau (right) from the combined fit to the three SRs. The observed confidence intervals obtained from the combined fit to data are displayed together with the best-fit value, at 68\% and 95\% confidence level. The observed 68\% CL interval is highly asymmetric for $\atau$ and has double minima for $\dtau$ due to the higher-than-expected observed yields. The expected interval from the combined fit, representing the median sensitivity assuming only SM processes occur, is also shown. The obtained results are compared with a selection of the most precise previous measurements from the OPAL, L3 and DELPHI experiments at LEP, and from ATLAS and CMS at the LHC.}
\label{fig:fits-atau_detlevel-POI}
\end{figure}

The measured best-fit value for \atau from the combined fit is $\atau=-0.037$ and the CIs are $(-0.048, -0.023)$ and $(-0.057, 0.035)$ at 68\% and 95\% CL, respectively. The measured best-fit value for \dtau from the combined fit is $|\dtau|=1.7 \times 10^{-16}~e\text{cm}$. The CIs for \dtau are $(-1.1, -2.2) \times 10^{-16}~e\text{cm}$ and $(1.1, 2.2) \times 10^{-16}~e\text{cm}$ at 68\% CL and $|\dtau|< 2.7 \times 10^{-16}~e\text{cm}$ at 95\% CL. The higher-than-expected observed yields lead to the highly asymmetric 68\% CL interval for $\atau$ and the double minima for $\dtau$. This arises from the nearly quadratic signal cross-section dependence on $\atau$ and the quadratic dependence on $\dtau$, caused by the interference of the SM and BSM amplitudes~\cite{Ackerstaff:1998mt,Acciarri:1998iv,Dyndal:2020yen,Beresford:2024dsc}. Figure~\ref{fig:fits-atau_detlevel-POI} shows the obtained constraints on $\atau$ and $\dtau$ for the combined fit together with a selection of the most precise existing constraints from $pp$ and \PbPb collisions at the LHC ~\cite{CMS:2024qjo,STDM-2019-19,CMS-HIN-21-009} and $e^+e^-$ collisions at LEP~\cite{Abdallah:2003xd,Ackerstaff:1998mt,Acciarri:1998iv}. Only constraints for which $q^2\sim0~\GeV^2$ are shown. The current strongest constraints on $\atau$ and $\dtau$ are set by CMS using $pp$ collisions. The constraints on $\atau$ obtained by this analysis using Pb+Pb collisions are of similar sensitivity to previous constraints obtained using Pb+Pb collisions and $e^+e^-$ collisions at LEP\@. For $|\dtau|$ the presented results are the first constraints obtained using Pb+Pb collisions and they are competitive with constraints from DELPHI, OPAL and L3.


%

%
%
\FloatBarrier

\section{Conclusion}

This paper presents the first differential fiducial measurements of photon-induced \tlep pair production in ultra-peripheral \PbPb collisions. The measurement uses the full Run 2 \PbPb data sample recorded by the ATLAS experiment at the LHC at $\sqrt{s_{_\text{NN}}} = 5.02$~\TeV with an integrated luminosity of 1.93 nb$^{-1}$. The phase space of the measurement is chosen to suppress breakup of the lead ions by selecting events without forward neutron emission. This ensures that both initial photons are almost on-shell with virtuality $q^2\sim 0~\GeV^2$. Fiducial cross-sections are measured as a function of seven different kinematic variables for three different final state configurations from the \tlep decays. The cross-sections are compared with several leading-order SM theoretical predictions and with a NLO electroweak prediction for the one muon plus one electron fiducial region. For the predictions, several different photon flux models are compared. The measured cross-sections are found to be largely compatible with the theoretical predictions within measurement uncertainties. The precision of the measurements is limited by statistical uncertainties. A fit to the detector-level muon-\pt distributions in each of the three SRs is used to set constraints on the \tlep electromagnetic dipole moments. A data-driven correction is applied to the photon flux for the \ggtt and \ggmmg predictions used in the fit. The correction is derived using a dedicated di-muon control region, exploiting the process independence of the initial-state photon flux. The observed 95\% CL intervals from the combined fit to the three SRs are $-0.057 <\atau< 0.035$ and $|\dtau|< 2.7 \times 10^{-16}~e\text{cm}$. For $\atau$ the achieved constraints are of similar precision to existing lepton-collider and \PbPb constraints. For $\dtau$ the achieved precision is competitive with existing lepton-collider precision and the presented results are the first constraints obtained using \PbPb data.


%
%
%
%
%
\section*{Acknowledgments}
%

%

%
%

%
%

We thank CERN for the very successful operation of the LHC and its injectors, as well as the support staff at
CERN and at our institutions worldwide without whom ATLAS could not be operated efficiently.

The crucial computing support from all WLCG partners is acknowledged gratefully, in particular from CERN, the ATLAS Tier-1 facilities at TRIUMF/SFU (Canada), NDGF (Denmark, Norway, Sweden), CC-IN2P3 (France), KIT/GridKA (Germany), INFN-CNAF (Italy), NL-T1 (Netherlands), PIC (Spain), RAL (UK) and BNL (USA), the Tier-2 facilities worldwide and large non-WLCG resource providers. Major contributors of computing resources are listed in Ref.~\cite{ATL-SOFT-PUB-2026-001}.

We gratefully acknowledge the support of ANPCyT, Argentina; YerPhI, Armenia; ARC, Australia; BMWFW and FWF, Austria; ANAS, Azerbaijan; CNPq and FAPESP, Brazil; NSERC, NRC and CFI, Canada; CERN; ANID, Chile; CAS, MOST, NSFC and FRFCU, China; Minciencias, Colombia; MEYS CR, Czech Republic; DNRF and DNSRC, Denmark; IN2P3-CNRS and CEA-DRF/IRFU, France; SRNSFG, Georgia; BMFTR, HGF and MPG, Germany; GSRI, Greece; RGC and Hong Kong SAR, China; ICHEP and Academy of Sciences and Humanities, Israel; INFN, Italy; MEXT and JSPS, Japan; CNRST, Morocco; NWO, Netherlands; RCN, Norway; MNiSW, Poland; FCT, Portugal; MNE/IFA, Romania; MSTDI, Serbia; MSSR, Slovakia; ARIS and MVZI, Slovenia; DSI/NRF, South Africa; MICIU/AEI, Spain; SRC and Wallenberg Foundation, Sweden; SERI, SNSF and Cantons of Bern and Geneva, Switzerland; NSTC, Taipei; TENMAK, T\"urkiye; STFC/UKRI, United Kingdom; DOE and NSF, United States of America.

Individual groups and members have received support from BCKDF, CANARIE, CRC and DRAC, Canada; CERN-CZ, FORTE and PRIMUS, Czech Republic; COST, ERC, ERDF, Horizon 2020 and Marie Sk{\l}odowska-Curie Actions, European Union; Investissements d'Avenir Labex, Investissements d'Avenir Idex and ANR, France; DFG and AvH Foundation, Germany; Herakleitos, Thales and Aristeia programmes co-financed by EU-ESF and the Greek NSRF, Greece; BSF-NSF and MINERVA, Israel; NCN and NAWA, Poland; La Caixa Banking Foundation, CERCA and AGAUR programs from Generalitat de Catalunya and PROMETEO and GenT Programmes Generalitat Valenciana, Spain; G\"{o}ran Gustafssons Stiftelse, Sweden; The Royal Society and Leverhulme Trust, United Kingdom; Eric and Wendy Schmidt Fund for Strategic Innovation, United States of America.

In addition, individual members wish to acknowledge support from Chile: Agencia Nacional de Investigaci\'on y Desarrollo (ANID FONDECYT reg. 1230987, FONDECYT 1230812, FONDECYT 1240864, Fondecyt 3240661, Fondecyt Regular 1240721); China: Fundamental Research Funds for the Central Universities (010-63263105), Chinese Ministry of Science and Technology (MOST-2023YFA1605700, MOST-2023YFA1609300), National Natural Science Foundation of China (NSFC 12275265, NSFC-W2543005); Czech Republic: Czech Science Foundation (GACR - 24-11373S), Ministry of Education Youth and Sports (ERC-CZ-LL2327, FORTE CZ.02.01.01/00/22\_008/0004632); EU: H2020 European Research Council (ERC - 101002463); European Union: European Research Council (BARD No. 101116429, ERC - 101219398, ERC 101089007), European Regional Development Fund (HE COFUND GA No.101081355, ERDF), Marie Sklodowska-Curie Actions (GAP-101168829); France: Agence Nationale de la Recherche (ANR-21-CE31-0013, ANR-22-EDIR-0002, ANR-24-CE31-0504-01); Germany: Deutsche Forschungsgemeinschaft (DFG - 469666862); China: Research Grants Council (GRF); Italy: Istituto Nazionale di Fisica Nucleare (LHC-MIUR - 28003/2025), Ministero dell'Università e della Ricerca (NextGenEU I53D23000820006 M4C2.1.1, SOE2024\_0000023); Japan: Japan Society for the Promotion of Science (JSPS KAKENHI  JP25H0063, JSPS KAKENHI JP22H04944, JSPS KAKENHI JP24K23939, JSPS KAKENHI JP24KK0251, JSPS KAKENHI JP25H00650, JSPS KAKENHI JP25H01291, JSPS KAKENHI JP25K01011, JSPS KAKENHI JP25K01023, JSPS KAKENHI JP25KK0047); Poland: Polish National Science Centre (NCN 2021/42/E/ST2/00350, NCN OPUS 2023/51/B/ST2/02507, NCN OPUS nr 2022/47/B/ST2/03059, NCN UMO-2019/34/E/ST2/00393, UMO-2022/47/O/ST2/00148, UMO-2023/49/B/ST2/04085, UMO-2023/51/B/ST2/00920, UMO-2024/53/N/ST2/00869); Spain: Agència de Gestió d'Ajuts Universitaris i de Recerca. (AGAUR - 2023 BP 00141), Ministry of Science and Innovation (RYC2019-028510-I, RYC2020-030254-I, RYC2021-031273-I, RYC2022-038164-I), Ministerio de Ciencia, Innovación y Universidades/Agencia Estatal de Investigaci\'on (EU NextGenerationEU (PRTR-C17.I1), PID2022-142604OB-C22); Sweden: Carl Trygger Foundation (Carl Trygger Foundation CTS 22:2312), Swedish Research Council (Swedish Research Council 2023-04654, VR 2021-03651, VR 2022-03845, VR 2022-04683, VR 2023-03403, VR 2024-05451, VR 2025-05940), Knut and Alice Wallenberg Foundation (KAW 2023.0366); United Kingdom: The Binks Trust, Royal Society (NIF-R1-231091); United States of America: U.S. Department of Energy (ECA DE-AC02-76SF00515), John Templeton Foundation (John Templeton Foundation 63206), Neubauer Family Foundation.

%
%


%
\clearpage
\appendix
\part*{Appendix}
\addcontentsline{toc}{part}{Appendix}

\section{Event-level reweighting for non-zero \atau and \dtau predictions}
\label{app:reweight}

Predictions with non-zero \atau or \dtau values are obtained through an event-level reweighting procedure, applied to the nominal signal MC prediction. In this procedure, the photon-lepton vertex is parameterized using form factors $F_{2}\left(q^2\right)$ and $F_{3}\left(q^2\right)$ as shown in Eq.~(\ref{eq:gttvertex}) in Section~\ref{sec:dipoles}, where the asymptotic form factor values for the squared momentum transfer $q^2\to0~\GeV^2$ are related to \atau and \dtau, respectively. The event-level reweighting procedure provides predictions both for non-zero \atau and non-zero \dtau values.

The event-level weights are determined from tree-level QED matrix element calculations of the \ggtt process.
For this, the four-momenta of the particles are defined as follows:
\begin{align}
p_{\gamma_{1}}^{\mu}&=\left(E,E\hat{z}\right), &
p_{\tau^{+}}^{\mu}&=\left(E,\vec{p}\right), \\
p_{\gamma_{2}}^{\mu}&=\left(E,-E\hat{z}\right), & p_{\tau^{-}}^{\mu}&=\left(E,-\vec{p}\right),
\end{align}
where $\vec{p}=\left(0,\beta E\sin\theta,\beta E\cos\theta\right)$ is the 3-momentum of the \tleps with $\beta=\sqrt{1-\frac{m_{\tau}^{2}}{E^{2}}}$ being the boost and $\theta$ being the longitudinal angle. Since the physics of interest in this $2\rightarrow 2$ process is invariant under coordinate transformations, the kinematics are defined in the center-of-mass frame with $p_{x}=0$ for convenience. By definition, the virtuality of the photons in the initial-state is $q^2=0~\GeV^2$.

The matrix element of the \ggtt process is calculated using spin-polarization sums, and factorized in the form factors $F_{2,3}\left(0\right)$ as:
\begin{equation}
\left|\mathcal{M}\left(\beta,\theta,F_{2}\left(0\right),F_{3}\left(0\right)\right)\right|^{2}=\sum_{i,j=0}^{4}\left[F_{2}\left(0\right)\right]^{i}\left[F_{3}\left(0\right)\right]^{j}C_{ij}
\end{equation}
where the non-zero parts of $C_{ij}$ are:
\begin{align}
C_{00} &=\frac{-\beta^{4}\left(\sin^{4}\theta+1\right)+2\beta^{2}\sin^{2}\theta+1}{\left(\beta^{2}\cos^{2}\theta-1\right)^{2}}\\
C_{10} &=-\frac{4}{\beta^{2}\cos^{2}\theta-1}\\
C_{20} &=\frac{2\left(-2\beta^{2}\cos2\theta-3\beta^{2}+5\right)}{\left(\beta^{2}-1\right)\left(\beta^{2}\cos2\theta+\beta^{2}-2\right)}\\
C_{30} &=\frac{4\left(-\beta^{2}\cos2\theta+1\right)}{\left(\beta^{2}-1\right)\left(\beta^{2}\cos2\theta+\beta^{2}-2\right)}\\
C_{40} &=\frac{\beta^{4}\left(-\cos^{4}\theta-2\cos^{2}\theta+2\right)+2\beta^{2}\cos2\theta-1}{4\left(\beta^{2}-1\right)^{2}\left(\beta^{2}\cos^{2}\theta-1\right)}\\
C_{02} &=\frac{2\left(-2\beta^{2}\cos2\theta-\beta^{2}+3\right)}{\left(\beta^{2}-1\right)\left(\beta^{2}\cos2\theta+\beta^{2}-2\right)}
\end{align}
\begin{align}
C_{12} &=\frac{4\left(-\beta^{2}\cos2\theta+1\right)}{\left(\beta^{2}-1\right)\left(\beta^{2}\cos2\theta+\beta^{2}-2\right)}\\
C_{22} &=\frac{\beta^{4}\left(-\cos^{4}\theta-2\cos^{2}\theta+2\right)+2\beta^{2}\cos2\theta-1}{2\left(\beta^{2}-1\right)^{2}\left(\beta^{2}\cos^{2}\theta-1\right)}\\
C_{04} &=\frac{\beta^{4}\left(-\cos^{4}\theta-2\cos^{2}\theta+2\right)+2\beta^{2}\cos2\theta-1}{4\left(\beta^{2}-1\right)^{2}\left(\beta^{2}\cos^{2}\theta-1\right)}
\end{align}

The $C_{00}$ term corresponds to the tree-level SM case (i.e., $a_{\tau}=d_{\tau}=0$). The predicted small non-zero \atau and \dtau values are neglected in the used leading-order SM signal samples. For \dtau, the approach above means that the linear dependence of the cross-section on \dtau is zero, and only the quadratic dependence on \dtau is non-zero. In the determination of the $C_{ij}$ above, this is visible through $C_{01}=0$.

The determined weights are applied based on the generator-level $\tau\tau$ kinematics. For this purpose, the four-vectors of the \tleps $p_{\tau^{\pm}}$ in event $i$ in the simulation are found, and the system four-vector $p_{\mathrm{system}}^{\mu}=p_{\tau^{+}}^{\mu}+p_{\tau^{-}}^{\mu}$ is calculated. The system is boosted back to the center-of-mass frame, and the boost $\beta$ and longitudinal angle $\theta$ of the system are calculated. Weights are calculated as the ratio of the squared matrix elements:
\begin{equation}
w_{i}\left(\beta,\theta,F_{2}\left(0\right)=x,F_{3}\left(0\right)=y\right)=\frac{\left|\mathcal{M}_i\left(\beta,\theta,x,y\right)\right|^{2}}{\left|\mathcal{M}_i\left(\beta,\theta,0,0\right)\right|^{2}}
\end{equation}
for chosen values of $x$ and $y$ for $F_{2}\left(0\right)$ and $F_{3}\left(0\right)$, respectively, and are multiplied to other MC signal weights. %


\FloatBarrier

\printbibliography

@article{CMS:2024qjo,
    author = "{CMS Collaboration}",
    title = "{Observation of $\gamma\gamma\to\tau\tau$ in proton-proton collisions and limits on the anomalous electromagnetic moments of the $\tau$ lepton}",
    eprint = "2406.03975",
    archivePrefix = "arXiv",
    primaryClass = "hep-ex",
    reportNumber = "CMS-SMP-23-005, CERN-EP-2024-127",
    doi = "10.1088/1361-6633/ad6fcb",
    journal = "Rept. Prog. Phys.",
    volume = "87",
    number = "10",
    pages = "107801",
    year = "2024"
}

@article{Biondi:2017pzs,
    author = "Biondi, Silvia",
    editor = "Foka, Y. and Brambilla, N. and Kovalenko, V.",
    collaboration = "ATLAS",
    title = "{Experience with using unfolding procedures in ATLAS}",
    reportNumber = "ATL-PHYS-PROC-2016-189",
    doi = "10.1051/epjconf/201713711002",
    journal = "EPJ Web Conf.",
    volume = "137",
    pages = "11002",
    year = "2017"
}

@techreport{DeNardo2002Propagation,
  author      = {De Nardo, Lara},
  title       = {On the Propagation of Statistical Errors},
  institution = {HERMES Collaboration / PHENIX},
  year        = {2002},
  month       = {March},
  type        = {Technical Note},
  number      = {02-008},
  url         = {https://www.phenix.bnl.gov/WWW/publish/elke/EIC/Files-for-Wiki/lara.02-008.errors.pdf}
}

@article{Akesson:2006sh,
    author = "Akesson, F. and Atkinson, T. and Costa, M. J. and Elsing, M. and Fleischmann, S. and Gaponenko, A. and Liebig, W. and Moyse, E. and Salzburger, A. and Siebel, M.",
    title = "{ATLAS tracking event data model}",
    reportNumber = "ATL-SOFT-PUB-2006-004",
    month = "7",
    year = "2006",
    url = {https://cds.cern.ch/record/973401}
}

@article{Cornelissen:2007aca,
    author = "Cornelissen, T. G. and Van Eldik, N and Elsing, M. and Liebig, W. and Moyse, E. and Piacquadio, N. and Prokofiev, K. and Salzburger, A. and Wildauer, A.",
    title = "{Updates of the ATLAS Tracking Event Data Model (Release 13)}",
    reportNumber = "ATL-SOFT-PUB-2007-003, ATL-COM-SOFT-2007-008",
    month = "6",
    year = "2007",
    url = {https://cds.cern.ch/record/1038095}
}

@article{L3:1997acq,
    author = "Acciarri, M. and others",
    collaboration = "L3",
    title = "{Production of $e$, $\mu$ and $\tau$ pairs in untagged two photon collisions at LEP}",
    reportNumber = "CERN-PPE-97-043, CERN-PPE-97-43",
    doi = "10.1016/S0370-2693(97)00731-4",
    journal = "Phys. Lett. B",
    volume = "407",
    pages = "341--350",
    year = "1997"
}

@article{L3:2004ful,
    author = "Achard, P. and others",
    collaboration = "L3",
    title = "{Muon-pair and tau-pair production in two-photon collisions at LEP}",
    eprint = "hep-ex/0402037",
    archivePrefix = "arXiv",
    reportNumber = "CERN-EP-2003-080",
    doi = "10.1016/j.physletb.2004.02.012",
    journal = "Phys. Lett. B",
    volume = "585",
    pages = "53--62",
    year = "2004"
}

@article{Bernlochner:2022oiw,
    %author = "Bernlochner, Florian Urs and Fry, Daniel C. and Menary, Stephen Burns and Persson, Eric",
    author = "{F. Urs Bernlochner, D. C. Fry, S. Burns Menary, and E. Persson}",
    title = "{Cover your bases: asymptotic distributions of the profile likelihood ratio when constraining effective field theories in high-energy physics}",
    eprint = "2207.01350",
    archivePrefix = "arXiv",
    primaryClass = "physics.data-an",
    doi = "10.21468/SciPostPhysCore.6.1.013",
    journal = "SciPost Phys. Core",
    volume = "6",
    pages = "013",
    year = "2023"
}

@article{10.1214/aoms/1177732360,
author = {S. S. Wilks},
title = {{The Large-Sample Distribution of the Likelihood Ratio for Testing Composite Hypotheses}},
volume = {9},
journal = {The Annals of Mathematical Statistics},
number = {1},
publisher = {Institute of Mathematical Statistics},
pages = {60 -- 62},
year = {1938},
doi = {10.1214/aoms/1177732360},
%URL = {https://doi.org/10.1214/aoms/1177732360}
}

@article{ATLAS:2025yww,
    author = "{ATLAS Collaboration}",
    title = "{Measurement of high-mass $t\bar{t}\ell^{+}\ell^{-}$ production and lepton flavour universality-inspired effective field theory interpretations at $\sqrt{s}=13$ TeV with the ATLAS detector}",
    eprint = "2504.05919",
    archivePrefix = "arXiv",
    primaryClass = "hep-ex",
    reportNumber = "CERN-EP-2025-053",
    doi = "10.1140/epjc/s10052-025-14695-9",
    journal = "Eur. Phys. J. C",
    volume = "85",
    number = "12",
    pages = "1434",
    year = "2025"
}

@article{ARGUS:2000riz,
    %author = "Albrecht, H. and others",
    %collaboration = "ARGUS",
    author = "{ARGUS Collaboration}",
    title = "{A search for the electric dipole moment of the $\tau-$lepton}",
    eprint = "hep-ex/0004031",
    archivePrefix = "arXiv",
    reportNumber = "DESY-00-064",
    doi = "10.1016/S0370-2693(00)00630-4",
    journal = "Phys. Lett. B",
    volume = "485",
    pages = "37--44",
    year = "2000"
}

@article{Davidson:2010ew,
    author = "Davidson, N. and Przedzinski, T. and Was, Z.",
    title = "{PHOTOS interface in C++: Technical and Physics Documentation}",
    eprint = "1011.0937",
    archivePrefix = "arXiv",
    primaryClass = "hep-ph",
    reportNumber = "CERN-PH-TH-2010-261, IFJPAN-IV-2010-6",
    doi = "10.1016/j.cpc.2015.09.013",
    journal = "Comput. Phys. Commun.",
    volume = "199",
    pages = "86--101",
    year = "2016"
}

@article{Dittmaier:2025ikh,
    author = "Dittmaier, Stefan and Engel, Tim and Ariza, Jose Luis Hernando and Pellen, Mathieu",
    title = "{Electroweak corrections to $\tau^{+}\tau^{-}$ production in ultraperipheral heavy-ion collisions at the LHC}",
    eprint = "2504.11391",
    archivePrefix = "arXiv",
    primaryClass = "hep-ph",
    reportNumber = "FR-PHENO-2025-004",
    doi = "10.1007/JHEP08(2025)051",
    journal = "JHEP",
    volume = "08",
    pages = "051",
    year = "2025"
}

@article{Korchin:2025vzx,
    author = "{A. Yu. Korchin, E. Richter-Was, and Z. Was}",
    title = "{TauSpinner Algorithms for Including Spin and New Physics Effects in \(\gamma \gamma \rightarrow \tau \tau \) Process}",
    eprint = "2506.15213",
    archivePrefix = "arXiv",
    primaryClass = "hep-ph",
    reportNumber = "IFJPAN-IV-2025-12",
    doi = "10.5506/APhysPolB.56.10-A4",
    journal = "Acta Phys. Polon. B",
    volume = "56",
    number = "10",
    pages = "10--A4",
    year = "2025"
}

@article{Przedzinski:2018ett,
    author = "Przedzinski, T. and Richter-Was, E. and Was, Z.",
    title = "{Documentation of $TauSpinner$ algorithms: program for simulating spin effects in $\tau$-lepton production at LHC}",
    eprint = "1802.05459",
    archivePrefix = "arXiv",
    primaryClass = "hep-ph",
    reportNumber = "IFJPAN-IV-2018-3",
    doi = "10.1140/epjc/s10052-018-6527-0",
    journal = "Eur. Phys. J. C",
    volume = "79",
    number = "2",
    pages = "91",
    year = "2019"
}

@article{Loizides:2017ack,
    author = "Loizides, Constantin and Kamin, Jason and d'Enterria, David",
    title = "{Improved Monte Carlo Glauber predictions at present and future nuclear colliders}",
    eprint = "1710.07098",
    archivePrefix = "arXiv",
    primaryClass = "nucl-ex",
    doi = "10.1103/PhysRevC.97.054910",
    journal = "Phys. Rev. C",
    volume = "97",
    number = "5",
    pages = "054910",
    year = "2018",
    note = "[Erratum: Phys.Rev.C 99, 019901 (2019)]"
}

@article{Klos:2007is,
    author = "Kłos, B. and others",
    title = "{Neutron density distributions from antiprotonic $^{208}$Pb and $^{209}$Bi atoms}",
    eprint = "nucl-ex/0702016",
    archivePrefix = "arXiv",
    doi = "10.1103/PhysRevC.76.014311",
    journal = "Phys. Rev. C",
    volume = "76",
    pages = "014311",
    year = "2007"
}

@article{Tarbert:2013jze,
    author = "{Crystal Ball at MAMI and A2 Collaboration}",
    title = "{Neutron skin of $^{208}$Pb from Coherent Pion Photoproduction}",
    eprint = "1311.0168",
    archivePrefix = "arXiv",
    primaryClass = "nucl-ex",
    doi = "10.1103/PhysRevLett.112.242502",
    journal = "Phys. Rev. Lett.",
    volume = "112",
    number = "24",
    pages = "242502",
    year = "2014"
}

@article{ALICE:2012aa,
    author = "{ALICE Collaboration}",
    title = "{Measurement of the Cross Section for Electromagnetic Dissociation with Neutron Emission in Pb-Pb Collisions at $\sqrt{s_{NN}}$ = 2.76 TeV}",
    eprint = "1203.2436",
    archivePrefix = "arXiv",
    primaryClass = "nucl-ex",
    reportNumber = "CERN-PH-EP-2012-065",
    doi = "10.1103/PhysRevLett.109.252302",
    journal = "Phys. Rev. Lett.",
    volume = "109",
    pages = "252302",
    year = "2012"
}

@article{Czyczula:2012ny,
    author = "Czyczula, Z. and Przedzinski, T. and Was, Z.",
    title = "{TauSpinner Program for Studies on Spin Effect in tau Production at the LHC}",
    eprint = "1201.0117",
    archivePrefix = "arXiv",
    primaryClass = "hep-ph",
    reportNumber = "CERN-PH-TH-20011-307",
    doi = "10.1140/epjc/s10052-012-1988-z",
    journal = "Eur. Phys. J. C",
    volume = "72",
    pages = "1988",
    year = "2012"
}

@article{Iguro:2023rom,
    author = "Iguro, Syuhei and Kitahara, Teppei",
    title = "{Electric dipole moments as probes of the $R_{D^{(*)}}$ anomaly}",
    eprint = "2307.11751",
    archivePrefix = "arXiv",
    primaryClass = "hep-ph",
    reportNumber = "P3H-23-046, TTP23-026, CHIBA-EP-263",
    doi = "10.1103/PhysRevD.110.075008",
    journal = "Phys. Rev. D",
    volume = "110",
    number = "7",
    pages = "075008",
    year = "2024"
}

@article{AtauUnparticles,
  title = {Weak properties of the $\ensuremath{\tau}$ lepton via a spin-0 unparticle},
  author = {Moyotl, A. and Tavares-Velasco, G.},
  journal = {Phys. Rev. D},
  volume = {86},
  issue = {1},
  pages = {013014},
  numpages = {9},
  year = {2012},
  month = {Jul},
  publisher = {American Physical Society},
  doi = {10.1103/PhysRevD.86.013014},
  %url = {https://link.aps.org/doi/10.1103/PhysRevD.86.013014}
}

@article{Sakharov:1967dj,
    author = "Sakharov, A. D.",
    title = "{Violation of CP Invariance, C asymmetry, and baryon asymmetry of the universe}",
    doi = "10.1070/PU1991v034n05ABEH002497",
    journal = "Pisma Zh. Eksp. Teor. Fiz.",
    volume = "5",
    pages = "32--35",
    year = "1967"
}

@article{Bhide:2024nje,
    author = "Bhide, Kartik and Lang, Valerie",
    title = "{Probing optimal measurements of the electromagnetic dipole moments of the $\tau$ lepton}",
    eprint = "2410.23070",
    archivePrefix = "arXiv",
    primaryClass = "hep-ph",
    month = "10",
    year = "2024",
}

@article{Shao:2023bga,
    author = "Shao, Dingyu and Yan, Bin and Yuan, Shu-Run and Zhang, Cheng",
    title = "{Spin asymmetry and dipole moments in \ensuremath{\tau}-pair production with ultraperipheral heavy ion collisions}",
    eprint = "2310.14153",
    archivePrefix = "arXiv",
    primaryClass = "hep-ph",
    doi = "10.1007/s11433-024-2389-y",
    journal = "Science China Physics, Mechanics \& Astronomy",
    volume = "67",
    number = "8",
    pages = "281062",
    year = "2024",
}

@article{Burmasov:2023cwv,
    author = {Burmasov, N. and Kryshen, E. and B\"uhler, P. and Lavicka, R.},
    title = "{Feasibility Studies of Tau-Lepton Anomalous Magnetic Moment Measurements in Ultraperipheral Collisions at the LHC}",
    doi = "10.1134/S1063779623040111",
    journal = "Physics of Particles and Nuclei",
    volume = "54",
    number = "4",
    pages = "590--594",
    year = "2023",
}

@article{Verducci:2023cgx,
    author = "Verducci, Monica and Roda, Chiara and Cavasinni, Vincenzo and Vignaroli, Natascia",
    title = "{Study of the measurement of the \ensuremath{\tau} lepton anomalous magnetic moment in high energy lead-lead collisions at the LHC}",
    eprint = "2307.15160",
    archivePrefix = "arXiv",
    primaryClass = "hep-ph",
    doi = "10.1103/PhysRevD.110.052001",
    journal = "Phys. Rev. D",
    volume = "110",
    number = "5",
    pages = "052001",
    year = "2024",
}

@article{Burmasov:2022par,
    author = "Burmasov, N. A.",
    title = "{Search for New Physics in Ultraperipheral Collisions at the Large Hadron Collider}",
    doi = "10.1134/S1063778823010143",
    journal = "Physics of Atomic Nuclei",
    volume = "85",
    number = "6",
    pages = "942--950",
    year = "2022",
}

@article{Goncalves:2020btj,
    author = "Goncalves, Victor P. and Martins, Daniel E. and Rangel, Murilo S.",
    title = "{Exclusive dilepton production at forward rapidities in $PbPb$ collisions}",
    eprint = "2012.08923",
    archivePrefix = "arXiv",
    primaryClass = "hep-ph",
    doi = "10.1140/epjc/s10052-021-09015-w",
    journal = "Eur. Phys. J. C",
    volume = "81",
    number = "3",
    pages = "220",
    year = "2021",
}

@article{Beresford:2024dsc,
    author = "Beresford, Lydia and Clawson, Savannah and Liu, Jesse",
    title = "{Strategy to measure tau g-2 via photon fusion in LHC proton collisions}",
    eprint = "2403.06336",
    archivePrefix = "arXiv",
    primaryClass = "hep-ph",
    doi = "10.1103/PhysRevD.110.092016",
    journal = "Phys. Rev. D",
    volume = "110",
    number = "9",
    pages = "092016",
    year = "2024",
}

@article{Atag:2010ja,
    author = "Atağ, S. and Billur, A. A.",
    title = "{Possibility of determining $\tau$ lepton electromagnetic moments in ${\gamma\gamma \to \tau^{+}\tau^{-}}$ process at the CERN-LHC}",
    eprint = "1005.2841",
    archivePrefix = "arXiv",
    primaryClass = "hep-ph",
    doi = "10.1007/JHEP11(2010)060",
    journal = "JHEP",
    volume = "11",
    pages = "060",
    year = "2010",
}

@article{Haisch:2023upo,
    author = "Haisch, Ulrich and Schnell, Luc and Weiss, Joachim",
    title = "{LHC tau-pair production constraints on $a_\tau$ and $d_\tau$}",
    eprint = "2307.14133",
    archivePrefix = "arXiv",
    primaryClass = "hep-ph",
    reportNumber = "MPP-2023-154",
    doi = "10.21468/SciPostPhys.16.2.048",
    journal = "SciPost Phys.",
    volume = "16",
    number = "2",
    pages = "048",
    year = "2024",
}

@article{Galon:2016ngp,
    author = "Galon, Iftah and Rajaraman, Arvind and Tait, Tim M. P.",
    title = "{$H \rightarrow \tau^+\tau^-\gamma$ as a probe of the $\tau$ magnetic dipole moment}",
    eprint = "1610.01601",
    archivePrefix = "arXiv",
    primaryClass = "hep-ph",
    reportNumber = "UCI-HEP-TR-2016-16, MITP-16-104",
    doi = "10.1007/JHEP12(2016)111",
    journal = "JHEP",
    volume = "12",
    pages = "111",
    year = "2016",
}

@article{Cao2021,
    doi = {10.1088/1674-1137/ac0e88},
%    url = {https://dx.doi.org/10.1088/1674-1137/ac0e88},
    year = {2021},
    month = {sep},
    publisher = {Chinese Physical Society and the Institute of High Energy Physics of the Chinese Academy of Sciences and the Institute of Modern Physics of the Chinese Academy of Sciences and IOP Publishing Ltd},
   volume = {45},
   number = {9},
   pages = {093108},
   author = {Cao, Qing-Hong and Jiang, Hao-Ran and Li, Bin and Liu, Yandong and Zeng, Guojin},
   title = {Probing magnetic moment operators in ${\boldsymbol H \gamma }$ production and ${\boldsymbol H \to \tau^+ \tau^- \gamma }$ rare decay},
   journal = {Chinese Physics C},
   eprint = "2106.04143",
   archivePrefix = "arXiv",
   primaryClass = "hep-ph",
}

@article{BERNABEU2008160,
    title = {Tau anomalous magnetic moment form factor at super B/flavor factories},
    journal = {Nuclear Physics B},
    volume = {790},
    number = {1},
    pages = {160-174},
    year = {2008},
    issn = {0550-3213},
    doi = {https://doi.org/10.1016/j.nuclphysb.2007.09.001},
   % url = {https://www.sciencedirect.com/science/article/pii/S0550321307006396},
    author = "{J. Bernabéu, G.A. González-Sprinberg, J. Papavassiliou and J. Vidal}",
}

@article{TauSpinCorrsInee,
    doi = {10.1088/1126-6708/2009/01/062},
    %url = {https://dx.doi.org/10.1088/1126-6708/2009/01/062},
    year = {2009},
    month = {jan},
    publisher = {},
    volume = {2009},
    number = {01},
    pages = {062},
    author = "{J. Bernabéu, G.A. González-Sprinberg and J. Vidal}",
    title = {Tau spin correlations and the anomalous magnetic moment},
journal = {Journal of High Energy Physics},
}

@article{Eidelman:2016aih,
    author = "Eidelman, S. and Epifanov, D. and Fael, M. and Mercolli, L. and Passera, M.",
    title = "{$\tau$ dipole moments via radiative leptonic $\tau$ decays}",
    eprint = "1601.07987",
    archivePrefix = "arXiv",
    primaryClass = "hep-ph",
    doi = "10.1007/JHEP03(2016)140",
    journal = "JHEP",
    volume = "03",
    pages = "140",
    year = "2016",
}

@article{Fomin:2018ybj,
    author = "Fomin, A. S. and Korchin, A. Yu and Stocchi, A. and Barsuk, S. and Robbe, P.",
    title = "{Feasibility of $\tau$ -lepton electromagnetic dipole moments measurement using bent crystal at the LHC}",
    eprint = "1810.06699",
    archivePrefix = "arXiv",
    primaryClass = "hep-ph",
    doi = "10.1007/JHEP03(2019)156",
    journal = "JHEP",
    volume = "03",
    pages = "156",
    year = "2019",
}

@article{BentCrystal,
  title = {Novel Method for the Direct Measurement of the $\ensuremath{\tau}$ Lepton Dipole Moments},
  author = {Fu, J. and Giorgi, M. A. and Henry, L. and Marangotto, D. and Vidal, F. Mart\'{\i}nez and Merli, A. and Neri, N. and Vidal, J. Ruiz},
  journal = {Phys. Rev. Lett.},
  volume = {123},
%  issue = {1},
  pages = {011801},
  numpages = {5},
  year = {2019},
  month = {Jul},
  publisher = {American Physical Society},
  doi = {10.1103/PhysRevLett.123.011801},
  %url = {https://link.aps.org/doi/10.1103/PhysRevLett.123.011801},
  archivePrefix = "arXiv",
  primaryClass = "hep-ex",
  eprint = "1901.04003",
}

@misc{dagostini2010improvediterativebayesianunfolding,
      title={Improved iterative Bayesian unfolding}, 
      author={G. D'Agostini},
      year={2010},
      eprint={1010.0632},
      archivePrefix={arXiv},
      primaryClass={physics.data-an},
      %url={https://arxiv.org/abs/1010.0632}, 
}

@inproceedings{Malaescu:2011yg,
    author = "Malaescu, Bogdan",
    title = "{An Iterative, Dynamically Stabilized (IDS) Method of Data Unfolding}",
    booktitle = "{PHYSTAT 2011}",
    eprint = "1106.3107",
    archivePrefix = "arXiv",
    primaryClass = "physics.data-an",
    reportNumber = "CERN-PH-EP-2011-111",
    doi = "10.5170/CERN-2011-006.271",
    pages = "271--275",
    year = "2011"
}

@article{Malaescu:2009dm,
    author = "Malaescu, Bogdan",
    title = "{An Iterative, dynamically stabilized method of data unfolding}",
    eprint = "0907.3791",
    archivePrefix = "arXiv",
    primaryClass = "physics.data-an",
    reportNumber = "LAL-09-107",
    month = "7",
    year = "2009"
}

@article{Eidelman:2007sb,
      author         = "Eidelman, S. and Passera, M.",
      title          = "{Theory of the $\tau$ lepton anomalous magnetic moment}",
      journal        = "Mod. Phys. Lett.",
      volume         = "A22",
      year           = "2007",
      pages          = "159-179",
      doi            = "10.1142/S0217732307022694",
      eprint         = "hep-ph/0701260",
      archivePrefix  = "arXiv",
      primaryClass   = "hep-ph",
      SLACcitation   = "%%CITATION = HEP-PH/0701260;%%"
}

@article{Belle:2021ybo,
    %author = "Inami, K. and others",
    %collaboration = "Belle",
    author = "{Belle Collaboration}",
    title = "{An improved search for the electric dipole moment of the $\tau$ lepton}",
    eprint = "2108.11543",
    archivePrefix = "arXiv",
    primaryClass = "hep-ex",
    reportNumber = "Belle Preprint 2021-22, KEK Preprint 2021-26",
    doi = "10.1007/JHEP04(2022)110",
    journal = "JHEP",
    volume = "04",
    pages = "110",
    year = "2022"
}

@article{Helenius:2017aqz,
    author = "Helenius, Ilkka",
    title = "{Photon-photon and photon-hadron processes in Pythia 8}",
    eprint = "1708.09759",
    archivePrefix = "arXiv",
    primaryClass = "hep-ph",
    doi = "10.23727/CERN-Proceedings-2018-001.119",
    journal = "CERN Proc.",
    volume = "1",
    pages = "119",
    year = "2018"
}

@article{Harland-Lang:2021ysd,
    author = "Harland-Lang, L. A. and Khoze, V. A. and Ryskin, M. G.",
    title = "{Elastic photon-initiated production at the LHC: the role of hadron-hadron interactions}",
    eprint = "2104.13392",
    archivePrefix = "arXiv",
    primaryClass = "hep-ph",
    reportNumber = "IPPP/20/97",
    doi = "10.21468/SciPostPhys.11.3.064",
    journal = "SciPost Phys.",
    volume = "11",
    pages = "064",
    year = "2021"
}

@Booklet{ATL-DAQ-PUB-2019-001-customRef,
    author         = "{ATLAS Collaboration}",
    title          = "{Trigger menu in 2018}",
    howpublished   = "{ATL-DAQ-PUB-2019-001}",
    url            = "https://cds.cern.ch/record/2693402",
    year           = "2019",
}

@article{Jadach:1993hs,
    author = "Jadach, S. and Was, Z. and Decker, R. and K{\"u}hn, Johann H.",
    title = "{The $\tau$ decay library TAUOLA, version 2.4}",
    reportNumber = "CERN-TH-6793-93",
    doi = "10.1016/0010-4655(93)90061-G",
    journal = "Comput. Phys. Commun.",
    volume = "76",
    pages = "361--380",
    year = "1993"
}

@article{Davidson:2010rw,
    author = "Davidson, N. and Nanava, G. and Przedzinski, T. and Richter-Was, E. and Was, Z.",
    title = "{Universal interface of TAUOLA: Technical and physics documentation}",
    eprint = "1002.0543",
    archivePrefix = "arXiv",
    primaryClass = "hep-ph",
    reportNumber = "IFJPAN-IV-2009-10",
    doi = "10.1016/j.cpc.2011.12.009",
    journal = "Comput. Phys. Commun.",
    volume = "183",
    pages = "821--843",
    year = "2012"
}

@article{Fermi1,
    author = "Fermi, E.",
    title = "{Über die Theorie des Stoßes zwischen Atomen und elektrisch geladenen Teilchen}",
    doi = "10.1007/BF03184853",
    journal = "Zeitschrift für Physik",
    volume = "29",
    pages = "315–-327",
    year = "1924"
}

@article{Fermi2,
    author = "Fermi, Enrico",
    editor = "Marciano, W. J. and White, S.",
    %title = "{On the theory of collisions between atoms and electrically charged particles}",
    title = "{Sulla teoria dell’ urto tra atomi e corpuscoli elettrici}", 
    eprint = "hep-th/0205086",
    archivePrefix = "arXiv",
    doi = "10.1007/BF02961914",
    %journal = "Nuovo Cim.",
    journal = "Il Nuovo Cimento (1924-1942)",
    volume = "2",
    pages = "143--158",
    year = "1925"
}

@article{Weizsaecker,
    author = "Weizsäcker, C.F.v.",
    title = "{Ausstrahlung bei Stößen sehr schneller Elektronen}",
    doi = "10.1007/BF01333110",
    journal = "Zeitschrift für Physik",
    volume = "88",
    pages = "612–-625",
    year = "1934"
}

@article{Williams,
    author = "Williams, E. J.",
    title = "{Correlation of certain collision problems with radiation theory}",
    journal = "Kong. Dan. Vid. Sel. Mat. Fys. Med.",
    volume = "13N4",
    number = "4",
    pages = "1--50",
    year = "1935",
    url = {https://gymarkiv.sdu.dk/MFM/kdvs/mfm%2010-19/mfm-13-4.pdf}
}

@misc{zolotorevMcDonald,
      title={Classical Radiation Processes in the Weizsacker-Williams Approximation}, 
      author={Max S. Zolotorev and Kirk T. McDonald},
      year={2000},
      eprint={physics/0003096},
      archivePrefix={arXiv},
      primaryClass={physics.class-ph},
      %url={https://arxiv.org/abs/physics/0003096}, 
}

@inproceedings{Roesler:2000he,
    author = "Roesler, Stefan and Engel, Ralph and Ranft, Johannes",
    title = {The Monte Carlo Event Generator DPMJET-III},
    booktitle = "{International Conference on Advanced Monte Carlo for Radiation Physics, Particle Transport Simulation and Applications (MC 2000)}",
    eprint = "hep-ph/0012252",
    archivePrefix = "arXiv",
    reportNumber = "SLAC-PUB-8740",
    doi = "10.1007/978-3-642-18211-2_166",
    pages = "1033--1038",
    month = "12",
    year = "2001"
}

@article{BreitWheeler1934,
  title = {Collision of Two Light Quanta},
  author = {Breit, G. and Wheeler, John A.},
  journal = {Phys. Rev.},
  volume = {46},
  pages = {1087--1091},
  numpages = {0},
  year = {1934},
  month = {12},
  publisher = {American Physical Society},
  doi = {10.1103/PhysRev.46.1087},
}

@article{HeisenbergEuler1936,
    author = "Heisenberg, W. and Euler, H",
    title = "{Folgerungen aus der Diracschen Theorie des Positrons}",
    journal = "{Zeitschrift für Physik}",
    volume = "{98}",
    pages = {714--732},
    year = "1936",
    doi = {10.1007/BF01343663},
    eprint         = "physics/0605038",
    archivePrefix  = "arXiv",
}

@article{SchwingerLimit,
  title = {On Gauge Invariance and Vacuum Polarization},
  author = {Schwinger, Julian},
  journal = {Phys. Rev.},
  volume = {82},
  pages = {664--679},
  numpages = {0},
  year = {1951},
  month = {6},
  publisher = {American Physical Society},
  doi = {10.1103/PhysRev.82.664},
}

@article{PhysRevD.7.3485,
  title = {Lepton Pair Production from Two-Photon Processes},
  author = {Chen, Min-Shih and Muzinich, I. J. and Terazawa, Hidezumi and Cheng, T. P.},
  journal = {Phys. Rev. D},
  volume = {7},
  pages = {3485--3502},
  numpages = {0},
  year = {1973},
  month = {6},
  publisher = {American Physical Society},
  doi = {10.1103/PhysRevD.7.3485},
}

@article{Klein:2020fmr,
    author = "{S. R. Klein and P. Steinberg}",
    title = "{Photonuclear and Two-Photon Interactions at High-Energy Nuclear Colliders}",
    eprint = "2005.01872",
    archivePrefix = "arXiv",
    primaryClass = "nucl-ex",
    doi = "10.1146/annurev-nucl-030320-033923",
    journal = "Ann. Rev. Nucl. Part. Sci.",
    volume = "70",
    pages = "323--354",
    year = "2020"
}

@article{Budnev:1974de,
      author         = "Budnev, V. M. and Ginzburg, I. F. and Meledin, G. V. and
                        Serbo, V. G.",
      title          = "{The two-photon particle production mechanism. Physical
                        problems. Applications. Equivalent photon approximation}",
      journal        = "Phys. Rept.",
      volume         = "15",
      year           = "1975",
      pages          = "181-281",
      doi            = "10.1016/0370-1573(75)90009-5",
      SLACcitation   = "%%CITATION = PRPLC,15,181;%%"
}

@article{Baltz:2007kq,
    author = "Baltz, A. J. and others",
    title = "{The physics of ultraperipheral collisions at the LHC}",
    eprint = "0706.3356",
    archivePrefix = "arXiv",
    primaryClass = "nucl-ex",
    doi = "10.1016/j.physrep.2007.12.001",
    journal = "Phys. Rept.",
    volume = "458",
    pages = "1--171",
    year = "2008"
}

@article{deFavereaudeJeneret:2009db,
      author         = "{de Favereau de Jeneret}, J. and Lemaître, V. and Liu, Y.
                        and Ovyn, S. and Pierzchala, T. and Piotrzkowski, K. and
                        Rouby, X. and Schul, N. and Vander Donckt, M.",
      title          = "{High energy photon interactions at the LHC}",
      year           = "2009",
      eprint         = "0908.2020",
      archivePrefix  = "arXiv",
      primaryClass   = "hep-ph",
      reportNumber   = "CP3-09-37",
      SLACcitation   = "%%CITATION = ARXIV:0908.2020;%%"
}

@article{Acciarri:1998iv,
      author         = "{L3 Collaboration}",
      title          = "{Measurement of the anomalous magnetic and electric
                        dipole moments of the tau lepton}",
      collaboration  = "L3",
      journal        = "Phys. Lett. B",
      volume         = "434",
      year           = "1998",
      pages          = "169-179",
      doi            = "10.1016/S0370-2693(98)00736-9",
      reportNumber   = "CERN-EP-98-045, CERN-EP-98-45",
      SLACcitation   = "%%CITATION = PHLTA,B434,169;%%"
}

@techreport{Cranmer:1456844,
      %author        = "Cranmer, Kyle and Lewis, George and Moneta, Lorenzo and Shibata, Akira and Verkerke, Wouter",
      author        = "{ROOT Collaboration}",
      title         = "{HistFactory: A tool for creating statistical models for
                       use with RooFit and RooStats}",
      institution   = "New York U.",
      collaboration = "ROOT Collaboration",
      address       = "New York",
      reportNumber  = "CERN-OPEN-2012-016",
      month         = "1",
      year          = "2012",
      url           = "https://cds.cern.ch/record/1456844",
}

@Booklet{ATLAS-TDR-ZDC,
    author         = "{ATLAS Collaboration}",
    title          = "{Zero Degree Calorimeters for ATLAS}",
    howpublished   = "{CERN-LHCC-2007-001}",
    url            = "https://cds.cern.ch/record/1009649",
    year           = "2007",
}

@article{dtauTheo,
  title = {Large Long-Distance Contributions to the Electric Dipole Moments of Charged Leptons in the Standard Model},
  author = {Yamaguchi, Yasuhiro and Yamanaka, Nodoka},
  journal = {Phys. Rev. Lett.},
  volume = {125},
  issue = {24},
  pages = {241802},
  numpages = {7},
  year = {2020},
  month = {Dec},
  publisher = {American Physical Society},
  doi = {10.1103/PhysRevLett.125.241802},
  %url = {https://link.aps.org/doi/10.1103/PhysRevLett.125.241802}
}

@article{Abdallah:2003xd,
      title          = "{Study of tau-pair production in photon-photon collisions
                        at LEP and limits on the anomalous electromagnetic moments
                        of the tau lepton}",
      author         = "{DELPHI Collaboration}",
      journal        = "Eur. Phys. J. C",
      volume         = "35",
      year           = "2004",
      pages          = "159-170",
      doi            = "10.1140/epjc/s2004-01852-y",
      eprint         = "hep-ex/0406010",
      archivePrefix  = "arXiv",
      primaryClass   = "hep-ex",
      reportNumber   = "CERN-EP-2003-058",
      SLACcitation   = "%%CITATION = HEP-EX/0406010;%%"
}

@article{Ackerstaff:1998mt,
      title          = "{An upper limit on the anomalous magnetic moment of the
                        $\tau$ lepton}",
      author         = "{OPAL Collaboration}",
      journal        = "Phys. Lett. B",
      volume         = "431",
      year           = "1998",
      pages          = "188-198",
      doi            = "10.1016/S0370-2693(98)00520-6",
      eprint         = "hep-ex/9803020",
      archivePrefix  = "arXiv",
      primaryClass   = "hep-ex",
      reportNumber   = "CERN-EP-98-033",
      SLACcitation   = "%%CITATION = HEP-EX/9803020;%%"
}

@article{DELAGUILA1991256,
    title = "The possibility of using a large heavy-ion collider for measuring the electromagnetic properties of the tau lepton",
    journal = "Phys. Lett. B",
    volume = "271",
    number = "1",
    pages = "256 - 260",
    year = "1991",
    doi = "10.1016/0370-2693(91)91309-J",
    author = "F. del Aguila and F. Cornet and J. I. Illana"
}

@article{Beresford:2019gww,
    author = "Beresford, L. and Liu, J.",
    title = "{New physics and tau $g-2$ using LHC heavy ion collisions}",
    eprint = "1908.05180",
    archivePrefix = "arXiv",
    primaryClass = "hep-ph",
    doi = "10.1103/PhysRevD.102.113008",
    journal = "Phys. Rev. D",
    volume = "102",
    number = "11",
    pages = "113008",
    year = "2020"
}

@article{Dyndal:2020yen,
    author = "Dyndał, Mateusz and Kłusek-Gawenda, Mariola and Szczurek, Antoni and Schott, Matthias",
    title = "{Anomalous electromagnetic moments of $\tau$ lepton in $\gamma \gamma \to \tau^+ \tau^-$ reaction in Pb+Pb collisions at the LHC}",
    eprint = "2002.05503",
    archivePrefix = "arXiv",
    primaryClass = "hep-ph",
    doi = "10.1016/j.physletb.2020.135682",
    journal = "Phys. Lett. B",
    volume = "809",
    pages = "135682",
    year = "2020"
}

@article{Klein:2016yzr,
    author = "Klein, Spencer R. and Nystrand, Joakim and Seger, Janet and Gorbunov, Yuri and Butterworth, Joey",
    title = "{STARlight: A Monte Carlo simulation program for ultra-peripheral collisions of relativistic ions}",
    eprint = "1607.03838",
    archivePrefix = "arXiv",
    primaryClass = "hep-ph",
    doi = "10.1016/j.cpc.2016.10.016",
    journal = "Comput. Phys. Commun.",
    volume = "212",
    pages = "258--268",
    year = "2017"
}

@article{Baltz:2009jk,
    author = "Baltz, Anthony J. and Gorbunov, Yuri and Klein, Spencer R. and Nystrand, Joakim",
    title = "{Two-Photon Interactions with Nuclear Breakup in Relativistic Heavy Ion Collisions}",
    eprint = "0907.1214",
    archivePrefix = "arXiv",
    primaryClass = "nucl-ex",
    doi = "10.1103/PhysRevC.80.044902",
    journal = "Phys. Rev. C",
    volume = "80",
    pages = "044902",
    year = "2009"
}

@article{Harland-Lang:2018iur,
    author = "Harland-Lang, L. A. and Khoze, V. A. and Ryskin, M. G.",
    title = "{Exclusive LHC physics with heavy ions: SuperChic 3}",
    eprint = "1810.06567",
    archivePrefix = "arXiv",
    primaryClass = "hep-ph",
    reportNumber = "IPPP/18/90",
    doi = "10.1140/epjc/s10052-018-6530-5",
    journal = "Eur. Phys. J. C",
    volume = "79",
    number = "1",
    pages = "39",
    year = "2019"
}

@article{gammaUPC,
    author = "Shao, Hua-Sheng and d'Enterria, David",
    title = "{gamma-UPC: automated generation of exclusive photon-photon processes in ultraperipheral proton and nuclear collisions with varying form factors}",
    eprint = "2207.03012",
    archivePrefix = "arXiv",
    primaryClass = "hep-ph",
    doi = "10.1007/JHEP09(2022)248",
    journal = "JHEP",
    volume = "09",
    pages = "248",
    year = "2022"
}

@article{Helenius:2019gbd,
    author = "Helenius, Ilkka and Rasmussen, Christine O.",
    title = "{Hard diffraction in photoproduction with Pythia 8}",
    eprint = "1901.05261",
    archivePrefix = "arXiv",
    primaryClass = "hep-ph",
    reportNumber = "LU TP 19-06, MCNET-19-01",
    doi = "10.1140/epjc/s10052-019-6914-1",
    journal = "Eur. Phys. J. C",
    volume = "79",
    number = "5",
    pages = "413",
    year = "2019"
}

@article{Schafer:2020bnm,
    author = {Sch\"afer, Wolfgang},
    title = "{Photon induced processes: from ultraperipheral to semicentral heavy ion collisions}",
    doi = "10.1140/epja/s10050-020-00231-8",
    journal = "The European Physical Journal A",
    volume = "56",
    number = "9",
    pages = "231",
    year = "2020"
}

@software{trexfitter,
  author       = {{ATLAS Collaboration}},
  title        = {TRExFitter},
  publisher    = {Zenodo},
  doi          = {10.5281/zenodo.14845712},
  url          = {https://doi.org/10.5281/zenodo.14845712},
}

@article{Sjostrand:2014zea,
      author         = "Sj{\"o}strand, Torbj{\"o}rn and Ask, Stefan and Christiansen,
                        Jesper R. and Corke, Richard and Desai, Nishita and Ilten,
                        Philip and Mrenna, Stephen and Prestel, Stefan and
                        Rasmussen, Christine O. and Skands, Peter Z.",
      title          = "{An introduction to PYTHIA 8.2}",
      journal        = "Comput. Phys. Commun.",
      volume         = "191",
      year           = "2015",
      pages          = "159",
      doi            = "10.1016/j.cpc.2015.01.024",
      eprint         = "1410.3012",
      archivePrefix  = "arXiv",
      primaryClass   = "hep-ph",
      reportNumber   = "LU-TP-14-36, MCNET-14-22, CERN-PH-TH-2014-190,
                        FERMILAB-PUB-14-316-CD, DESY-14-178, SLAC-PUB-16122,
                        --FERMILAB-PUB-14-316-CD",
      SLACcitation   = "%%CITATION = ARXIV:1410.3012;%%"
}

@Article{Alwall:2014hca,
      author         = "Alwall, J. and Frederix, R. and Frixione, S. and Hirschi,
                        V. and Maltoni, F. and Mattelaer, O. and Shao, H. -S. and
                        Stelzer, T. and Torrielli, P. and Zaro, M.",
      title          = "{The automated computation of tree-level and
                        next-to-leading order differential cross sections, and
                        their matching to parton shower simulations}",
      journal        = "JHEP",
      volume         = "07",
      year           = "2014",
      pages          = "079",
      doi            = "10.1007/JHEP07(2014)079",
      eprint         = "1405.0301",
      archivePrefix  = "arXiv",
      primaryClass   = "hep-ph",
      reportNumber   = "CERN-PH-TH-2014-064, CP3-14-18, LPN14-066, MCNET-14-09,
                        ZU-TH-14-14",
      SLACcitation   = "%%CITATION = ARXIV:1405.0301;%%"
}

@article{Agostinelli:2002hh,
      author         = "Agostinelli, S. and others",
      title          = "{\textsc{Geant4} -- a simulation toolkit}",
      journal        = "Nucl. Instrum. Meth. A",
      volume         = "506",
      year           = "2003",
      pages          = "250",
      doi            = "10.1016/S0168-9002(03)01368-8",
      reportNumber   = "SLAC-PUB-9350, FERMILAB-PUB-03-339",
      SLACcitation   = "%%CITATION = NUIMA,A506,250;%%"
}

@Article{Cowan:2010js,
   Author = {Cowan, Glen and Cranmer, Kyle and Gross, Eilam and Vitells, Ofer},
   Title = "{Asymptotic formulae for likelihood-based tests of new physics}",
   Journal = {Eur. Phys. J. C},
   Volume = {71},
   Year = {2011},
   Pages = {1554},
   doi = "10.1140/epjc/s10052-011-1554-0",
   Eprint = "1007.1727",
   Archiveprefix = {arXiv},
   Primaryclass = {physics.data-an},
   related = "Cowan:2010js-err",
   relatedstring  = "Erratum:",
}

@Inproceedings{Adye:2011gm,
    author = "Adye, Tim",
    title = "{Unfolding algorithms and tests using RooUnfold}",
    booktitle = "{Proceedings, 2011 Workshop on Statistical Issues Related to Discovery Claims in Search Experiments and Unfolding (PHYSTAT 2011)}",
    venue = "CERN, Geneva, Switzerland",
    eventdate = "2011-01-17/2011-01-20",
    pages = "313-318",
    eprint = "1105.1160",
    archivePrefix = "arXiv",
    primaryClass = "physics.data-an",
    doi = "10.5170/CERN-2011-006.313",
}

@article{DAgostini:1994fjx,
      author         = "D'Agostini, G.",
      title          = "{A multidimensional unfolding method based on Bayes' theorem}",
      journal        = "Nucl. Instrum. Meth. A",
      volume         = "362",
      year           = "1995",
      pages          = "487-498",
      doi            = "10.1016/0168-9002(95)00274-X",
      reportNumber   = "DESY-94-099",
      SLACcitation   = "%%CITATION = NUIMA,A362,487;%%"
}

@Article{PERF-2007-01,
    author         = "{ATLAS Collaboration}",
    title          = "{The ATLAS Experiment at the CERN Large Hadron Collider}",
    journal        = "JINST",
    volume         = "3",
    year           = "2008",
    pages          = "S08003",
    doi            = "10.1088/1748-0221/3/08/S08003",
    primaryClass   = "hep-ex",
}

@Article{SOFT-2010-01,
    author         = "{ATLAS Collaboration}",
    title          = "{The ATLAS Simulation Infrastructure}",
    journal        = "Eur. Phys. J. C",
    volume         = "70",
    year           = "2010",
    pages          = "823",
    doi            = "10.1140/epjc/s10052-010-1429-9",
    eprint         = "1005.4568",
    archivePrefix  = "arXiv",
    primaryClass   = "physics.ins-det",
}

@Article{STDM-2011-01,
    author         = "{ATLAS Collaboration}",
    title          = "{Rapidity gap cross sections measured with the ATLAS detector in \(pp\) collisions at \(\sqrt{s} = 7\,\text{TeV}\)}",
    journal        = "Eur. Phys. J. C",
    volume         = "72",
    year           = "2012",
    pages          = "1926",
    doi            = "10.1140/epjc/s10052-012-1926-0",
    reportNumber   = "CERN-PH-EP-2011-220",
    eprint         = "1201.2808",
    archivePrefix  = "arXiv",
    primaryClass   = "hep-ex",
}

@Article{PERF-2014-07,
    author         = "{ATLAS Collaboration}",
    title          = "{Topological cell clustering in the ATLAS calorimeters and its performance in LHC Run~1}",
    journal        = "Eur. Phys. J. C",
    volume         = "77",
    year           = "2017",
    pages          = "490",
    doi            = "10.1140/epjc/s10052-017-5004-5",
    reportNumber   = "CERN-PH-EP-2015-304",
    eprint         = "1603.02934",
    archivePrefix  = "arXiv",
    primaryClass   = "hep-ex",
}

@Article{PERF-2015-07,
    author         = "{ATLAS Collaboration}",
    title          = "{Study of the material of the ATLAS inner detector for Run~2 of the LHC}",
    journal        = "JINST",
    volume         = "12",
    year           = "2017",
    pages          = "P12009",
    doi            = "10.1088/1748-0221/12/12/P12009",
    reportNumber   = "CERN-EP-2017-081",
    eprint         = "1707.02826",
    archivePrefix  = "arXiv",
    primaryClass   = "hep-ex",
}

@Article{PERF-2015-08,
    author         = "{ATLAS Collaboration}",
    title          = "{Performance of the ATLAS track reconstruction algorithms in dense environments in LHC Run~2}",
    journal        = "Eur. Phys. J. C",
    volume         = "77",
    year           = "2017",
    pages          = "673",
    doi            = "10.1140/epjc/s10052-017-5225-7",
    reportNumber   = "CERN-EP-2017-045",
    eprint         = "1704.07983",
    archivePrefix  = "arXiv",
    primaryClass   = "hep-ex",
}

@Article{PERF-2015-10,
    author         = "{ATLAS Collaboration}",
    title          = "{Muon reconstruction performance of the ATLAS detector in proton--proton collision data at \(\sqrt{s} = 13\,\text{TeV}\)}",
    journal        = "Eur. Phys. J. C",
    volume         = "76",
    year           = "2016",
    pages          = "292",
    doi            = "10.1140/epjc/s10052-016-4120-y",
    reportNumber   = "CERN-EP-2016-033",
    eprint         = "1603.05598",
    archivePrefix  = "arXiv",
    primaryClass   = "hep-ex",
}

@Article{HION-2016-02,
    author         = "{ATLAS Collaboration}",
    title          = "{Exclusive dimuon production in ultraperipheral Pb+Pb collisions at \(\sqrt{s_{\text{NN}}} = 5.02\,\text{TeV}\) with ATLAS}",
    journal        = "Phys. Rev. C",
    volume         = "104",
    year           = "2021",
    pages          = "024906",
    doi            = "10.1103/PhysRevC.104.024906",
    reportNumber   = "CERN-EP-2020-138",
    eprint         = "2011.12211",
    archivePrefix  = "arXiv",
    primaryClass   = "nucl-ex",
}

@Article{TRIG-2016-01,
    author         = "{ATLAS Collaboration}",
    title          = "{Performance of the ATLAS trigger system in 2015}",
    journal        = "Eur. Phys. J. C",
    volume         = "77",
    year           = "2017",
    pages          = "317",
    doi            = "10.1140/epjc/s10052-017-4852-3",
    reportNumber   = "CERN-EP-2016-241",
    eprint         = "1611.09661",
    archivePrefix  = "arXiv",
    primaryClass   = "hep-ex",
}

@Article{DAPR-2018-01,
    author         = "{ATLAS Collaboration}",
    title          = "{ATLAS data quality operations and performance for 2015--2018 data-taking}",
    journal        = "JINST",
    volume         = "15",
    year           = "2020",
    pages          = "P04003",
    doi            = "10.1088/1748-0221/15/04/P04003",
    reportNumber   = "CERN-EP-2019-207",
    eprint         = "1911.04632",
    archivePrefix  = "arXiv",
    primaryClass   = "physics.ins-det",
}

@Article{EGAM-2018-01,
    author         = "{ATLAS Collaboration}",
    title          = "{Electron and photon performance measurements with the ATLAS detector using the 2015--2017 LHC proton--proton collision data}",
    journal        = "JINST",
    volume         = "14",
    year           = "2019",
    pages          = "P12006",
    doi            = "10.1088/1748-0221/14/12/P12006",
    reportNumber   = "CERN-EP-2019-145",
    eprint         = "1908.00005",
    archivePrefix  = "arXiv",
    primaryClass   = "hep-ex",
}

@Article{MUON-2018-03,
    author         = "{ATLAS Collaboration}",
    title          = "{Muon reconstruction and identification efficiency in ATLAS using the full Run~2 \(pp\) collision data set at \(\sqrt{s} = 13\,\text{TeV}\)}",
    journal        = "Eur. Phys. J. C",
    volume         = "81",
    year           = "2021",
    pages          = "578",
    doi            = "10.1140/epjc/s10052-021-09233-2",
    reportNumber   = "CERN-EP-2020-199",
    eprint         = "2012.00578",
    archivePrefix  = "arXiv",
    primaryClass   = "hep-ex",
}

@Article{PIX-2018-001,
    author         = "Abbott, B. and others",
    title          = "{Production and integration of the ATLAS Insertable B-Layer}",
    journal        = "JINST",
    volume         = "13",
    year           = "2018",
    pages          = "T05008",
    doi            = "10.1088/1748-0221/13/05/T05008",
    eprint         = "1803.00844",
    archivePrefix  = "arXiv",
    primaryClass   = "physics.ins-det",
}

@Article{TRIG-2018-01,
    author         = "{ATLAS Collaboration}",
    title          = "{Performance of the ATLAS muon triggers in Run~2}",
    journal        = "JINST",
    volume         = "15",
    year           = "2020",
    pages          = "P09015",
    doi            = "10.1088/1748-0221/15/09/p09015",
    reportNumber   = "CERN-EP-2020-031",
    eprint         = "2004.13447",
    archivePrefix  = "arXiv",
    primaryClass   = "physics.ins-det",
}

@Article{HION-2019-08,
    author         = "{ATLAS Collaboration}",
    title          = "{Measurement of light-by-light scattering and search for axion-like particles with \(2.2\,\text{nb}^{-1}\) of Pb+Pb data with the ATLAS detector}",
    journal        = "JHEP",
    volume         = "03",
    year           = "2021",
    pages          = "243",
    doi            = "10.1007/JHEP03(2021)243",
    reportNumber   = "CERN-EP-2020-135",
    eprint         = "2008.05355",
    archivePrefix  = "arXiv",
    primaryClass   = "hep-ex",
    related        = "HION-2019-08-err",
    relatedstring  = "Erratum:",
}

@Article{STDM-2019-19,
    author         = "{ATLAS Collaboration}",
    title          = "{Observation of the \(\gamma\gamma \rightarrow \tau\tau\) Process in Pb+Pb Collisions and Constraints on the \(\tau\)-Lepton Anomalous Magnetic Moment with the ATLAS Detector}",
    journal        = "Phys. Rev. Lett.",
    volume         = "131",
    year           = "2023",
    pages          = "151802",
    doi            = "10.1103/PhysRevLett.131.151802",
    reportNumber   = "CERN-EP-2022-079",
    eprint         = "2204.13478",
    archivePrefix  = "arXiv",
    primaryClass   = "hep-ex",
}

@Article{TRIG-2019-04,
    author         = "{ATLAS Collaboration}",
    title          = "{Operation of the ATLAS trigger system in Run~2}",
    journal        = "JINST",
    volume         = "15",
    year           = "2020",
    pages          = "P10004",
    doi            = "10.1088/1748-0221/15/10/P10004",
    reportNumber   = "CERN-EP-2020-109",
    eprint         = "2007.12539",
    archivePrefix  = "arXiv",
    primaryClass   = "physics.ins-det",
}

@Article{DAPR-2021-01,
    author         = "{ATLAS Collaboration}",
    title          = "{Luminosity determination in \(pp\) collisions at \(\sqrt{s} = 13\,\text{TeV}\) using the ATLAS detector at the LHC}",
    journal        = "Eur. Phys. J. C",
    volume         = "83",
    year           = "2023",
    pages          = "982",
    doi            = "10.1140/epjc/s10052-023-11747-w",
    reportNumber   = "CERN-EP-2022-281",
    eprint         = "2212.09379",
    archivePrefix  = "arXiv",
    primaryClass   = "hep-ex",
}

@Article{HION-2021-16,
    author         = "{ATLAS Collaboration}",
    title          = "{Exclusive dielectron production in ultraperipheral Pb+Pb collisions at \(\sqrt{s_{\text{NN}}} = 5.02\,\text{TeV}\) with ATLAS}",
    journal        = "JHEP",
    volume         = "06",
    year           = "2023",
    pages          = "182",
    doi            = "10.1007/JHEP06(2023)182",
    reportNumber   = "CERN-EP-2022-130",
    eprint         = "2207.12781",
    archivePrefix  = "arXiv",
    primaryClass   = "nucl-ex",
}

@Article{EXOT-2022-42,
    author         = "{ATLAS Collaboration}",
    title          = "{A measurement of the high-mass \(\tau\bar{\tau}\) production cross-section at \(\sqrt{s} = 13\,\text{TeV}\) with the ATLAS detector and constraints on new particles and couplings}",
    journal        = "JHEP",
    volume         = "10",
    year           = "2025",
    pages          = "054",
    doi            = "10.1007/JHEP10(2025)054",
    reportNumber   = "CERN-EP-2025-063",
    eprint         = "2503.19836",
    archivePrefix  = "arXiv",
    primaryClass   = "hep-ex",
}

@Article{HION-2023-13,
    author         = "{ATLAS Collaboration}",
    title          = "{Measurement of coincident photon-initiated processes in ultra-peripheral Pb+Pb collisions with the ATLAS detector}",
    year           = "2025",
    reportNumber   = "CERN-EP-2025-071",
    eprint         = "2504.07795",
    archivePrefix  = "arXiv",
    primaryClass   = "nucl-ex",
}

@Booklet{ATL-SOFT-PUB-2021-001,
    author         = "{ATLAS Collaboration}",
    title          = "{The ATLAS Collaboration Software and Firmware}",
    howpublished   = "{ATL-SOFT-PUB-2021-001}",
    url            = "https://cds.cern.ch/record/2767187",
    year           = "2021",
}

@Booklet{ATL-SOFT-PUB-2026-001,
    author         = "{ATLAS Collaboration}",
    title          = "{ATLAS Computing Acknowledgements}",
    howpublished   = "{ATL-SOFT-PUB-2026-001}",
    url            = "https://cds.cern.ch/record/2952666",
    year           = "2026",
}

@Report{ATLAS-TDR-2010-19,
    author         = "{ATLAS Collaboration}",
    title          = "{ATLAS Insertable B-Layer Technical Design Report}",
    type           = "ATLAS-TDR-19; CERN-LHCC-2010-013",
    year           = "2010",
    url            = "https://cds.cern.ch/record/1291633",
    related        = "ATLAS-TDR-2010-19-addm",
    relatedstring  = "Addendum:",
}

@Article{LUCID2,
    author         = "Avoni, G. and others",
    title          = "{The new LUCID-2 detector for luminosity measurement and monitoring in ATLAS}",
    journal        = "JINST",
    volume         = "13",
    year           = "2018",
    pages          = "P07017",
    doi            = "10.1088/1748-0221/13/07/P07017",
}

@Article{CMS-HIN-21-009,
    author         = "{CMS Collaboration}",
    title          = "{Observation of \(\tau\) Lepton Pair Production in Ultraperipheral Pb--Pb Collisions at \(\sqrt{s_{\text{NN}}} = 5.02\,\text{TeV}\)}",
    journal        = "Phys. Rev. Lett.",
    volume         = "131",
    year           = "2023",
    pages          = "151803",
    doi            = "10.1103/PhysRevLett.131.151803",
    reportNumber   = "CERN-EP-2022-098",
    eprint         = "2206.05192",
    archivePrefix  = "arXiv",
    primaryClass   = "nucl-ex",
}

@Booklet{ATL-PHYS-PUB-2015-051,
    author         = "{ATLAS Collaboration}",
    title          = "{Early Inner Detector Tracking Performance in the 2015 Data at \(\sqrt{s} = 13\,\text{TeV}\)}",
    howpublished   = "{ATL-PHYS-PUB-2015-051}",
    url            = "https://cds.cern.ch/record/2110140",
    year           = "2015",
}

@Booklet{ATL-PHYS-PUB-2021-011,
    author         = "{ATLAS Collaboration}",
    title          = "{Evaluating statistical uncertainties and correlations using the bootstrap method}",
    howpublished   = "{ATL-PHYS-PUB-2021-011}",
    url            = "https://cds.cern.ch/record/2759945",
    year           = "2021",
}
\clearpage
 
\begin{flushleft}
\hypersetup{urlcolor=black}
{\Large The ATLAS Collaboration}

\bigskip

\AtlasOrcid[0000-0002-6665-4934]{G.~Aad}$^\textrm{\scriptsize 102}$,
\AtlasOrcid[0000-0001-7616-1554]{E.~Aakvaag}$^\textrm{\scriptsize 17}$,
\AtlasOrcid[0000-0002-5888-2734]{B.~Abbott}$^\textrm{\scriptsize 121}$,
\AtlasOrcid[0000-0002-0287-5869]{S.~Abdelhameed}$^\textrm{\scriptsize 83b}$,
\AtlasOrcid[0000-0002-1002-1652]{K.~Abeling}$^\textrm{\scriptsize 54}$,
\AtlasOrcid[0000-0001-5763-2760]{N.J.~Abicht}$^\textrm{\scriptsize 48}$,
\AtlasOrcid[0000-0002-8496-9294]{S.H.~Abidi}$^\textrm{\scriptsize 30}$,
\AtlasOrcid[0009-0003-6578-220X]{M.~Aboelela}$^\textrm{\scriptsize 44}$,
\AtlasOrcid[0000-0002-9987-2292]{A.~Aboulhorma}$^\textrm{\scriptsize 36e}$,
\AtlasOrcid[0000-0001-5329-6640]{H.~Abramowicz}$^\textrm{\scriptsize 154}$
% ...
\bigskip
\\

% ...
$^{17}$Department for Physics and Technology, University of Bergen, Bergen; Norway.\\
% ...
$^{30}$Physics Department, Brookhaven National Laboratory, Upton NY; United States of America.\\
% ...
$^{36}$$^{(a)}$Facult\'e des Sciences Ain Chock, Universit\'e Hassan II de Casablanca;$^{(b)}$Facult\'{e} des Sciences, Universit\'{e} Ibn-Tofail, K\'{e}nitra;$^{(c)}$Facult\'e des Sciences Semlalia, Universit\'e Cadi Ayyad, LPHEA-Marrakech;$^{(d)}$LPMR, Facult\'e des Sciences, Universit\'e Mohamed Premier, Oujda;$^{(e)}$Facult\'e des sciences, Universit\'e Mohammed V, Rabat;$^{(f)}$Institute of Applied Physics, Mohammed VI Polytechnic University, Ben Guerir; Morocco.\\
% ...
$^{44}$Physics Department, Southern Methodist University, Dallas TX; United States of America.\\
% ...
$^{48}$Fakult\"{a}t Physik, Technische Universit{\"a}t Dortmund, Dortmund; Germany.\\
% ...
$^{54}$II. Physikalisches Institut, Georg-August-Universit\"{a}t G\"ottingen, G\"ottingen; Germany.\\
% ...
$^{83}$$^{(a)}$Khalifa University of Science and Technology, Abu Dhabi;$^{(b)}$New York University Abu Dhabi, Abu Dhabi;$^{(c)}$United Arab Emirates University, Al Ain;$^{(d)}$University of Sharjah, Sharjah; United Arab Emirates.\\
% ...
$^{102}$CPPM, Aix-Marseille Universit\'e, CNRS/IN2P3, Marseille; France.\\
% ...
$^{121}$Homer L. Dodge Department of Physics and Astronomy, University of Oklahoma, Norman OK; United States of America.\\
% ...
$^{154}$Raymond and Beverly Sackler School of Physics and Astronomy, Tel Aviv University, Tel Aviv; Israel.\\
% ...
$^{a}$ Also at Affiliated with an institute formerly covered by a cooperation agreement with CERN.\\
$^{b}$ Also at An-Najah National University, Nablus; Palestine.\\
$^{c}$ Also at Borough of Manhattan Community College, City University of New York, New York NY; United States of America.\\
$^{d}$ Also at Center for Interdisciplinary Research and Innovation (CIRI-AUTH), Thessaloniki; Greece.\\
$^{e}$ Also at Centre of Physics of the Universities of Minho and Porto (CF-UM-UP); Portugal.\\
$^{f}$ Also at CERN, Geneva; Switzerland.\\
% ...

\end{flushleft}


%

%
%

%

%
%
%
%

\end{document}